\documentclass[12pt,british,pra,reprint]{revtex4-1}
\usepackage[T1]{fontenc}
\usepackage[latin9]{inputenc}
\usepackage{babel}
\usepackage{color}
\usepackage{mathrsfs}
\usepackage{amsmath}
\usepackage{amssymb}
\usepackage{esint}
\usepackage[unicode=true,pdfusetitle,
bookmarks=true,bookmarksnumbered=false,bookmarksopen=false,
breaklinks=false,pdfborder={0 0 1},backref=false,colorlinks=false]
{hyperref}
\usepackage{breakurl}

\makeatletter
\@ifundefined{textcolor}{}
{%
	\definecolor{BLACK}{gray}{0}
	\definecolor{WHITE}{gray}{1}
	\definecolor{RED}{rgb}{1,0,0}
	\definecolor{GREEN}{rgb}{0,1,0}
	\definecolor{BLUE}{rgb}{0,0,1}
	\definecolor{CYAN}{cmyk}{1,0,0,0}
	\definecolor{MAGENTA}{cmyk}{0,1,0,0}
	\definecolor{YELLOW}{cmyk}{0,0,1,0}
}

\makeatother
\usepackage[matrix,frame,arrow]{xy}
\usepackage{amsmath}
\newcommand{\bra}[1]{\left\langle{#1}\right\vert}
\newcommand{\ket}[1]{\left\vert{#1}\right\rangle}
\newcommand{\qw}[1][-1]{\ar @{-} [0,#1]}

\newcommand{\gate}[1]{*{\xy *+<.6em>{#1};p\save+LU;+RU **\dir{-}\restore\save+RU;+RD **\dir{-}\restore\save+RD;+LD **\dir{-}\restore\POS+LD;+LU **\dir{-}\endxy} \qw}

\newcommand{\measureD}[1]{*{\xy*+=+<.5em>{\vphantom{\rule{0em}{.1em}#1}}*\cir{r_l};p\save*!R{#1} \restore\save+UC;+UC-<.5em,0em>*!R{\hphantom{#1}}+L **\dir{-} \restore\save+DC;+DC-<.5em,0em>*!R{\hphantom{#1}}+L **\dir{-} \restore\POS+UC-<.5em,0em>*!R{\hphantom{#1}}+L;+DC-<.5em,0em>*!R{\hphantom{#1}}+L **\dir{-} \endxy} \qw}

\newcommand{\multigate}[2]{*+<1em,.9em>{\hphantom{#2}} \qw \POS[0,0].[#1,0];p !C *{#2},p \save+LU;+RU **\dir{-}\restore\save+RU;+RD **\dir{-}\restore\save+RD;+LD **\dir{-}\restore\save+LD;+LU **\dir{-}\restore}
\newcommand{\ghost}[1]{*+<1em,.9em>{\hphantom{#1}} \qw}

\newcommand{\Qcircuit}[1][0em]{\xymatrix @*[o] @*=<#1>}  
 \renewcommand{\Qcircuit}[1][0em]{\xymatrix @*=<#1>}

\newcommand{\pureghost}[1]{*+<1em,.9em>{\hphantom{#1}}}
\newcommand{\multiprepareC}[2]{*+<1em,.9em>{\hphantom{#2}}\save[0,0].[#1,0];p\save !C
  *{#2},p+RU+<0em,0em>;+LU+<+.8em,0em> **\dir{-}\restore\save +RD;+RU **\dir{-}\restore\save
  +RD;+LD+<.8em,0em> **\dir{-} \restore\save +LD+<0em,.8em>;+LU-<0em,.8em> **\dir{-} \restore \POS
  !UL*!UL{\cir<.9em>{u_r}};!DL*!DL{\cir<.9em>{l_u}}\restore}
\newcommand{\prepareC}[1]{*{\xy*+=+<.5em>{\vphantom{#1\rule{0em}{.1em}}}*\cir{l^r};p\save*!L{#1} \restore\save+UC;+UC+<.5em,0em>*!L{\hphantom{#1}}+R **\dir{-} \restore\save+DC;+DC+<.5em,0em>*!L{\hphantom{#1}}+R **\dir{-} \restore\POS+UC+<.5em,0em>*!L{\hphantom{#1}}+R;+DC+<.5em,0em>*!L{\hphantom{#1}}+R **\dir{-} \endxy}}
\newcommand{\poloFantasmaCn}[1]{{{}^{#1}_{\phantom{#1}}}}

\makeatletter

\newcommand{\R}{\mathbb{R}}
\newcommand{\Comp}{\mathbb{C}}

\newcommand{\set}[1]{\mathsf{#1}}

\newcommand{\spc}[1]{\mathcal{#1}}

\def\d{{\rm d}}

\newcommand{\Span}{{\mathsf{Span}}}

\def\>{\rangle}
\def\<{\langle}

\newcommand{\st}[1]{\mathbf{#1}}
\newcommand{\bs}[1]{\boldsymbol{#1}}     

\newcommand{\map}[1]{\mathcal{#1}}
\newcommand{\Tr}{\operatorname{Tr}}

\newcommand{\St}{{\mathsf{St}}}
\newcommand{\Eff}{{\mathsf{Eff}}}
\newcommand{\Pur}{{\mathsf{Pur}}}
\newcommand{\Transf}{{\mathsf{Transf}}}
\newcommand{\Obs}{{\mathsf{Obs}}}

\newtheorem{theo}{Theorem}
\newtheorem{lemma}{Lemma}
\newtheorem{cor}{Corollary}
\newtheorem{defi}{Definition}
\newtheorem{ax}{Axiom}
\newtheorem{ass}{Assumption}
\newtheorem{prop}{Proposition}

\newtheorem{example}{Example}

\newtheorem{conj}{Conjecture}

\def\Proof{{\bf Proof.~}}
\def\qed{$\blacksquare$ \newline}

\newcommand{\rA}{\mathrm{A}}
\newcommand{\rB}{\mathrm{B}}
\newcommand{\rC}{\mathrm{C}}

\newcommand{\rE}{\mathrm{E}}

\newcommand{\rS}{\mathrm{S}}

\newcommand{\cA}{\mathcal{A}}
\newcommand{\cB}{\mathcal{B}}
\newcommand{\cC}{\mathcal{C}}

\newcommand{\cU}{\mathcal{U}}

\makeatother

\addto\captionsamerican{}
\addto\captionsamerican{}
\addto\captionsamerican{}
\addto\captionsamerican{}
\addto\captionsamerican{}
\addto\captionsamerican{}
\addto\captionsamerican{}
\addto\captionsamerican{}
\addto\captionsbritish{}
\addto\captionsbritish{}
\addto\captionsbritish{}
\addto\captionsbritish{}
\addto\captionsbritish{}
\addto\captionsbritish{}
\addto\captionsbritish{}
\addto\captionsbritish{}

\begin{document}

%
\title{Entanglement as an axiomatic   foundation for statistical mechanics}

\author{Giulio Chiribella}
\email{giulio@cs.hku.hk}
\affiliation{Department of Computer Science, University of Hong Kong, Pokfulam Road, Hong Kong}
\affiliation{Canadian Institute for Advanced Research,
CIFAR Program in Quantum Information Science, Toronto, ON M5G 1Z8}
\author{Carlo Maria Scandolo}
\email{carlomaria.scandolo@cs.ox.ac.uk}
\affiliation{Department of Computer Science, University of Oxford, Parks Road, Oxford, UK}

\begin{abstract}
We propose four information-theoretic axioms for the foundations of statistical mechanics   in general physical theories. 
   The  axioms---Causality, Purity Preservation, Pure Sharpness, and Purification---identify a class of theories where every mixed state can be modelled as the marginal of a pure entangled  state and   where every unsharp measurement can be  modelled as  a sharp measurement  on a composite system.  This class of theories---called \emph{sharp theories with purification}--- includes quantum theory both with complex and real amplitudes,  as well as a suitable extension of classical probability theory where classical systems can be entangled with other, non-classical systems. Theories satisfying our axioms  support well-behaved notions of majorization, entropy, and Gibbs states, allowing for    an information-theoretic derivation of Landauer's principle. We conjecture that every theory admitting a sensible thermodynamics must be extendable  to a sharp  theory with purification. 
    	\end{abstract}
\maketitle

\section{Introduction}

Thermodynamics is an extremely successful discipline, with   applications ranging from engineering to astrophysics, biology, computation, down to the microscopic regime of molecular engines. 
   When it was first formulated,  thermodynamics introduced  notions, such as heat, entropy, and temperature---which had no dynamical explanation.   One of the first questions arising was: are these notions fundamental or derived?  
The answer was far from obvious.   Many early theories leaned towards the view that heat and temperature are  primitive  notions: for example,  Lavoisier's caloric theory treated heat as a material fluid  and temperature as a sort of potential energy governing the dynamics of this fluid  \cite{Fox-caloric}.   This view was quite influential, leading e.g.\ to Carnot's result about the maximal efficiency of thermal cycles.  The very name thermodynamics is  a reminiscent of this early view.     
Later, the classic works of Maxwell  \cite{Maxwell-1,Maxwell-2}, Boltzmann \cite{Boltzmann}, and Gibbs \cite{Gibbs} undertook  a reduction of the laws of thermodynamics to the laws of an underlying dynamics of particles and fields.  This reduction led to the establishment of statistical mechanics as the foundations for  thermodynamics.

The statistical paradigm led to new applications  as well as new questions.  Statistical mechanics itself needed a foundation, the central question being   how to reconcile the use of statistical notions, associated with the incomplete knowledge of an agent, with the picture of nature provided by classical mechanics,  where there is no place for ignorance at the fundamental level.  Different proposals have been made for the foundations of classical statistical mechanics, the best known of which are ergodic theory \cite{Birkhoff-ergodic,vonNeumann-ergodic,Ergodic1,Ergodic2} and Jaynes maximum entropy  approach  \cite{Jaynes1,Jaynes2}.  Despite the successes of both approaches, however,  a tension remains between the statistical character of the thermodynamical laws and the deterministic character of classical kinematics.  Referring to this tension, Deffner and Zurek  described the classical attempts to bridge statistics and dynamics as a ``half way house, populated by fictitious but useful concepts such as \emph{ensembles}'' \cite{Zurek}.    In this scenario, quantum theory offers a radically new opportunity.  As originally noted by Schr\"odinger \cite{Schrodinger},  a system and its environment can be jointly in a pure state, whereas the system is individually in a mixed state. Here the mixed state does not represent an ensemble of identical systems, but rather the state of a single quantum system.      Based on this idea, Popescu, Short, and Winter \cite{Popescu-Short-Winter} and  Goldstein, Lebowitz, Tumulka, and Zangh\`i   \cite{Canonical-typicality} proposed that  entanglement can be the starting point for a new, genuinely quantum foundation of statistical mechanics.  The  idea was that, when the environment is large enough, the system is approximately in the equilibrium state for the typical  joint pure states of the system and the environment.   This idea  has been explored in a variety of settings \cite{Bocchieri,Seth-Lloyd,Lubkin,Gemmer-Otte-Mahler,Mahler-book,Concentration-measure,BrandaoQIP2015,Thermo-recent}, including also some potential post-quantum scenarios \cite{Dahlsten,Muller-blackhole}.      More recently, Deffner and Zurek put forward  an alternative approach  proposing  that the equilibrium state of the system should  be derived from symmetries of entanglement  enforced at the dynamical level \cite{Zurek}. Besides the differences, both the approach based on symmetries and the approach based on typicality share a common inspiration in the idea that  quantum entanglement  can provide a new foundation to statistical mechanics, and ultimately, thermodynamics.  

In this paper we   turn entanglement into an axiomatic foundation for statistical mechanics  in general  physical theories.       We explore the hypothesis that the physical systems admitting a well-behaved statistical mechanics are exactly those where---\emph{at least in principle}---mixed states can be modelled as the local  states of  larger systems, globally in a pure  state.    This modelling is possible in quantum theory, where it  provides the  stepping stone for the derivation of the microcanonical and canonical states in Refs.~\cite{Popescu-Short-Winter,Canonical-typicality} and \cite{Zurek}.    But the foundational role of entanglement is not limited to quantum theory.  We will show in this paper that  even classical statistical mechanics, where entanglement is absent,  can find a new foundation  if   classical  theory is regarded as part of a larger physical theory where classical mixed states can be obtained as marginals of pure  states of non-classical composite systems.  Remarkably, the mere fact that classical systems \emph{could} be entangled with some other physical systems   determines some of their properties and opens the way to the use of typicality/symmetry arguments as in the quantum case~\cite{Popescu-Short-Winter,Canonical-typicality,Zurek}.   The same approach is  applicable to several extensions of quantum theory, including quantum theory with superselection rules \cite{Preskill-superselection,Fermionic1,Fermionic2,Purity}, and variants of quantum theory with real amplitudes  \cite{Stuckelberg,Araki-real,Wootters-real,Hardy-real,QuantumFromPrinciples}.

Let us give an overview of the methodology and  contents of  the paper.    Our results are obtained in the framework  of general probabilistic theories (GPTs) \cite{Hardy-informational-1,D'Ariano,Barrett,Barnum-1,Chiribella-purification,Chiribella-informational,Barnum-2,Hardy-informational-2,hardy2011,hardy2013,Chiribella14,QuantumFromPrinciples,chiribella2016quantum},  which allows one to treat  quantum theory, classical theory, and a variety of hypothetical post-quantum theories in a unified way.     In this framework we demand the validity of four  information-theoretic axioms, informally stated as follows:
 \begin{enumerate}
 \item \emph{Causality.}  No signal can be sent from the future to the past. 
 \item \emph{Purity Preservation.}  The composition of two pure transformations is a pure transformation.
 \item \emph{Pure Sharpness.}    Every system has at least one pure   sharp observable. 
 \item \emph{Purification.}  Every state can be modelled as the marginal of a pure state. Such a modelling is unique up to local reversible transformations.  
 \end{enumerate}    
This   combination of axioms has been first used in our earlier work \cite{QPL15}, where we proved that axioms 1--4 imply that every state can be diagonalized---i.e.\ decomposed into a random mixture of perfectly distinguishable pure states.    Here we propose axioms 1--4 as an   axiomatic platform for the information-theoretic foundation of statistical mechanics.    

We call a GPT satisfying axioms 1--4 a \emph{sharp theory with purification}.   The central part of our paper is the study of the kinematics and measurement theory in  sharp theories with purification.   Two crucial results are \emph{i)}   the validity of the No Disturbance Without Information  property \cite{Chiribella-informational,Wehner-Pfister,Chiribella-Yuan2014,Chiribella-Yuan2015}, stating that measurements that extract no information about certain sources of states can be performed in a non-disturbing way, and \emph{ii)} the existence of a one-to-one correspondence between states and effects.  These results are used to prove an operational version of the Schmidt decomposition,  to prove that  that the probabilities in the diagonalization of  a state depend only on the state and not on the specific diagonalization,  and to formulate a majorization criterion for the thermodynamical conversion of mixed states.  Using these properties, we introduce a family of well-behaved entropy measures that are monotonically increasing under thermodynamic transformations. Among them, a GPT version of the von Neumann entropy \cite{Entropy-Barnum,Entropy-Short,Entropy-Kimura,Chiribella-Scandolo-entanglement,Krumm-Muller,Krumm-thesis} turns out to have appealing properties, such as the validity of a subadditivity relation.   Leveraging  these properties, we formulate a GPT version of the \emph{maximum entropy principle} and we introduce a GPT version of the Gibbs state.    As a concrete thermodynamic application, we produce an information-theoretic derivation of Landauer's principle, valid in arbitrary sharp theories with purification.

It is important to emphasize what we are \emph{not} assuming in this paper:  First, we do  not assume the Local Tomography axiom \cite{Hardy-informational-1,D'Ariano,Chiribella-purification,QuantumFromPrinciples}, stating that the state of a generic composite system is completely determined by the correlations of the local measurements performed on its subsystems.  Quantum theory with real amplitudes is an easy example of theory that satisfies our axioms but violates Local Tomography.   Our extended version of classical theory will be another example of non-locally-tomographic sharp theory with purification.      Second, we do not assume the  Strong Symmetry axiom  \cite{Barnum-interference}---the requirement that every maximal set of perfectly distinguishable states can be transformed reversibly into any other such set---nor do we assume the Permutability axiom \cite{Hardy-informational-2,hardy2013}---corresponding to the weaker requirement that every permutation of the states in a maximal set can be implemented reversibly.   Strong Symmetry and Permutability are generally different requirements, but turn out to be equivalent in the context of sharp theories with purification \cite{Purity}.  The implications of Strong Symmetry/Permutability for the foundations of thermodynamics  have been discussed in  a series of recent works  \cite{colleagues,Purity, QPL15, Krumm-Muller, Scandolo, Krumm-thesis}.  In this respect, the contribution of this paper is to identify a number of results that are logically independent of Strong Symmetry/Permutability.   Not having  is Strong Symmetry/Permutability among our axioms is important to our programme because it gives us the freedom to regard classical theory as part of a larger theory including both classical and non-classical bits.  Such theory must necessarily violate Strong Symmetry  (and hence Permutability, if we assume our axioms), because it is impossible to reversibly convert a non-classical bit into a classical bit, despite the two systems have maximal sets of the same size.    In the same vein, we do not assume Hardy's subspace axiom \cite{Hardy-informational-1, Brukner, Masanes-physical-derivation}, which in the context of sharp theories with purification is equivalent to Permutability/Strong Symmetry, nor do we assume the  Ideal Compression axiom of Ref.~\cite{Chiribella-informational}, whose relation with our axioms is still an open question.

The paper is structured as follows: in section~\ref{sec:Framework} we give a quick introduction to the formalism of GPTs, and in section~\ref{sec:Axioms} we present the axioms defining a sharp theory with purification. In section~\ref{sec:codit} we show how to accommodate classical theory in the framework of sharp theories with purification by introducing the theory of coherent dits.   The kinematics and measurement theory of sharp theory with purification is developed in section~\ref{sec:pre-diagonalization}, while section \ref{sec:puriresource}  is dedicated to the notion of majorization in sharp theories with purification.    In section \ref{sec:monotones} we define the mixedness monotones (functions of state that are non-decreasing under thermodynamical transformations) and derive their properties.  The special case of the von Neumman entropy is discussed in section \ref{sec:vN} and applied in section \ref{sec:landauer} to the  formulation of the maximum entropy principle, the definition of the Gibbs state, and the derivation of Landauer's principle.  
The conclusions are drawn in section \ref{sec:conclusion}.

\section{Framework\label{sec:Framework}}

Our results are expressed in  the framework of general probabilistic
theories, adopting the specific variant of Refs.~\cite{Chiribella-purification,Chiribella-informational,Chiribella14,QuantumFromPrinciples},
known as  \emph{operational-probabilistic theories}
(\emph{OPTs}), and based on a circuit framework that generalizes the framework of quantum circuits to general theories.  
  Here we give a quick recap
of the framework, referring the reader to the original papers and
to the related work by Hardy \cite{Hardy-informational-2,hardy2013}
for a more in-depth presentation.

\subsection{States, transformations, and effects}

Physical processes can be combined in sequence or in parallel, giving
rise to circuits like the following
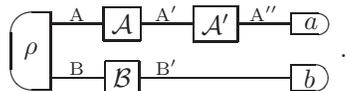
\[
\begin{aligned}\Qcircuit @C=1em @R=.7em @!R { & \multiprepareC{1}{\rho} & \qw \poloFantasmaCn{\rA} & \gate{\cA} & \qw \poloFantasmaCn{\rA'} & \gate{\cA'} & \qw \poloFantasmaCn{\rA''} &\measureD{a} \\ & \pureghost{\rho} & \qw \poloFantasmaCn{\rB} & \gate{\cB} & \qw \poloFantasmaCn{\rB'} &\qw &\qw &\measureD{b} }\end{aligned}~.
\]Here, $\mathrm{A}$, $\mathrm{A}'$, $\mathrm{A}''$, $\mathrm{B}$,
$\mathrm{B}'$ are \emph{systems}, $\rho$ is a bipartite \emph{state},
$\mathcal{A}$, $\mathcal{A}'$ and $\mathcal{B}$ are \emph{transformations},
$a$ and $b$ are \emph{effects}. Circuits with no external wires,
like the one in the above example, are associated with probabilities.
We denote by 
\begin{itemize}
	\item $ \mathsf{Sys} $ the set of all physical systems
	\item $\mathsf{St}\left(\mathrm{A}\right)$ the set of states of system
	$\mathrm{A}$ 
	\item $\mathsf{Eff}\left(\mathrm{A}\right)$ the set of effects on $\mathrm{A}$ 
	\item $\mathsf{Transf}\left(\mathrm{A},\mathrm{B}\right)$ the set of transformations
	from $\mathrm{A}$ to $\mathrm{B}$ 
	\item $\mathrm{A}\otimes\mathrm{B}$ the composition of systems $\mathrm{A}$
	and $\mathrm{B}$. 
	\item $\mathcal{A}\otimes\mathcal{B}$ the parallel composition of the transformations
	$\mathcal{A}$ and $\mathcal{B}$. 
\end{itemize}
Among the list of valid physical systems, OPTs include a particular system, the trivial system $ \mathrm{I} $, corresponding to the degrees of freedom ignored by the theory. The trivial system acts as a unit for the tensor product: for every system $ \mathrm{A} $, one has $\mathrm{I}\otimes\mathrm{A}=\mathrm{A}\otimes\mathrm{I}=\mathrm{A} $. States (resp.\ effects) are transformations
with the trivial system as input (resp.\ output). We will often make
use of the short-hand notation $\left(a|\rho\right)$ to denote the
scalar\[
\left(a|\rho\right)~:=\!\!\!\!\begin{aligned}\Qcircuit @C=1em @R=.7em @!R { & \prepareC{\rho}    & \qw \poloFantasmaCn{\rA}  &\measureD{a}}\end{aligned}~,
\]and of the notation $\left(a\right|\mathcal{C}\left|\rho\right)$
to mean\[
\left(a\right|\cC\left|\rho\right)~:=\!\!\!\!\begin{aligned}\Qcircuit @C=1em @R=.7em @!R { & \prepareC{\rho}    & \qw \poloFantasmaCn{\rA}  &\gate{\cC}  &\qw \poloFantasmaCn{\rB} &\measureD{a}}\end{aligned}~.
\]     A rigorous  mathematical foundation of the circuit framework in GPTs is provided by  the graphical languages of  symmetric monoidal categories \cite{Abramsky2004,cqm,Coecke-Picturalism,Categories-physicist,Selinger,Coecke2016}.   

We identify the scalar $\left(a|\rho\right)$ with a real number in
the interval $\left[0,1\right]$, representing the probability of
a joint occurrence of the state $\rho$ and the effect $a$ in a circuit
where suitable non-deterministic elements are put in place. The fact
that scalars are real numbers induces a notion of sum for transformations,
whereby the sets $\mathsf{St}\left(\mathrm{A}\right)$, $\mathsf{Transf}\left(\mathrm{A},\mathrm{B}\right)$,
and $\mathsf{Eff}\left(\mathrm{A}\right)$ become spanning sets of
vector spaces over the real numbers, denoted as $\mathsf{St}_{\mathbb{R}}\left(\mathrm{A}\right)$,
$\mathsf{Transf}_{\mathbb{R}}\left(\mathrm{A},\mathrm{B}\right)$,
and $\mathsf{Eff}_{\mathbb{R}}\left(\mathrm{A}\right)$ respectively.
In this paper we will restrict our attention to finite systems, i.e.\ systems
$\mathrm{A}$ for which the vector spaces $\mathsf{St}_{\mathbb{R}}\left(\mathrm{A}\right)$
and $\mathsf{Eff}_{\mathbb{R}}\left(\mathrm{A}\right)$ are finite-dimensional.
Also, it will be assumed as a default that the sets $\mathsf{St}\left(\mathrm{A}\right)$,
$\mathsf{Transf}\left(\mathrm{A},\mathrm{B}\right)$, and $\mathsf{Eff}\left(\mathrm{A}\right)$
are compact in the topology induced by probabilities, by which one
has $\lim_{n\rightarrow+\infty}\mathcal{C}_{n}=\mathcal{C}$, where
$\mathcal{C}_{n},\mathcal{C}\in\mathsf{Transf}\left(\mathrm{A},\mathrm{B}\right)$,
if and only if 
\[
\lim_{n\rightarrow+\infty}\left(E\right|\mathcal{C}_{n}\otimes\mathcal{I}_{\mathrm{R}}\left|\rho\right)=\left(E\right|\mathcal{C}\otimes\mathcal{I}_{\mathrm{R}}\left|\rho\right),
\]
for all systems $\mathrm{R}  $, all states $ \rho\in\mathsf{St}\left(\mathrm{A}\otimes\mathrm{R}\right) $, and all effects $ E\in\mathsf{Eff}\left(\mathrm{B}\otimes\mathrm{R}\right) $.

\subsection{Tests}

A \emph{test} from $\mathrm{A}$ to $\mathrm{B}$ is a collection
of transformations $\left\{ \mathcal{C}_{i}\right\} _{i\in\mathsf{X}}$
from $\mathrm{A}$ to $\mathrm{B}$, which can occur in an experiment
with outcomes in $\mathsf{X}$. If $\mathrm{A}$ (resp.\ $\mathrm{B}$)
is the trivial system, the test is called a \emph{preparation-test}
(resp.\ \emph{observation-test}). We stress that not all the collections
of transformations are tests: the specification of the collections
that are to be regarded as tests is part of the theory, the only requirement
being that the set of test is closed under parallel and sequential
composition.

If $\mathsf{X}$ contains a single outcome, we say that the test is
\emph{deterministic}. We will refer to deterministic transformations
as \emph{channels}. Following the most recent version of the formalism
\cite{Chiribella14}, we assume as part of the framework that every
test arises from an observation-test performed on one of the outputs
of a channel. The motivation for such an assumption is the idea that
the readout of the outcome could be interpreted physically as a measurement
allowed by the theory. Precisely, the assumption is the following.
\begin{ass}[Physicalization of Readout \cite{Chiribella14}]
	\label{assu:read}For every pair of systems $\mathrm{A}$, $\mathrm{B}$,
	and every test $\left\{ \mathcal{M}_{i}\right\} _{i\in\mathsf{X}}$
	from $\mathrm{A}$ to $\mathrm{B}$, there exist a system $\mathrm{C}$,
	a channel $\mathcal{M}\in\mathsf{Transf}\left(\mathrm{A},\mathrm{B}\otimes\mathrm{C}\right)$,
	and an observation-test $\left\{ c_{i}\right\} _{i\in\mathsf{X}}\subset\mathsf{Eff}\left(\mathrm{C}\right)$
	such that\[
	\begin{aligned} \Qcircuit @C=1em @R=.7em @!R { & \qw \poloFantasmaCn{\rA}& \gate{\mathcal{M}_{i} } & \qw \poloFantasmaCn{\rB} &\qw }\end{aligned} ~=~\begin{aligned} \Qcircuit @C=1em @R=.7em @!R { & \qw \poloFantasmaCn{\rA}& \multigate{1}{\mathcal{M}} & \qw \poloFantasmaCn{\rB} & \qw \\ & & \pureghost{\mathcal M} & \qw \poloFantasmaCn{\rC} & \measureD{c_i}} \end{aligned} \qquad \forall i\in\mathsf{X}.
	\]
\end{ass}
A channel $\mathcal{U}$ from $\mathrm{A}$ to $\mathrm{B}$ is called
\emph{reversible} if there exists a channel $\mathcal{U}^{-1}$ from
$\mathrm{B}$ to $\mathrm{A}$ such that $\mathcal{U}^{-1}\mathcal{U}=\mathcal{I}_{\mathrm{A}}$
and $\mathcal{U}\mathcal{U}^{-1}=\mathcal{I}_{\mathrm{B}}$, where
$\mathcal{I}_{\mathrm{S}}$ is the identity channel on a generic system
$\mathrm{S}$. If there exists a reversible channel transforming $\mathrm{A}$
into $\mathrm{B}$, we say that $\mathrm{A}$ and $\mathrm{B}$ are
\emph{operationally equivalent}, denoted by $\mathrm{A}\simeq\mathrm{B}$.
The composition of systems is required to be \emph{symmetric}, meaning
that $\mathrm{A}\otimes\mathrm{B}\simeq\mathrm{B}\otimes\mathrm{A}$.
Physically, this means that for every pair of systems there exists a reversible channel that swaps them. 

A state $\chi\in\mathsf{St}\left(\mathrm{A}\right)$ is called \emph{invariant}
if $\mathcal{U}\chi=\chi$, for every reversible channel $\mathcal{U}$.
Note that, in general, invariant states may not exist. In this paper
their existence will be a consequence of the axioms and of a standing
assumption of finite-dimensionality, adopted throughout our work.

\subsection{Pure transformations}

The probabilistic structure offers an easy way to define pure
transformations. The definition is based on the notion of \emph{coarse-graining},
i.e.\ the operation of joining two or more outcomes of a test into
a single outcome. More precisely, a test $\left\{ \mathcal{C}_{i}\right\} _{i\in\mathsf{X}}$
is a \emph{coarse-graining} of the test $\left\{ \mathcal{D}_{j}\right\} _{j\in\mathsf{Y}}$
if there is a partition $\left\{ \mathsf{Y}_{i}\right\} _{i\in\mathsf{X}}$
of $\mathsf{Y}$ such that $\mathcal{C}_{i}=\sum_{j\in\mathsf{Y}_{i}}\mathcal{D}_{j}$
for every $i\in\mathsf{X}$. In this case, we say that $\left\{ \mathcal{D}_{j}\right\} _{j\in\mathsf{Y}}$
is a \emph{refinement} of $\left\{ \mathcal{C}_{i}\right\} _{i\in\mathsf{X}}$.
The refinement of a given transformation is defined via the refinement
of a test: if $\left\{ \mathcal{D}_{j}\right\} _{j\in\mathsf{Y}}$
is a refinement of $\left\{ \mathcal{C}_{i}\right\} _{i\in\mathsf{X}}$,
then the transformations $\left\{ \mathcal{D}_{j}\right\} _{j\in\mathsf{Y}_{i}}$
are a refinement of the transformation $\mathcal{C}_{i}$.

A transformation $\mathcal{C}\in\mathsf{Transf}(\mathrm{A},\mathrm{B})$
is called \emph{pure} if it has only trivial refinements, namely for
every refinement $\left\{ \mathcal{D}_{j}\right\} $ one has $\mathcal{D}_{j}=p_{j}\mathcal{C}$,
where $\left\{ p_{j}\right\} $ is a probability distribution. Pure
transformations are those for which the experimenter has maximal information
about the evolution of the system. We denote the set of pure transformations
from $\mathrm{A}$ to $\mathrm{B}$ as $\mathsf{PurTransf}\left(\mathrm{A},\mathrm{B}\right)$.
In the special case of states (resp.\ effects) of system $\mathrm{A}$
we use the notation $\mathsf{PurSt}\left(\mathrm{A}\right)$ (resp.\ $\mathsf{PurEff}\left(\mathrm{A}\right)$). The set of pure observation-tests of system $ \mathrm{A} $ will be denoted as $\mathsf{PurObs}\left(\mathrm{A}\right)$.
As usual, non-pure states are called \emph{mixed}. 

The pairing between states and effects leads naturally to a notion
of norm. We define the norm of a state $\rho$ as $\left\Vert \rho\right\Vert :=\sup_{a\in\mathsf{Eff}\left(\mathrm{A}\right)}\left(a|\rho\right)$.  Similarly,
the norm of an effect $a$ is defined as $\left\Vert a\right\Vert :=\sup_{\rho\in\mathsf{St}\left(\mathrm{A}\right)}\left(a|\rho\right)$.
We will use a subscript 1 to denote the set of normalised (i.e.\ with unit norm) states and effects. For instance, the set of normalised states of $  \mathrm{A} $ will be denoted by $ \mathsf{St}_{1}\left(\mathrm{A}\right) $, and so on.
\begin{defi}
	Let $\rho$ be a normalized state. We say that a normalised state $\sigma$ is
	\emph{contained} in $\rho$ if we can write $\rho=p\sigma+\left(1-p\right)\tau$,
	where $p\in\left(0,1\right]$ and $\tau$ is another normalised state. 
\end{defi}
It is clear that no states are contained in a pure state, except the
pure state itself. We say that a state is  \emph{complete}  if  it contains every other state. 

\begin{defi}
	\label{def:upon input}We say that two transformations $\mathcal{A},\mathcal{A}'\in\mathsf{Transf}\left(\mathrm{A},\mathrm{B}\right)$
	are \emph{equal upon input} of the state $\rho\in\mathsf{St}_{1}\left(\mathrm{A}\right)$
	if $\mathcal{A}\sigma=\mathcal{A}'\sigma$ for every state $\sigma$
	contained in $\rho$. In this case we will write $\mathcal{A}=_{\rho}\mathcal{A}'$.
\end{defi}

\section{Sharp theories with purification\label{sec:Axioms}}
We propose four axioms for the foundation of  thermodynamics.  
  The first axioms is  Causality.  The axiom forbids signalling from
the future to the past:
\begin{ax}[Causality \cite{Chiribella-purification,Chiribella-informational}]
The outcome probabilities of a test do not depend on the choice of
other tests performed later in the circuit.
\end{ax}
Causality is equivalent to the requirement that, for every system
$\mathrm{A}$, there exists a unique deterministic effect $u_{\mathrm{A}}$
(or simply $u$, when no ambiguity can arise). In a causal theory
(i.e.\ a theory satisfying Causality), observation-tests are normalized
as follows (cf.\ corollary 3 of Ref.~\cite{Chiribella-purification}):
\begin{prop}
\label{prop:characterization observation-tests}In a causal theory,
if $\left\{ a_{i}\right\} _{i\in\mathsf{X}}$ is an observation-test,
then $\sum_{i\in\mathsf{X}}a_{i}=u$.\end{prop}

Thanks to the uniqueness of the deterministic effect, it is possible
to define the \emph{marginal state} of a bipartite state $\rho_{\mathrm{AB}}$
on system $\mathrm{A}$ as $\rho_{\mathrm{A}}:=\left(\mathcal{I}_{\mathrm{A}}\otimes u_{\mathrm{B}}\right)\rho_{\mathrm{AB}}=:\mathrm{Tr}_{\mathrm{B}}   [\rho_{\mathrm{AB}}]$, where we have chosen the ``trace notation''   in  
formal analogy with the notation in quantum theory.
In diagrams,\[
\mathrm{Tr}_{\mathrm{B}}  [\rho_{\mathrm{AB}}]~=\!\!\!\!\begin{aligned}\Qcircuit @C=1em @R=.7em @!R { & \multiprepareC{1}{\rho}    & \qw \poloFantasmaCn{\rA} &  \qw   \\  & \pureghost{\rho}    & \qw \poloFantasmaCn{\rB}  &   \measureD{u} }\end{aligned}~.
\]
The operation of taking the marginal of a state is widely used in thermodynamics as the procedure of restricting ourselves to a subsystem of a larger system. Causality ensures that such an operation is uniquely  defined, and this justifies the choice of Causality as one of the    axioms for  a sensible theory of thermodynamics.

In a causal theory the norm of a state can be defined as $ \left\| \rho\right\| = \left( u|\rho\right) =\mathrm{Tr}\,[\rho]$. In causal theories it is easy to prove that physical transformations are norm-non-increasing.

\begin{prop}	\label{prop:norm-transformations}In a causal theory, if $ \mathcal{A}\in \mathsf{Transf}\left(\mathrm{A},\mathrm{B} \right)  $, then for any state $ \rho\in\mathsf{St}\left(\mathrm{A} \right)  $, we have $  \left\|\mathcal{A} \rho\right\|\leq \left\|\rho\right\| $, and one has the equality if and only if $ \mathcal{A} $ is a channel.
	
	In particular $ \mathcal{A} $ is a channel if and only if $ u_{\mathrm{B}}\mathcal{A}=u_{\mathrm{A}} $.
	\end{prop}
\Proof 
The proof is an easy adaptation of lemma 1 of Ref.~\cite{Chiribella-purification}. 
\qed

Furthermore, Causality guarantees that it is consistent to assume that the choice
of a test can depend on the outcomes of previous tests---namely that
it is possible to perform \emph{conditional tests} \cite{Chiribella-purification}.  
Combined with the assumption of compactness, the ability to perform
conditional tests implies that every state is proportional to a normalized
state. Another consequence is that all the sets $\mathsf{St}\left(\mathrm{A}\right)$,
$\mathsf{Transf}\left(\mathrm{A},\mathrm{B}\right)$, and $\mathsf{Eff}\left(\mathrm{A}\right)$
are \emph{convex}. In the following we will take for granted the ability
to perform conditional tests, the fact that every state is proportional
to a normalized state, and the convexity of all the sets of transformations.

The second axiom in our list is Purity Preservation.
\begin{ax}[Purity Preservation \cite{Scandolo14}]
Sequential and parallel compositions of pure transformations are
pure transformations.
\end{ax}
We consider Purity Preservation as a fundamental requirement. Considering
the theory as an algorithm to make deductions about physical processes,
Purity Preservation ensures that, when presented with maximal information
about two processes, the algorithm outputs maximal information about
their composition \cite{Scandolo14}. Purity Preservation is very
close to a slightly weaker axiom introduced by D'Ariano in \cite{D'Ariano}, and used
in the axiomatization of \cite{Chiribella-informational}.   The axiom 
 is called Atomicity of Composition and stipulates that the \emph{sequential} composition of two pure transformations
is a pure transformation. Purity Preservation is stronger, in that
it requires the preservation of purity also for the \emph{parallel}
composition.

The third axiom is  Pure Sharpness \cite{QPL15}. This axiom ensures that there exists at least one elementary
property associated with every system.
\begin{ax}[Pure Sharpness \cite{QPL15}]
\label{axm:pure effect}For every system  there exists
at least one pure effect 
that occurs with unit probability  on some state.
\end{ax}
Pure Sharpness  is reminiscent of the Sharpness axiom used in Hardy's
2011 axiomatization \cite{Hardy-informational-2,hardy2013}, which
requires a one-to-one correspondence between pure states and effects
that distinguish maximal sets of states.
A similar axiom also appeared in  works by Wilce \cite{Wilce-spectral,Royal-road}, where he stipulates that for every effect there exists a \emph{unique} state on which it occurs with probability 1.

The last axiom is Purification. This principle characterizes
the physical theories admitting a description where all deterministic
processes are pure and reversible at the fundamental level. Essentially,
Purification expresses a strengthened version of the principle of
conservation of information \cite{Chiribella-educational,Scandolo14}, demanding  not only  that information be conserved, but also that randomness can always be modelled as due to the presence of some inaccessible degree of freedom.  
In its simplest form, Purification is phrased as a requirement about
\emph{causal} theories, where the marginal of a bipartite state is
defined in a canonical way. In this case, we say that a state $\rho\in\mathsf{St}_{1}\left(\mathrm{A}\right)$
can be purified if there exists a pure state $\Psi\in\mathsf{PurSt}_{1}\left(\mathrm{A}\otimes\mathrm{B}\right)$
that has $\rho$ as its marginal on system $\mathrm{A}$. In this
case, we call $\Psi$ a \emph{purification} of $\rho$ and $\mathrm{B}$
a \emph{purifying system}. The axiom is as follows.
\begin{ax}[Purification \cite{Chiribella-purification,Chiribella-informational}]
Every state can be purified. Every two purifications of the same
state, with the same purifying system, differ by a reversible channel
on the purifying system.
\end{ax}
The second part of the axiom states that, if $\Psi,\Psi'\in\mathsf{PurSt}_{1}\left(\mathrm{A}\otimes\mathrm{B}\right)$
are such that $\mathrm{Tr}_{\mathrm{B}}  [\Psi_{\mathrm{AB}}]=\mathrm{Tr}_{\mathrm{B}}[\Psi'_{\mathrm{AB}}]$,
then
\begin{align*}
\begin{aligned}\Qcircuit @C=1em @R=.7em @!R { & \multiprepareC{1}{\Psi'}    & \qw \poloFantasmaCn{\rA} &  \qw   \\  & \pureghost{\Psi'}    & \qw \poloFantasmaCn{\rB}  &   \qw }\end{aligned}~=\!\!\!\!\begin{aligned}\Qcircuit @C=1em @R=.7em @!R { & \multiprepareC{1}{\Psi}    & \qw \poloFantasmaCn{\rA} &  \qw & \qw & \qw  \\  & \pureghost{\Psi}    & \qw \poloFantasmaCn{\rB}  &   \gate{\cU}& \qw \poloFantasmaCn{\rB}& \qw  }\end{aligned}~,
\end{align*}
where $\mathcal{U}$ is a reversible channel on $\mathrm{B}$.

Recently also Spekkens' Toy Model has been shown to satisfy Purification \cite{Disilvestro}. Moreover, recently Purification has been applied to the construction of  controlled  transformations and  generalized phase kick-backs  \cite{Control-reversible}, and to Grover's algorithm in theories admitting higher-order interference \cite{Lee-Selby-Grover}. 
In quantum theory, the validity of Purification lies at the foundation
of dilation theorems, such Stinespring's \cite{Stinespring}, Naimark's
\cite{Naimark}, and Ozawa's \cite{Ozawa}. In the finite-dimensional
setting, these theorems have been
reconstructed axiomatically in \cite{Chiribella-purification} for states and channels and  have been postulated in \cite{Chiribella-Yuan2014,Chiribella-Yuan2015} for measurements. It is interesting to mention that, under our axioms 1-4 the operational version of Naimark's theorem for measurements can be explicitly derived, as we will show later in the paper.

Having presented our axioms, we are ready to define the class of theories studied in this paper 
\begin{defi}[Sharp theories with purification]
A theory satisfying Causality, Purity Preservation, Pure Sharpness, and Purification will be called a \emph{sharp theory with purification}.
\end{defi}

All the results in the rest of the paper apply to  sharp theories with purification. 
\section{Example: Extended Classical Theory\label{sec:codit}}

In this section we show that classical probability theory can be regarded as a part of a larger theory that includes non-classical systems, called \emph{coherent bits} in analogy with the similar notion in quantum Shannon theory \cite{Harrow}.  

\subsection{Coherent bits}  

Classical probability theory   can be operationally characterized as the theory where   
\begin{enumerate}
\item all the pure states of every system are jointly distinguishable, deterministically and without error
\item the pure states of a composite system  are the products of pure states of the component systems
\item all random mixtures of pure states are valid mixed states
\item all permutations of the set of pure states can be implemented reversibly.  
\end{enumerate}
It is relatively easy to check that classical theory satisfies Causality, Purity Preservation, and Pure Sharpness.    On the other hand,  classical theory     
 violates Purification in an obvious way:  no mixed state can be the marginal of a pure state.  
  
We now show that, despite the appearance, classical theory has a deep relation with Purification: unlike many other GPTs, classical theory can be extended to a larger theory that satisfies Purification.    
This is done by adding  some non-classical systems that can be entangled with classical systems, thus providing the  desired purifications.     In the two-dimensional case, we call the additional systems \emph{coherent bits}  (or \emph{cobits}) in analogy with the related notion introduced by Harrow in the context of quantum Shannon theory \cite{Harrow}.          
  For  systems of dimension $d$ we use the expression \emph{coherent dits} (or \emph{codits}).

Let us illustrate  the extension of classical theory in the two-dimensional case first.   Using the Hilbert space notation, the states of a classical bit (cbit) are  represented as  
\begin{equation}\label{cbitstate}  \rho   =    p  \ket{ 0} \bra{ 0}   \oplus    \left( 1-p\right)    \ket{ 1} \bra{ 1}   , \end{equation}
where $p$ is a probability and the direct sum sign   is a reminder that the off-diagonal elements are forbidden.   The composite system of  two classical bits $\rA$ and $\rB$ contains only states of the form  
\begin{align*} \rho_{\rA\rB}    & =   p_{00} \ket{ 0} \bra{ 0}    \otimes \ket{ 0} \bra{ 0}     \oplus   p_{01} \ket{ 0} \bra{ 0}    \otimes \ket{1} \bra{ 1}    \\
  & \quad    \oplus  p_{10} \ket{1} \bra{ 1}    \otimes \ket{ 0} \bra{ 0}     \oplus  p_{11} \ket{1} \bra{ 1}    \otimes \ket{1} \bra{ 1}          ,
  \end{align*}
where $p_{ij}$,  $i,j  \in  \left\{0,1\right\}$ are probabilities.    Now, the states of a cobit are the same as the states of the classical bit---they are still of the form of Eq.~\eqref{cbitstate}.   But the composition of a cbit $\rA$ with a cobit $\rB$ yields a composite system   with many more  states than  the composition of two classical bits:  we allow every joint state of the form
\begin{equation}\label{cbitcobit} 
\rho_{\rA\rB}  =       p_{\rm even}    \rho_{\rm even}  \oplus p_{\rm odd}  \rho_{\rm odd}   \,,     
\end{equation} 
where $  p_{\rm even}$ and $p_{\rm odd}$ are two probabilities and $\rho_{\rm even}$   (respectively, $\rho_{\rm odd}$) is a generic density matrix with support in the subspace 
\[
 \spc H_{\rm even}    : =  \Span  \left\{  \ket{0}\otimes \ket{0}  ,  \ket{1}\otimes \ket{1}\right\}  
     \]
(resp.\ $ \spc H_{\rm odd}     :=\Span  \left\{  \ket{0}\otimes \ket{1} ,  \ket{1}\otimes \ket{0}\right\}$). In particular, the pure states of the cbit-cobit composite can be represented as unit vectors, either  of the form  
\[  \ket{\Phi}  =   \alpha  \ket{0} \otimes \ket{0} +  \beta   \ket{1}\otimes \ket{1}   , \qquad \alpha, \beta \in \Comp  ,       \]
or of the form 
\[  \ket{\Psi}   =   \alpha  \ket{0}  \otimes \ket{1}  +  \beta    \ket{1}\otimes \ket{0}   , \qquad \alpha, \beta \in \Comp  .       \]

As joint operations on the composite system we allow all quantum channels (completely positive trace-preserving maps) that send states of the block-diagonal form~\eqref{cbitcobit} into states of the same form.    In particular, the reversible channels can be represented by unitary operators of the form  
 \[ U_{\rA\rB}    =     \left( X^m \otimes X^n\right)    \mathrm{e}^{- i \Theta}  ,  \]   
 where $X  =  \begin{pmatrix}    0  & 1 \\  1 &  0  \end{pmatrix}$ is the bit flip,  $m,n  \in \left\{  0, 1\right\}$, and $\Theta$ is an diagonal  operator of the form 
 \[   \Theta  =   \sum_{k,l}      \theta_{kl}   \ket{k} \bra{k}  \otimes \ket{l} \bra{l}   .  \]
 Note that the above expression includes unitary operators of the form  
 \begin{align}\nonumber       U_\rA\otimes I_\rB     &=  X^m   \mathrm{e}^{  -  i \theta     Z}  \otimes I_\rB   \\  
  \label{local}    I_\rA\otimes U_\rB   &=   I_\rA \otimes   X^n    \mathrm{e}^{  -  i \theta     Z}                      ,     \end{align}
  where $\theta \in  \left[   0,2\pi \right) $   is an angle, and  $Z  =  \begin{pmatrix}    1 & 0 \\   0   &  -1   \end{pmatrix}$ is the phase flip. 
  We interpret the unitary operators $ U_\rA $ and $U_\rB$ in Eq.~\eqref{local} as \emph{local operations} performed on systems $\rA$ and $\rB$, respectively.   Note that the phase $\mathrm{e}^{-i\theta Z}$  cannot be detected locally, but only when applied on one side of the composite  system $\rA\otimes\rB$. In other words, the operations $U_\rA$ and $U_\rB$ are locally implementable, but not locally distinguishable.  This situation is possible because the theory we are defining violates the Local Tomography axiom, thus leaving  room for operations that are indistinguishable  from their action on local states, but  are still different from one another.   Another example of this  situation arises in real vector space quantum theory, where there exist   local operations that are indistinguishable globally   \cite{Chiribella-purification}.       

With the above settings, it is easy to see that every state of a classical bit can be purified, using a coherent bit as purifying system.  For example, the generic  cbit state $\rho  =    p  \ket{0} \bra{0}  \oplus  \left( 1-p\right)   \ket{1} \bra{1}$  has the purification  
\[ \ket{\Phi} =  \sqrt p    \ket{0}\otimes\ket{0}     + \sqrt {1-p}   \ket{1}  \otimes \ket{1}  .\] 
In addition, it is possible to show that every two purifications of the same state differ by a local unitary operation on the purifying system. This includes, for example, the purification   
\[ \ket{\Psi}  =  \sqrt p   \ket{0}\otimes \ket{1}     + \sqrt {1-p}  \ket{1}  \otimes \ket{0}   ,\] 
    obtained from $|\Phi\>$ through the application of a bit flip on system $\rB$.  

The coherent bits defined here are related to the original notion of coherent bits as a  communication resource \cite{Harrow}, in that the transmission of a coherent bit in our sense is equivalent to the coherent communication of one bit in the sense of Harrow.  
    
\subsection{Coherent dits}  

The definition of cobit can be easily generalized to the case of systems with $d$ perfectly distinguishable states. The state of a single codit are the same of the state of a cdit (classical $d$-level system), namely  
\[  \rho  = \bigoplus_{x=0}^{d-1}   p_x  \ket{x} \bra{x}   .\]      
When a cdit and a codit are combined, the states of the composite systems are density matrices of the  block-diagonal form 
\[   \rho   =  \bigoplus_{\delta  =0}^{d-1}    q_\delta    \rho_\delta  ,\]
  where each $\rho_\delta$ is a density matrix with support in the subspace  
  \[\spc H_\delta   : = \Span \left\{   \ket{x} \otimes \ket{\left( x  +\delta\right)   \mod d} ~|~  x =    0,\dots,  d -1  \right\} \, .  \]
The joint evolutions on the tensor product are the quantum channels that preserve the block-diagonal states.  With this definition, every mixed state of a classical $d$-dimensional system can be purified to a state of a composite system, and the purification is unique up to local transformations on the coherent system used as purifying system.   
\subsection{The other composites}

So far we have defined how to compose classical bits with coherent bits. In order to have a full theory we need to specify all the other composites.  

We do it as follows:    The composte of a $d_\rA$-dimensional cbit $\rA$ with a $d_\rB$-dimensional codit $\rB$ has all the states  of the form  
\begin{align}\label{generalproduct} \rho  =  \bigoplus_{\delta =  0  }^{\max  \left\{d_\rA-1,  d_\rB-1\right\} }  q_\delta  \rho_\delta ,\end{align} 
with $\rho_\delta$ a generic density matrix with support in 
\begin{align*} \spc H_{\delta}    =     \Span  \left\{  \ket{x}  \otimes  \ket{\left( x+\delta\right)   \mod d_\rB  }    ~|~  x  =  0,\dots,  d_\rA  -1\right\}  
    \end{align*}
    for $d_\rA\le d_\rB$, and 
    \begin{align*}
 \spc H_\delta  =    \Span  \left\{  \ket{\left( y- \delta \right)  \mod  d_\rA  } \otimes  \ket{y}    ~|~  y  =  0,\dots,  d_\rB  -1\right\}  
    \end{align*}  
    for $d_\rB\le d_\rA$.   The allowed transformations are all the transformations that preserve states of the block-diagonal form.  
  
  The composite of two codits is defined in the same way as the composite of a cdit and a codit.  Note that     the product of a state space of the form~\eqref{generalproduct} can be written as  
  \[   \rho   \simeq    \sum_{\delta  }    q_\delta    \rho_{\delta}  \otimes   \ket{\delta} \bra{\delta}      ,     \]
  and interpreted as states of a composite system consisting of a quantum system and a  classical system.   This observation allows one to define further products by using the ordinary tensor product of quantum systems for the quantum part and the aforementioned rules for the classical part.    
  The allowed transformations are all the completely positive maps that preserve the block-diagonal form of the corresponding systems.

  \section{Kinematics and measurement theory in sharp theories with purification\label{sec:pre-diagonalization}}

Here we list a few properties of sharp theories with purification that are not directly related to thermodynamics, but provide the substrate on which the thermodynamic results of our paper are based.

\subsection{Transitivity and pure steering}

The easiest consequence of Purification is that reversible channels
act transitively on the set of pure states (cf.\ lemma 20 in Ref.~\cite{Chiribella-purification}): 
\begin{prop}[Transitivity on pure states]\label{prop:transitivity}
	For any pair of pure states $\psi,\psi'\in\mathsf{PurSt}_{1}\left(\mathrm{A}\right)$
	there exists a reversible channel $\mathcal{U}$ on $\mathrm{A}$
	such that $\psi'=\mathcal{U}\psi$.
\end{prop}

As a consequence, every finite-dimensional system possesses one invariant
state (cf.\ corollary 34 of Ref.~\cite{Chiribella-purification}):
\begin{prop}[Uniqueness of the invariant state]
	For every system $\mathrm{A}$, there exists a unique invariant state
	$\chi_{\mathrm{A}}$. 
	The invariant state is  complete.
\end{prop}

Also, transitivity implies that the set of pure states is compact
for every system (corollary 32 of Ref.~\cite{Chiribella-purification}).
This property is generally a non-trivial property---cf.\ Ref.~\cite{Entropy-Barnum}
for a counterexample of a state space with a non-closed set of pure
states.

A crucial consequence of Purification is the \emph{steering property}:
\begin{theo}[Pure Steering]
	Let $\rho\in\mathsf{St}_{1}\left(\mathrm{A}\right)$ and let $\Psi\in\mathsf{PurSt}_{1}\left(\mathrm{A}\otimes\mathrm{B}\right)$
	be a purification of $\rho$. Then $\sigma$ is contained in $\rho$
	if and only if there exist an effect $b_{\sigma}$ on the purifying
	system $\mathrm{B}$ and a non-zero probability $p$ such that\[
	p\!\!\!\!\begin{aligned}\Qcircuit @C=1em @R=.7em @!R { & \prepareC{\sigma} & \qw \poloFantasmaCn{\rA} & \qw }\end{aligned}~=\!\!\!\!\begin{aligned}\Qcircuit @C=1em @R=.7em @!R { & \multiprepareC{1}{\Psi} & \qw \poloFantasmaCn{\rA} & \qw \\ & \pureghost{\Psi} & \qw \poloFantasmaCn{\rB} & \measureD{b_{\sigma}} }\end{aligned}~.
	\]\end{theo}
\Proof 
	The proof follows the same lines of theorem 6 and corollary 9 in Ref.~\cite{Chiribella-purification},
	with the only difference that here we do not assume the existence
	of distinguishable states. In its place, we use the framework assumption~\ref{assu:read}  (Physicalization of Readout),
	which guarantees that the outcome of every test can be read out from
	a physical system.
 \qed
 
Purification also enables us to link equality upon input (as defined in definition~\ref{def:upon input})
to equality on purifications (cf.\ theorem 7 of Ref.~\cite{Chiribella-purification}): 
\begin{prop}
	\label{prop:purifications -> input}
	Let $\rho$ be a state of system $\rA$  and let $\Psi\in\mathsf{St}_{1}\left(\mathrm{A}\otimes\mathrm{B}\right)$
	be a purification of $\rho$.   Then, for every pair of transformations  $\mathcal{A}$ and $\mathcal{A}'$, transforming  $\mathrm{A}$ into $\mathrm{C}$, if
	\[  \begin{aligned}\Qcircuit @C=1em @R=.7em @!R { & \multiprepareC{1}{\Psi} & \qw \poloFantasmaCn{\rA} & \gate{\mathcal A}  &  \qw \poloFantasmaCn{\rC}&\qw  \\ & \pureghost{\Psi} & \qw \poloFantasmaCn{\rB} & \qw &\qw&\qw}\end{aligned} ~= \!\!\!\! \begin{aligned}\Qcircuit @C=1em @R=.7em @!R { & \multiprepareC{1}{\Psi} & \qw \poloFantasmaCn{\rA} & \gate{\mathcal A'} &  \qw \poloFantasmaCn{\rC} &\qw \\ & \pureghost{\Psi} & \qw \poloFantasmaCn{\rB} & \qw &\qw &\qw}\end{aligned}~,  \]
	then $ \mathcal{A}=_{\rho}\mathcal{A'} $.
	
	If system $\rC$ is trivial, then one has the full equivalence:   for every pair of effects $a$ and $a'$ 
	\[  \begin{aligned}\Qcircuit @C=1em @R=.7em @!R { & \multiprepareC{1}{\Psi} & \qw \poloFantasmaCn{\rA} & \measureD{a}   \\ & \pureghost{\Psi} & \qw \poloFantasmaCn{\rB} & \qw }\end{aligned} ~= \!\!\!\! \begin{aligned}\Qcircuit @C=1em @R=.7em @!R { & \multiprepareC{1}{\Psi} & \qw \poloFantasmaCn{\rA} & \measureD{a'} \\ & \pureghost{\Psi} & \qw \poloFantasmaCn{\rB} & \qw}\end{aligned}  \]
	if and only if $a=_{\rho} a'$
\end{prop}

Pure Steering guarantees the existence of pure dynamically faithful states, in the following sense:
\begin{defi}
	A state $\rho\in\mathsf{St}_{1}\left(\mathrm{A}\otimes\mathrm{B}\right)$
	is \emph{dynamically faithful} on system $\mathrm{A}$ if  for every system $\rC$ and for every pair of transformations $\mathcal A$ and $\mathcal A'$ transforming $\rA$ into $\rC$  \[
	\begin{aligned}\Qcircuit @C=1em @R=.7em @!R { & \multiprepareC{1}{\rho} & \qw \poloFantasmaCn{\rA} & \gate{\mathcal A}  &  \qw \poloFantasmaCn{\rC}&\qw  \\ & \pureghost{\rho} & \qw \poloFantasmaCn{\rB} & \qw &\qw&\qw}\end{aligned} ~= \!\!\!\! \begin{aligned}\Qcircuit @C=1em @R=.7em @!R { & \multiprepareC{1}{\rho} & \qw \poloFantasmaCn{\rA} & \gate{\mathcal A'} &  \qw \poloFantasmaCn{\rC} &\qw \\ & \pureghost{\rho} & \qw \poloFantasmaCn{\rB} & \qw &\qw &\qw}\end{aligned} \] 
	implies $ \mathcal A  =  \mathcal A' $.
	\end{defi}
Thanks to Pure Steering, we have the following characterization
\begin{prop}[Dynamical faithfulness]
	\label{prop:faithful-effects} A pure state $\Psi_{\mathrm{AB}}$ is
	dynamically faithful on system $\mathrm{A}$ if and only if its marginal
	$\omega_{\mathrm{A}}$ on $\mathrm{A}$ is complete.
\end{prop}
\Proof The proof is an adaptation of the arguments of theorems 8 and 9 of Ref.~\cite{Chiribella-purification}, which is valid even without invoking the Local Tomography axiom used therein. \qed

An indirect consequence of Pure Steering is a simple condition for a set of transformations to be a test (cf.\ theorem 18 of Ref.~\cite{Chiribella-purification}):  
\begin{prop}[Characterization of tests]
	\label{prop:sufficientfortest} 
	A set of transformations  $\left\{\mathcal A_i\right\}_{i=1}^n  \subset  \mathsf{Transf} \left(\rA,\rB\right)$ is a test if and only if $  \sum_{i=1}^{n} u_\rB\mathcal{A}_{i} =   u_\rA$.   Specifically, a set of effects $\left\{a_i\right\}_{i=1}^n$ is an observation-test if and only if  $\sum_{i=1}^n  a_i  =  u_\rA$. 
\end{prop}

\subsection{Certification  of pure states}
Pure Sharpness stipulates that for every system there is a pure effect occurring with probability 1 on some state. We can easily show such a state must be pure: 

\begin{prop}
\label{prop:uniqueness of state} Let $a$ be a normalized  pure effect.
Then, there exists a pure state $\alpha$
such that $\left(a|\alpha\right)=1$. Furthermore, such a pure state is unique: for every $\rho\in\mathsf{St}\left(\mathrm{A}\right)$, one has the implication  
\[
\left(a|\rho\right)=1  \quad \Longrightarrow \quad \rho=\alpha .\]
\end{prop}
\Proof 
See lemma 26 and theorem 7 of Ref.~\cite{Chiribella-informational}
for the proof idea. 
 \qed

Combining the above result with our Pure Sharpness axiom, we derive
the following (cf.\ proposition 9 of Ref.~\cite{QPL15})
\begin{prop}\label{prop:identifyingeffect}
For every pure state $\alpha\in  \Pur\St_{1}\left(\rA\right)$ there exists at least one pure effect $a\in \Pur\Eff\left(\rA\right)$ such that  $\left(a|\alpha\right)=1$.  
\end{prop}

 In summary, for every normalized pure effect $a \in \mathsf{PurEff}_{1}\left( \mathrm{A}\right)  $, we can associate a \emph{unique} pure state $ \alpha \in \mathsf{PurSt}_{1}\left( \mathrm{A}\right) $ with it such that $ \left( a|\alpha\right) =1 $. Conversely, given a pure state $ \alpha $, there always exists at least one pure effect $ a $ such that $ \left( a|\alpha\right) =1 $. This shows that there is a surjective correspondence between normalized pure effects and normalized pure states.
 
 \subsection{Probability balance of pure bipartite states\label{sub:probability balance pure}}  
 We recall that in a general theory, perfectly distinguishable states are defined
 as follows: 
 \begin{defi}
 	The normalized states $\left\{ \rho_{i}\right\} $ are \emph{perfectly
 		distinguishable} if there exists an observation-test $\left\{ a_{j}\right\} $
 	such that $\left(a_{j}|\rho_{i}\right)=\delta_{ij}$. $\left\{ a_{j}\right\} $
 	is called \emph{perfectly distinguishing test}.
 \end{defi}
 Given a normalized state $\rho \in \mathsf{ St}_1\left( \rA\right) $, we define the probability $p_{*}$  as the maximum probability that a pure state can have  in a convex decomposition of $\rho$, namely \footnote{Note that the maximum is well defined because the set of pure states
 	is compact, thanks to transitivity.}
 \[
 p_{*} :=\max_{\alpha\in\mathsf{PurSt}_{1}\left(\mathrm{A}\right)} \left\{ p:\rho=p\alpha+\left(1-p\right)\sigma,\sigma\in\mathsf{St}_{1}\left(\mathrm{A}\right)\right\} .
 \]
 We call $p_*$ the \emph{maximum eigenvalue} of $\rho$ and say that the \emph{pure} state $\alpha$ is the corresponding \emph{eigenstate}.  The reason for this terminology will become clear once we prove our diagonalization theorem.

 A fundamental  consequence of our axioms is that the marginals of a bipartite state have the same maximum eigenvalue: 
 \begin{theo}[Probability balance]
 	\label{thm:probability balance}    Let $\Psi$ be a pure bipartite state of system $\rA\otimes \rB$, let $\rho_\rA$ and $\rho_\rB$ be its marginals on systems $\rA$ and $\rB$ respectively. Then, $\rho_\rA$ and $\rho_\rB$ have the same maximum eigenvalue, namely 
 	\[  p_{*,\mathrm{A}} =  p_{*,\mathrm{B}}  =:  p_* , \]
 	where $p_{*,\mathrm{A}}$ and $p_{*,\mathrm{B}}$ are the maximum eigenvalues of $\rho_\rA$ and $\rho_\rB$ respectively.     
 	Moreover, when $\rho_\rA $ (or, equivalently $ \rho_{\mathrm{B}} $) is decomposed as $\rho_\rA   =  p_*   \alpha  +  \left( 1-p_*\right) \sigma$ for some \emph{pure} state $\alpha$ and some state $\sigma$, the states $\alpha$ and $\sigma$ are perfectly distinguishable.   
 	Specifically, they can be distinguished by the observation-test $\left\{a,  u_\rA  -a\right\}$, where $a$ is any pure  effect such that $\left( a|\alpha\right) =1$.     
 \end{theo}
 \Proof The fact that both marginals have the same maximum eigenvalue was proved in theorem 2 and corollary 1 of Ref.~\cite{QPL15}. Write $ \rho_{\mathrm{A}} $ as 
 \begin{equation}\label{eq:rho}\rho_{\mathrm{A}}=p_{*} \alpha+\left(1-p_{*}\right) \sigma,\end{equation}
 where $\alpha$ is an eigenstate with maximum eigenvalue of $ \rho $ and $\sigma$ is possibly mixed. By proposition 11 of Ref.~\cite{QPL15}, if $a$ is a pure effect such that  $\left(a| \alpha\right) =1$ we have $\left( a|\rho_\rA\right)   =  p_*$. Combining this equality with Eq.~\eqref{eq:rho} we finally obtain 
 \[  p_*   =  \left( a|\rho_{\mathrm{A}}\right)    =    p_*  +  \left( 1-p_*\right)     (a|\sigma) ,  \]  
 which implies $\left( a|\sigma\right) =0$ (unless $ p_*=1 $, but in this case the state $ \rho_{\mathrm{A}} $ is pure).  Hence, $\alpha$ and $\sigma$ are perfectly distinguishable by the test $  \left\{a,  u_\rA  - a\right\}$. \qed

 Now we have managed to decompose every given state into a mixture of two perfectly distinguishable states. The probability balance has many other consequences.   The first is that every non-trivial system has at least two perfectly distinguishable pure states. However, first of all, we must note that for the invariant state $ \chi $,  due to its invariance under the action of reversible channels, every pure state is an eigenstate with maximum eigenvalue.  Indeed, if $\chi$ is decomposed as 
 \[\chi  =  p_* \alpha+  \left( 1-p_*\right)  \sigma,  \] 
 it can also be decomposed as 
 \[\chi  =  p_*  \mathcal U \alpha+  \left( 1-p_*\right)  \mathcal U\sigma , \]   
 where $\mathcal U$ is a reversible channel.   Owing to transitivity, every pure state $\alpha'$ can be obtained as $\mathcal U\alpha$ for some suitable reversible channel---meaning that every pure state can be an eigenstate with maximum eigenvalue.  
 \begin{cor}
 	\label{cor:existence pure perfectly}  If $\rA\not =  \mathrm{I}$, then every pure state of $\rA$ is perfectly
 	distinguishable from some other  pure state.\end{cor}
 \Proof 
The proof is an application of theorem~\ref{thm:probability balance} to the case of the invariant state, and it has already appeared in corollary 3 of Ref.~\cite{QPL15}. \qed

 It is quite remarkable that the existence of perfectly distinguishable
 states pops out from the axioms, without being assumed from
 the start. In principle, the general theories considered in our framework
 might not have had any perfectly distinguishable states at all.

 Another consequence of the probability balance  is the following
 \begin{cor}\label{cor:pupstar}
 	Let $\rho$ be a mixed state of system $\rA$.      Then, the following are equivalent 
 	\begin{enumerate}
 		\item $\alpha$ is an eigenstate of $\rho$ with maximum eigenvalue $p_*$
 		\item $\left( a|\rho\right)   =  p_*$ for every  pure effect $ a $   such that $ \left( a|\alpha\right) =1 $.    
 	\end{enumerate}
 \end{cor}
 The proof is reported in appendix~\ref{app:pupstar}.

For every state we can also define another probability 
\begin{equation}\label{eq:p^*}
p^{*}:=\sup_{a\in\mathsf{PurEff} \left(\mathrm{A}\right)}     \left(a|\rho\right)  .
\end{equation}  
We will  see soon that  the supremum in the definition of $p^*$ is in fact a maximum. Note that for pure states $ p^{*}=1 $, thanks to the Pure Sharpness axiom  (cf.\ Proposition~\ref{prop:identifyingeffect}).

Thanks to the results of corollary~\ref{cor:pupstar} for every state one has the bound $ p_*  \le p^*  $. We can in fact prove that one has the equality. 
\begin{prop}\label{prop:pstar=pstar}
	For every state $\rho\in\St_{1}\left( \rA\right) $ one has $p^*=p_*$ 
\end{prop} 
The proof is presented in appendix~\ref{app:pstar=pstar}.

As a consequence we have that $ p^* $ is achieved by applying any pure effect $ a $ such that $ \left(a|\alpha \right) =1 $ to $ \rho $, where $ \alpha $ is an eigenstate of $ \rho $ with maximum eigenvalue. Therefore, the supremum in the definition of $ p_* $ (Eq.~\eqref{eq:p^*}) is in fact a maximum.

\subsection{Probability balance for purifications of the invariant state} 

Recall that the invariant state $ \chi \in \mathsf{St}_{1}\left(\mathrm{A} \right)  $ can be written as 
\[\chi  =  p_* \alpha+  \left( 1-p_*\right)  \sigma  \] for every \emph{pure} state $ \alpha \in \mathsf{PurSt}_{1}\left( \mathrm{A}\right)  $, where $ \sigma $ is a suitable state. 
In addition, we have the following
\begin{prop}\label{prop:steerpstar}
	Let $\Phi  \in \mathsf {PurSt}_{1} \left( \rA\otimes \rB\right) $ be a purification of the invariant state $\chi_\rA$. Then, for every pure state   $\alpha \in  \mathsf{PurSt}_{1} \left( \rA\right) $ there exists a pure effect $  b  \in  \mathsf{PurEff}  \left( \rB\right) $ such that 
	\begin{equation}\label{daprovare}
	\begin{aligned}\Qcircuit @C=1em @R=.7em @!R { & \multiprepareC{1}{\Phi}    & \qw \poloFantasmaCn{\rA} &  \qw   \\  & \pureghost{\Phi}    & \qw \poloFantasmaCn{\rB}  &   \measureD{b} }\end{aligned}~=p_{*}\!\!\!\!\begin{aligned}\Qcircuit @C=1em @R=.7em @!R { & \prepareC{\alpha}    & \qw \poloFantasmaCn{\rA} &  \qw   }\end{aligned} ~,
	\end{equation}
	where $p_*$ is the maximum eigenvalue of $\chi_\rA$.  As a consequence,  every  normalized pure effect $a\in  \mathsf{PurEff}_1  \left( \rA\right) $ satisfies the condition $\left( a|\chi_\rA\right)   = p_*$.
\end{prop}
The proof is presented in appendix~\ref{app:steerpstar}.
	
	\subsection{The state-effect duality\label{sub:duality}}
	
	Using the results of the previous subsection, one can establish a one-to-one correspondence between normalized pure states and normalized pure effects.   We refer to this correspondence as  the \emph{dagger}  of states and effects.

	\begin{prop}\label{prop:atmostoneeffect}
		For every pure state $ \alpha \in \mathsf{PurSt}_{1}\left( \mathrm{A}\right)  $ there is a unique (normalized) pure effect $ a \in \mathsf{PurEff}_{1}\left( \mathrm{A}\right) $ such that $ \left(a|\alpha \right) =1 $.  
	\end{prop}	
			The proof is essentially the same as the proof of theorem 8 of Ref.~\cite{Chiribella-informational}, even though we are assuming fewer axioms.  We reproduce it  in  appendix~\ref{app:atmostoneeffect} for convenience of the reader.

	Putting propositions~\ref{prop:uniqueness of state}, \ref{prop:identifyingeffect}, and \ref{prop:atmostoneeffect} together we finally obtain the desired result:  
	\begin{theo}[State-effect duality]\label{thm:duality}
		For every system $\rA$ there exists a one-to-one correspondence between normalized pure states and normalized pure effects, called the {\em dagger} and denoted by $\dag:   \mathsf{PurSt}_1  \left( \rA\right)   \to  \mathsf {PurEff}_1  \left( \rA\right)  $.    The dagger satisfies the condition  
		\[  \left( \alpha^\dag| \alpha\right) =1  \qquad \forall \alpha\in  \mathsf{PurSt}_1  \left( \rA\right) .\]
	\end{theo}   	
	Therefore, for every normalised pure state $ \alpha $, $ \alpha^{\dagger} $ denotes the associated pure effect such that $ \left(\alpha^{\dagger}|\alpha \right) =1 $. With a little abuse of notation we will denote the unique normalised pure state associated with a normalised pure effect $ a $ also with a dagger, i.e.\ $ a^{\dagger} $: $ \left( a|a^{\dagger}\right)=1  $.
	
	An easy corollary of the state-effect duality is the following (cf.\ corollary 13 of Ref.~\cite{Chiribella-informational})
	\begin{cor}[Transitivity on pure   effects]\label{cor:transitivityeffects}
		For every pair of   pure normalized effects $a,a'\in\mathsf{PurEff}_{1}\left(\mathrm{A}\right)$,
		there exists a reversible channel $\mathcal{U}$ on $\mathrm{A}$
		such that $a'=a\mathcal{U}$.
	\end{cor}
		
	\subsection{No Disturbance Without Information}
	An important consequence of our axioms  is the possibility of constructing transformations that are ``minimally-disturbing'', in the following sense \cite{Chiribella-informational,Wehner-Pfister,Chiribella-Yuan2015}.  
	 
	\begin{prop}
		\label{prop:non-disturbing measurement}Let $a$ be an effect such
		that $\left(a|\rho\right)=1$, for some $\rho\in\mathsf{St}_{1}\left(\mathrm{A}\right)$.
		Then there exists a pure transformation $\mathcal{T}$ on $\mathrm{A}$
		such that $\mathcal{T}=_{\rho}\mathcal{I}$, where $\mathcal{I}$
		is the identity, and $\left(u \right| \mathcal{T}  \left| \sigma\right)\le\left(a|\sigma\right)  $,
		for every state $\sigma\in\mathsf{St}_{1}\left(\mathrm{A}\right)$.\end{prop}
		The proof can be found in appendix~\ref{app:non-disturbing measurement}.
		
	  Note that the pure transformation $ \mathcal{T} $ is non-disturbing on $ \rho $ because it acts as the identity on $ \rho $ and on all the states contained in it. In other words, whenever we have a (possibly mixed) effect occurring with unit probability on some state $ \rho $, we can always find a transformation that does not disturb $ \rho $ (i.e.\ a non-disturbing non-demolition measurement). Being non-disturbing  means that  $ \mathcal{T}  $ occurs with unit probability on all the states contained in $ \rho $.	  
	  The other notable result of this proposition is that the probability of $  \mathcal{T} $ occurring on a generic state $ \sigma $ is less than or equal to the probability of the original effect occurring on the same state. The implications of the existence of $  \mathcal{T} $ for the distinguishability of states are explored in appendix~\ref{app:distinguishability of states}.

\section{Diagonalization and Schmidt decomposition\label{sec:Diagonalization-of-states}}

\subsection{The diagonalization theorem}
A \emph{diagonalization} of $\rho$ is a convex decomposition of $\rho$
into perfectly distinguishable pure states. The probabilities in such
a convex decomposition will be called the \emph{eigenvalues} of $\rho$ and the perfectly distinguishable pure states the \emph{eigenstates} \cite{QPL15}.

\begin{theo}
	\label{thm:diago}In a theory satisfying Causality, Purity Preservation,
	Purification, and Pure Sharpness every state of every non-trivial system can be
	diagonalized.
\end{theo}
The proof is reported in appendix~\ref{app:diago}. It is a constructive procedure that returns
a diagonalization of $\rho$ where the eigenvalues are naturally listed
in decreasing order, namely $p_{i}\geq p_{i+1}$ for every $i$. In particular, one has $p_1   =  p_*$, which justifies why we called $p_*$ the ``maximum eigenvalue''.     Later in this section we will show that the vector of the eigenvalues $\st p$ is uniquely determined by the state $\rho$. In other words, all the diagonalizations of $\rho$ have the same eigenvalues.  

Before moving forward, it is important to note that the eigenvalues can be characterized as the outcome probabilities of a pure measurement  performed on the system:
 \begin{cor}\label{cor:computeegv}
	Let $\rho$ be a generic state, diagonalized as $\rho  =  \sum_{i=1}^r  p_i  \alpha_i$.   Then, one has the equality  
	\[
	p_i  =   \left(\alpha_i^\dag|\rho\right)   \qquad \forall i  \in  \left\{1,\dots, r\right\}  .
	\]  
\end{cor}
\Proof 
	Immediate from the combination of theorem~\ref{thm:diago}  and   corollary~\ref{cor:dagger distinguishable}, because  $ \left( \alpha_i ^{\dagger}| \alpha_{j}\right)=\delta_{ij} $.
 \qed
 
\subsection{The Schmidt decomposition}

Using the diagonalization theorem we can prove an operational version of the Schmidt decomposition of pure bipartite states \cite{Nielsen-Chuang}.  The intuitive content of the Schmidt decomposition is that for every state of a bipartite system there exist two perfectly correlated pure measurements on the component systems, a similar situation to having conjugates \cite{Wilce-spectral,Royal-road}.
More formally, this property is stated by the following 
\begin{theo}\label{theo:schmidt}
Let $\Psi$ be a pure state of the composite system $\rA\otimes\rB$.  Then, there exist a \emph{pure} observation-test on system $\rA$, say  $\left\{a_i\right\}_{i=1}^{n_\rA}$, and a \emph{pure} observation-test on system $\rB$, say $\left\{  b_j\right\}_{j=1}^{n_\rB}$,    such that   
  \begin{equation}\label{eq:schmidt}
  \begin{aligned}\Qcircuit @C=1em @R=.7em @!R { & \multiprepareC{1}{\Psi}    & \qw \poloFantasmaCn{\rA} &  \measureD{a_i}   \\  & \pureghost{\Psi}    & \qw \poloFantasmaCn{\rB}  &   \measureD{b_j} }\end{aligned}~=p_{i}\delta_{ij}\qquad \forall  i \in  \left\{1,\ldots, r\right\}
  \end{equation}
  where  $r  \le \min \left\{n_\rA, n_\rB\right\}$ is a suitable integer, here called the \emph{Schmidt rank},  $\left\{p_i\right\}_{i=1}^r$ is a probability distribution, with all non-vanishing entries ordered as $p_1  \ge p_2  \ge \cdots \ge p_r>0$.  
  
  Moreover, one has  the diagonalizations
  \[
  \rho_\rA  =  \sum_{i=1}^r   p_i a_i^\dag  \qquad {\textrm and}\qquad \rho_\rB   =  \sum_{i=1}^r  p_i  b_i^\dag , 
  \]    
  where $\rho_\rA$ and $\rho_\rB$ are the marginals of $\Psi$ on systems $\rA$ and $\rB$ respectively. 
\end{theo}
The proof is presented in appendix~\ref{app:schmidt}.

   \begin{defi}
 	Let $ \rho\in\mathsf{St}_1\left( \mathrm{A}\right)$ and let $ \Psi\in\mathsf{St}_1\left( \mathrm{A}\otimes\mathrm{B}\right) $ be one of its purifications. A \emph{complementary state} of $ \rho $ is the state $ \widetilde{\rho}\in\mathsf{St}_1\left( \mathrm{B}\right) $ defined as $ \widetilde{\rho}=\mathrm{Tr}_{\mathrm{A}}\Psi $.
 \end{defi}
 When we have a state $\rho$ we can relate its diagonalizations
 to those of any of its complementary states $\widetilde{\rho}$.   In the proof of
 corollary~\ref{cor:pupstar} we have shown that if $ \alpha $ is an eigenstate of $ \rho $ with maximum eigenvalue $ p_* $, the effect $ \alpha^\dagger $, when applied to $ \Psi $, prepares an eigenstate of $ \widetilde{\rho} $ on $ \mathrm{B} $ with maximum eigenvalue $ p_* $. Since every step in our diagonalization algorithm
 involves finding the maximum eigenvalue of $p_{*}$ of a particular mixed state (called $ \rho_i $ in the proof of theorem~\ref{thm:diago}), each step generates a diagonalization of $ \widetilde{\rho} $ with the same eigenvalues as $ \rho $. Therefore every diagonalization of $ \rho $ induces a diagonalization of $\widetilde{\rho}$ with the same eigenvalues.

\subsection{Diagonalization of the invariant state}

The diagonalization of the invariant state is special.   First of all, all the eigenvalues are equal:  
\begin{prop}
	\label{prop:diagonalization chi d-level}  
	For every non-trivial system, there exists a positive integer number $d$  such that 
	\begin{enumerate}
		\item every diagonalization of
		the invariant state  consists of exactly  $d$ pure states
		\item the eigenvalues of the invariant states are all equal to  $\frac{1}{d}$. 
	\end{enumerate}  
\end{prop}
\Proof 
	Let $ \chi=\sum_{i=1}^{r}p_i\alpha_i $ be a diagonalization of the invariant state $ \chi $. By corollary~\ref{cor:computeegv}, $ p_i=\left(\alpha_i^\dag|\chi\right) $, but by proposition~\ref{prop:steerpstar} we have $ \left(\alpha_i^\dag|\chi\right)=p_* $, whence $ p_i=p_* $ for every $ i $. It follows that $ p_*=\frac{1}{r}$. Now consider another diagonalization of $ \chi $: $ \chi= \sum_{i=1}^{r'}p'_i\alpha_i $. Repeating the same argument, we conclude that $ p'_i=p_*=\frac{1}{r'} $. This means that $ r=r'=:d $.
 \qed
 
We will refer to $d$ as the \emph{dimension} of the system, for reasons that will become clear soon.

Moreover, the set of  states $\left\{\alpha_i\right\}_{i=1}^d$ and the corresponding set of effects $\left\{\alpha_i^\dag\right\}_{i=1}^d$ should be \emph{maximal},    in the following sense  
\begin{defi}
	We say that a set of perfectly distinguishable states $\left\{ \rho_{i}\right\} _{i=1}^{n}$ is \emph{maximal}
	if there exists no  other state $\rho_{n+1}$ such that the states $\left\{ \rho_{i}\right\} _{i=1}^{n+1}$
	are perfectly distinguishable.
\end{defi}
Maximality for effects is defined as 
\begin{defi}
	Let $\left\{ a_{i}\right\} _{i=1}^{n}$ be a set of effects coexisting in an observation-test.   We say that $\left\{ a_{i}\right\} _{i=1}^{n}$ is \emph{maximal}
	if there exists no  other effect $a_{n+1}$ such that the effects $\left\{ a_{i}\right\} _{i=1}^{n+1}$ coexist in an observation-test.
\end{defi}
	Equivalently,  a set of coexisting effects $\left\{a_i\right\}_{i=1}^n$ is maximal if and only if the effects  form an observation-test.

With these definitions, we have the following  
\begin{prop}\label{prop:maximal}
	For every system $\rA$, the following are equivalent: 
		\begin{enumerate} 
		\item $\left\{\alpha_i\right\}_{i=1}^d  \in \Pur\St_1 (\rA)$ is a maximal set of perfectly distinguishable pure states
		\item   $\left\{\alpha_i^\dag\right\}_{i=1}^d  \in  \Pur\Eff_1 (\rA)  $  is a maximal set of coexisting pure effects.  
	\end{enumerate}
\end{prop}
The proof is presented in appendix~\ref{app:maximal}.

Propositions~\ref{prop:diagonalization chi d-level} and \ref{prop:maximal} imply that the invariant
state is a uniform mixture of a maximal set of perfectly distinguishable pure
states.   Remarkably, the converse holds too:  \emph{every} maximal
set of perfectly distinguishable pure states, mixed with equal
weights, yields the invariant state:  
\begin{prop}
	\label{prop:diagonalization chi d-level 2}   Let $\left\{ \alpha_{i}\right\} _{i=1}^{r}$
	be a maximal set of perfectly distinguishable pure states.    Then one
	has $r=d$ and $\chi=\frac{1}{d}\sum_{i=1}^{d}\alpha_{i}$.   Conversely, whenever $\chi$ is decomposed as a uniform mixture of $d$ pure states, the states must be perfectly distinguishable, and must form a maximal set. 
\end{prop}
The proof is reported in appendix~\ref{app:diagonalization chi d-level 2}

In summary, the above proposition guarantees that all sets of perfectly distinguishable pure states of a system have the same cardinality, equal to the dimension of the system. As a consequence, every state can have at most $ d $ terms in its diagonalizations. In previous works \cite{Barnum-interference,Scandolo,QPL15,Krumm-Muller}, this result and other properties of the diagonalizations  were derived from the Strong Symmetry axiom  \cite{Barnum-interference}, stating  that all pure maximal sets are connected by reversible channels. Our result shows that the properties of diagonalization can be derived from a very different set of axioms   (Causality, Purity Preservation,  Pure Sharpness, and Purification).

Finally, proposition \ref{prop:maximal}  implies an operational version of Naimark's theorem:  
\begin{prop}\label{prop:naimark}
Let $\left\{a_i\right\}_{i\in  \set X}$ be a generic observation-test. Then, there exists a system $\rE$, a pure state $\varphi_0  \in\Pur\St_1 \left( \rE\right) $, and a test $\left\{\Pi_i\right\}_{i\in\set X}$ consisting of pure transformations on the composite system $\rA\otimes \rE$ such that 
\[
\Pi_i  \Pi_j   =  \delta_{ij}  \Pi_i  ,\qquad \forall i,j\in\set X
 \]
and
\[
\begin{aligned}\Qcircuit @C=1em @R=.7em @!R { & & \qw \poloFantasmaCn{\rA} & \measureD{a_i}}\end{aligned} ~= \!\!\!\! \begin{aligned}\Qcircuit @C=1em @R=.7em @!R { & & \qw \poloFantasmaCn{\rA} & \multigate{1}{\Pi_i} & \qw \poloFantasmaCn{\rA} &\measureD{u} \\ & \prepareC{\varphi_0} & \qw \poloFantasmaCn{\rE} &\ghost{\Pi_i} & \qw \poloFantasmaCn{\rE} & \measureD{u} }\end{aligned}~.
\]
\end{prop}      
The proof is presented in  appendix~\ref{app:naimark}. It is worth noting that, thanks to the axioms assumed here, the observation-test on the composite system, defined by $\left\{   E_i   \right\}_{i\in\set X}$,  $E_i   :   =     \left(  u_\rA\otimes u_\rE\right)    \Pi_i$ is a sharp measurement \cite{Chiribella-Yuan2014, Chiribella-Yuan2015}, in the sense that it is repeatable and has minimal disturbance.

\subsection{Diagonalization of complete states}
As an aside, we show here that complete states  (states that  contain all other states in their convex decomposion) have exactly $ d $ non-zero eigenvalues.  In the quantum case, this amounts to saying that complete states are full-rank density matrices. 

\begin{prop}
	\label{prop:completely-mixed eigenvalues}Every complete state $ \omega $
	has precisely $d$  non-vanishing eigenvalues in every diagonalization. Consequently, the pure states arising in every diagonalization of $ \omega $ form a maximal set.\end{prop}
The proof is reported in appendix~\ref{app:completely-mixed eigenvalues}.
	
	The converse also holds: 
	\begin{prop}\label{prop:completely mixed converse}
	Let $ \left\lbrace \alpha_i\right\rbrace_{i=1}^{d}  $ be a maximal set of perfectly distinguishable pure states. Every convex combination of the $ \alpha_i $'s with \emph{non-vanishing} coefficients yields a complete state.\end{prop}

	The proof already appeared in Ref.~\cite{Chiribella-informational}, and does not make use of the stronger axioms therein assumed. We report it in appendix~\ref{app:completely mixed converse} for the convenience of the reader.

\subsection{Duality between maximal sets of pure  states and pure sharp measurements}  

The diagonalization of the invariant state induces a one-to-one correspondence between maximal sets of perfectly distinguishable pure states and \emph{pure sharp measurements}  \cite{Chiribella-Yuan2014,Chiribella-Yuan2015}, which can be characterized as follows:  
\begin{defi}  
	An observation-test $\left\{a_i\right\}_{i=1}^n$ is a \emph{pure sharp measurement} if every effect $a_i$ is pure and normalized.  
\end{defi}
Under the validity of our axioms, every pure sharp measurement can be  written as $\left\{\alpha_i^{\dag}\right\}_{i=1}^n$,  for some set of pure states $\left\{\alpha_i\right\}_{i=1}^n$  (cf.\ theorem~\ref{thm:duality}).

\begin{prop}
	\label{prop:pure test chi}For every  maximal set of perfectly distinguishable pure states $\left\{ \alpha_{i}\right\} _{i=1}^{d}$,
	the effects $\left\{ \alpha_{i}^{\dagger}\right\} _{i=1}^{d}$ form a pure sharp measurement.   Conversely, for  every pure sharp measurement  $\left\{ \alpha_{i}^{\dagger}\right\} _{i=1}^{n}$, the states  $\left\{ \alpha_{i} \right\} _{i=1}^{n}$  form a  maximal set of perfectly distinguishable pure states,  and therefore  $n=d$.  \end{prop}
The proof is presented in appendix~\ref{app:pure test chi}.

\subsection{Double stochasticity of the transition matrices}  
Given two maximal sets of perfectly distinguishable pure states, $\left\{ \alpha_{i}\right\} _{i=1}^{d}$
and $\left\{ \alpha'_{i}\right\} _{i=1}^{d}$, we call the matrix $  T_{ij}  =    \left(  {\alpha_i^\dag} |\alpha'_j\right)  $  a \emph{transition matrix}. With this definition, we have the following (cf.\ lemma 4 of Ref.~\cite{QPL15}).
\begin{lemma}
	\label{lem:doubly stochastic} 
	Under the validity of our axioms, all transition matrices are doubly stochastic.
	\end{lemma}

\subsection{Uniqueness of the diagonalization}
Thanks to our axioms, the diagonalization of a state is unique, up to the obvious freedom arising in the presence of degeneracy among the eigenvalues.  This is a non-trivial consequence of the axioms: notably, Refs.~\cite{Krumm-Muller,Krumm-thesis} exhibited  examples of  GPTs where states can be diagonalized, but the same state can have more than one diagonalization and more than one spectrum. 
To take   degeneracy into account, we define the \emph{spectrum of $\rho$}, denoted by $\mathsf{Sp} \left(\rho\right)=   \left( \lambda_1,\ldots, \lambda_s\right)$,  as the set of the \emph{distinct} eigenvalues of $\rho$, ordered in strictly decreasing order $\lambda_1 >  \lambda_2  >  \ldots  >\lambda_s$ and we rewrite the diagonalization  as 
\[   \rho  =   \sum_{k=1}^{s}  \lambda_k \Pi_k ,\]where
\[\Pi_k  :  =  \sum_{i: p_i   =   \lambda_k}   \alpha_i .     \]
When expressed in this form, the diagonalization is unique.  Now we present the main theorem.  
\begin{theo}\label{thm:uniqueness diago}
	Let $\rho  =     \sum_{k=1}^{s}  \lambda_k  \Pi_k  $ and $\rho  =     \sum_{l=1}^{s'}  \lambda'_l    \Pi'_l  $ be two diagonalizations of the same state.  Then, one has 
	\[  s=s' ,  \quad   \lambda_k  = \lambda_k'  , \quad \Pi_k  =  \Pi_k'  \qquad \forall k\in\{1,\dots, s\} .\] 
\end{theo}
The proof is presented in appendix~\ref{app:uniqueness diago}. 	
A very close result was proved by Wilce in the framework of probabilistic models with conjugates and Jordan algebras \cite{Royal-road}. 

This shows that the diagonalization is unique up to the choice of the eigenstates when we have degeneracy. Only then do we have the freedom of choice of the eigenstates relative to degenerate eigenvalues, i.e.\ eigenvalues arising more than once in a diagonalization of a state. Cf.\ Ref.~\cite{QPL15} for another proof of the uniqueness of the eigenvalues based on majorization and a further axiom.

\subsection{Extending the diagonalization to arbitrary vectors}

The diagonalization theorem, proved for normalized states, can be easily extended to arbitrary elements of the vector space $  \mathsf {St}_{\mathbb R}  \left(\rA\right)$: 
\begin{prop}\label{prop:diag1}
	For every  system $\rA$ and for every vector $\xi  \in   \mathsf {St}_{\mathbb R}  \left(\rA\right)$ there exists   a set of $ d $ real numbers $\left\{x_i\right\}_{i=1}^d$ and  a maximal  set of perfectly distinguishable states  $  \left\{\alpha_i\right\}_{i=1}^d$ such that 
	\[   \xi  =  \sum_{i=1}^d   x_i   \alpha_i .\]
\end{prop}   
We omit the proof, which  is the same as the proof of   corollary 21   in Ref.~\cite{Chiribella-informational}.  Again the $ x_i $'s are called the eigenvalues of $ \xi $, and the $ \alpha_i $'s are called the eigenstates of $\xi$. A similar result was obtained also in Ref.~\cite{Krumm-Muller} under different axioms.

Using Steering, one can convert the diagonalization result for the elements of $\mathsf {St}_{\mathbb R}\left(\rA\right)$  into  a diagonalization result for the elements of  $\mathsf {Eff}_{\mathbb R} \left(\rA\right)$ (see also \cite{Wilce-spectral,Royal-road} for a slightly different approach): 

\begin{prop}\label{prop:diag2}
	For every finite dimensional system $\rA$ and for every vector $X  \in   \mathsf {Eff}_{\mathbb R}  \left(\rA\right)$ there exists a set of $ d $ real numbers $\left\{x_i\right\}_{i=1}^d$ and a pure maximal set  of states $  \left\{\alpha_i\right\}_{i=1}^d$  
	such that 
	\[   X  =  \sum_{i=1}^d   x_i   \alpha^\dag_i .\]
\end{prop}   

\subsection{Extending the dagger map}  

Thanks to the diagonalization theorems, the dagger map $\dag:   \mathsf{PurSt}_1  \left(\rA\right)  \to  \mathsf{PurEff}_1 \left(\rA\right)$ can be extended to arbitrary vectors via the relation   
\[    \xi   =  \sum_{i=1}^d  x_i  \alpha_i      \quad \longmapsto \quad   \xi^\dag:  =   \sum_{i=1}^d   x_i \alpha_i^\dag  .     \]
Note that, since the diagonalization is unique (up to degeneracy), the vector $\xi^\dag$ is well-defined, i.e.\ it does not depend on the choice of the $ \alpha_i $'s as long as they are eigenstates of $ \xi $.  Writing $ \xi $ like in theorem~\ref{thm:uniqueness diago}, $ \xi= \sum_{k=1}^{s}  \lambda_k  \Pi_k  $, to prove this fact, it  suffices  to show the following 
\begin{prop}\label{prop:well-defined}
	Let $\left\{\alpha_i\right\}_{i=1}^r$ and $\left\{\alpha_j'\right\}_{j=1}^r$ be two sets of perfectly distinguishable pure states. Then, one has the implication  
	\[  \sum_{i=1}^r \alpha_i  =  \sum_{j=1}^r \alpha_j'  \quad \Longrightarrow  \quad \sum_{i=1}^r \alpha^\dag_i  =  \sum_{j=1}^r \alpha^{\prime \dag}_j .\] 
\end{prop}
The proof is reported in appendix~\ref{app:well-defined}.

Similarly to what we did in section~\ref{sub:duality}, with a little abuse of notation we will denote as $ \dagger $ even the inverse map, from $ \mathsf{Eff}_{\mathbb{R}} \left(\rA\right) $ to $  \mathsf{St}_{\mathbb{R}}  \left(\rA\right) $.
 
 Now we are ready to define observables and to introduce a functional calculus on them.
 
\subsection{Functional calculus on the observables\label{sub:functional calculus}} 

Thanks to the diagonalization theorem, the elements of the vector space $\mathsf {Eff}_{\mathbb R} \left(\rA\right)$ can be regarded as \emph{observables}, in a similar sense to the use of the term in quantum theory. Indeed, given a diagonalization  $X  =  \sum_{i=1}^d  x_i  \alpha^\dag_i$ one can think of the eigenvalues as the ``values'' associated with the outcomes of the sharp measurement $\left\{\alpha_i^\dag\right\}_{i=1}^d$.  In this way,  one can interpret 
\[\left\langle X\right\rangle_{\rho}:=\left( X | \rho\right)   = \sum_{i=1}^d  x_i   \left(\alpha_i^\dag | \rho   \right) \]  as the \emph{expectation value} of the observable $X$, because $ \left(\alpha_i^\dag | \rho   \right) $ are probabilities. 

Like in quantum theory, the spectral theorem allows one to  define a functional calculus on the observables (see also \cite{Royal-road} for a slightly different approach):   given an observable $X$ and a function $f  :    \mathbb R  \to \mathbb R  $, one can define the observable 
\[  f\left( X\right)    :  =   \sum_{i=1}^d   f\left( x_i\right)    \alpha_i^\dag  .\]
Note that the observable  $f(X)$ is well-defined, because the diagonalization of $X$ is unique and as a consequence of proposition~\ref{prop:well-defined}.   In particular, one can choose the observable $X$ to be the dagger of a state $ \rho   = \sum_{i=1}^d   p_i   \alpha_i $,  thus obtaining
\[  f\left(\rho^\dag\right)     =  \sum_{i=1}^d    f\left( p_i\right)    \alpha_i^\dag.\]
In the following we will use this notation for the logarithm function: defining the \emph{``surprisal observable''}    
\[ - \log \rho^\dag   =-  \sum_{i=1}^d \log p_i  \alpha_i^\dag     ,
\] 
the Shannon-von Neumann entropy as the expecation value of the surprisal observable
\begin{equation}\label{eq:surprisal Shannon}   S\left( \rho\right)    :  =    \left(-\log \rho^\dag  |  \rho  \right)  ,   \end{equation}
and the Kullback-Leibler divergence  
\[   S\left( \rho \parallel  \sigma\right)   :  =    \left(\log \rho^{\dagger}  -  \log\sigma^{\dagger} |  \rho\right)  .\]

Like  in classical and quantum theory, the Shannon-von Neumann entropy defined here is an important  indicator of the degree of ``mixedness'' of a given state. Likewise,  the Kullback-Leibler divergence is an indicator of the deviation of a state relative to another.      These two notions will be analysed in the next sections.

\section{Purity and majorization\label{sec:puriresource}}

Majorization is traditionally used as a criterion to compare the degree
of mixedness of probability distributions. Here we extend this approach
to general probabilistic theories satisfying Purity Preservation, Causality, Purification, and Pure Sharpness. 
In order to define  the degree of mixedness of a state operationally, we adopt the operational resource theory of purity formulated in our earlier work \cite{Chiribella-Scandolo-entanglement}, which considered the operational scenario  where an experimenter has limited control on the dynamics
of a closed system. In this scenario, the set of free operations are
 \emph{Random Reversible (RaRe) channels}, defined as random mixtures
of reversible transformations:
\begin{defi}
	A channel $\mathcal{R}$ is \emph{RaRe} if there exist a probability
	distribution $\left\{ p_{i}\right\} _{i\in\mathsf{X}}$ and a set
	of reversible channels $\left\{ \mathcal{U}_{i}\right\} _{i\in\mathsf{X}}$
	such that $\mathcal{R}=\sum_{i\in\mathsf{X}}p_{i}\mathcal{U}_{i}$.
\end{defi}
Under the present set of axioms, RaRe channels cannot increase the purity of a state.
If $\rho=\mathcal{R}\sigma$, where $\mathcal{R}$ is a RaRe channel,
we say that $\rho$ is \emph{more mixed} than $\sigma$ \cite{Chiribella-Scandolo-entanglement}  (see also definition 8 of the appendix of \cite{Muller3D}).
If $\rho$ is more mixed than $\sigma$ and $\sigma$ is more mixed
than $\rho$ we say that $\rho$ and $\sigma$ are \emph{equally mixed}.

We will now show that, under the validity of our axioms, the ordering of states according to their mixedness implies majorization of the eigenvalues, just as it happens in quantum theory \cite{Nielsen-Chuang}.
Let us start by recalling the definition of majorization. Let $x\in\mathbb{R}^{d}$
be a vector, and let $x_{\left[i\right]}$ be the $i$-th component
of the decreasing rearrangement of the entries of $x$, such that $x_{\left[i\right]}\geq x_{\left[j\right]}$ if $ i<j $.
\begin{defi}
	Let $\mathbf{x}$ and $\mathbf{y}$ be vectors in $\mathbb{R}^{d}$. 
	Then, $\mathbf{x}$
	is \emph{majorized} by $\mathbf{y}$ (or $\mathbf{y}$ \emph{majorizes}
	$\mathbf{x}$), and we write $\mathbf{x}\preceq\mathbf{y}$, if
	\begin{itemize}
		\item $\sum_{i=1}^{k}x_{\left[ i\right] }\leq\sum_{i=1}^{k}y_{\left[ i\right] }$, for every $k=1,\ldots,d-1$
		\item $\sum_{i=1}^{d}x_{\left[ i\right] }=\sum_{i=1}^{d}y_{\left[ i\right] }$.
	\end{itemize}
\end{defi}
It is well known that $\mathbf{x}\preceq\mathbf{y}$ if and only if $\mathbf{x}=D\mathbf{y}$,
where $D$ is a doubly stochastic matrix  \cite{Hardy-Littlewood-Polya1929,Olkin}.  

\subsection{Majorization as a necessary condition for convertibility}  
We now show that   the existence of a mixedness  ordering between two states implies a  majorization condition in terms of the  eigenvalues. For a state of a system of dimension $ d $, its eigenvalues can be arranged in a vector of $\mathbb{R}^d$.
\begin{theo}
	\label{thm:mixedness -> majorization}
	Let $\rho$ and $\sigma$ be two states of a generic system
	and let $\mathbf{p}$ and $\mathbf{q}$ be the vectors of the eigenvalues
	in the diagonalizations of $\rho$ and $\sigma$. If $\rho$ is more
	mixed than $\sigma$, then $\mathbf{p}\preceq\mathbf{q}$.\end{theo}
	The proof is based the fact that transition matrices are doubly stochastic (cf.\ lemma~\ref{lem:doubly stochastic}).    Earlier proofs    \cite{Scandolo,QPL15} (theorem 4) derived the double stochasticity from the Strong Symmetry axiom \cite{Barnum-interference}, which is not assumed in this paper.

An easy corollary of the majorization condition is the following: 
\begin{prop}
	If two states  are equally mixed, then  they have the same eigenvalues. \end{prop}
\Proof 
	If $\rho$ and $\sigma$ are equally mixed, then their eigenvalues should satisfy $\mathbf{p}\preceq\mathbf{q}$ and
	$\mathbf{q}\preceq\mathbf{p}$. 
	It is a well-known fact about majorization that this condition is satisfied only if ${\bf p} =  {\bf q}$, once $ \mathbf{p} $ and $ \mathbf{q} $ are ordered in decreasing order. 
 \qed

\subsection{Operational characterization of the eigenvalues}
Majorization  also provides an operational characterization of the eigenvalues of a state: the eivenvalues are  the least mixed probability distribution that can be generated by pure measurements. The proof is reported in appendix~\ref{app:operational eigenvalues}, and requires a lemma on the structure of pure observation-tests (lemma~\ref{lem:characterization of pure tests}). The proof line will be close to
the that in lemma B.1 of Ref.~\cite{Entropy-Short} for quantum theory. A similar result was also proved in GPTs in Refs.~\cite{Krumm-thesis,Krumm-Muller}. 
\begin{prop}\label{prop:majorization measurement}
	Consider a \emph{pure} observation-test $ \left\lbrace a_i\right\rbrace_{i=1}^{n}  $ and state $ \rho $. Let $ \mathbf{q_a} $ be the vector with entries $ q_{a,i}=\left( a_i|\rho\right) $ for $ i\in\left\lbrace 1,\ldots,n\right\rbrace  $. Then let $ \widetilde{\mathbf{p}} $ be the vector of the eigenvalues of $ \rho $ with $ n-d $ 0's appended. Then $ \mathbf{q_{a}}\preceq\widetilde{\mathbf{p}} $.
\end{prop}

\section{Mixedness monotones and generalized entropies}\label{sec:monotones}


\subsection{Definition} 

In thermodynamics and information theory  it is     convenient to have  numerical indicators of the amount of mixedness (or equivalently, the amount of purity) of a state. Such numerical indicators, hereafter called mixedness monotones \cite{Chiribella-Scandolo-entanglement}, can be defined as follows: 

\begin{defi}\label{def:monotone}
	A \emph{mixedness monotone} for
	system $\mathrm{A}$ is a function $M:\mathsf{St}_{1}\left(\mathrm{A}\right)\rightarrow\mathbb{R}$ that satisfies the condition $M\left( \rho\right)   \ge M\left( \sigma\right) $ whenever $\rho$ is more mixed than $\sigma$. \end{defi}
Similarly, a \emph{purity monotone} is a function $M:\mathsf{St}_{1}\left(\mathrm{A}\right)\rightarrow\mathbb{R}$ that satisfies the condition $M\left( \rho\right)   \ge M\left( \sigma\right) $ whenever $\rho$ is purer than $\sigma$. In practice, a purity monotone reverses the inequality in definition~\ref{def:monotone}: if $ M $ is a mixedness monotone, $ -M $ is a purity monotone.

It is immediate that a mixedness monotone assigns the same real number to equally mixed states.
Note that mixedness monotones give a simple criterion to check when a state is \emph{not} more mixed than another~\cite{Resource}. Indeed, if $ M\left( \rho\right) <M\left( \sigma\right) $, we can conclude that $ \rho $ is \emph{not} more mixed than $ \sigma $. Similarly, if $ M\left( \rho\right)\neq M\left( \sigma\right) $, $ \rho $ and $ \sigma $ are \emph{not} equally mixed.

Mixedness monotones are abundant \cite{Chiribella-Scandolo-entanglement}. A slightly more restrictive notion is the notion of \emph{generalized entropy}, defined as follows  
\begin{defi}
	For every system $\rA$, let $M: \mathsf{St}_1 \left( \rA\right) \to \mathbb R$  be a mixedness monotone.    We say that 
	$ M $ is a \emph{generalized entropy} if  it is additive on product states, that is
	\begin{equation}\label{additivity} M  \left( \rho_{\mathrm{A}}\otimes \sigma_{\mathrm{B}}\right)      =   M\left( \rho_{\mathrm{A}}\right)    +     M  \left( \sigma_{\mathrm{B}}\right)
	\end{equation}
	for all  $\rho_{\mathrm{A}} \in\mathsf{St}_1  \left( \rA\right) $,   and  all $\sigma_{\mathrm{B}}\in\mathsf{St}_1  \left( \rB\right) $. 
\end{defi}
Let us see now how the diagonalization theorem and the majorization condition can be used to construct   mixedness monotones  and generalized entropies.  
First of all, mixedness monotones can be obtained from Schur-concave functions \footnote{Recall that a function $f:  \mathbb R^d \to \mathbb R$ 
	is \emph{Schur-concave} if $f\left(\mathbf{x}\right)\geq f\left(\mathbf{y}\right)$,
	whenever $\mathbf{x}\preceq\mathbf{y}$.}: 
\begin{prop}\label{prop:mixedness monotones}
	Let  $f  : \mathbb R^d  \to \mathbb  R$  be a Schur-concave function and, for every state $\rho  \in  \mathsf {St}_1 \left( \rA\right) $, let ${\mathbf  p}  \in  \mathbb R$  be the vector of  its eigenvalues.   Then, the function   on the state space  $M_f  :  \mathsf {St}_1  \left( \rA\right)  \to \mathbb R$  defined as    $M_f \left( \rho\right)  : =  f \left(  {\bf  p}\right) $  is a mixedness monotone. 
\end{prop}
\Proof If $\rho$ is more mixed than $\sigma$, then the majorization criterion implies that $\bf p$  (the  vector of the eigenvalues of $\sigma$) is majorized by $\bf q$   (the vector of the eigenvalues of $\rho$).  Therefore, one has 
\[  M_f\left( \rho\right)    =  f\left(   {\bf p}\right)   \ge f\left(   {\bf q}\right)   =  M_f\left( \sigma\right) ,\]
which proves that $M$ is a mixedness monotone. \qed 
A  similar result can be obtained for generalized entropies:  
\begin{cor}
	Let    $f:\mathbb{R}^d\rightarrow \mathbb{R}$ be a  Schur-concave function for all $ d $,  satisfying the additivity property $  f\left(  { \bf p} \otimes {\bf q}\right)    =    f\left(  { \bf p}\right)  +  f \left( {\bf q}\right) $, where ${\bf p} \otimes {\bf q}$ denotes the Kronecker product. Then, the corresponding mixedness monotone $ M\left( \rho\right) =f\left( \mathbf{p}\right)  $, where $ \mathbf{p} $ is the vector of the eigenvalues of $ \rho $, is a generalized entropy. 
\end{cor} 
Similarly, purity monotones can be obtained from Schur-convex functions.

Two important examples of additive Schur-concave functions are   the \emph{R\'enyi entropies} \cite{Renyi}
\[  H_\alpha \left( {\bf p}\right)   =\frac{1}{1-\alpha} \log  \left(  \sum_{i=1}^{d}    p_i^\alpha \right),  \]
for $ \alpha\geq0 $, and the \emph{Shannon-von Neumann entropy} \cite{vonNeumann,Shannon} 
\[H\left( {\bf p}\right)   =  -\sum_{i=1}^{d}  p_i  \log p_i    \equiv  \lim_{\alpha  \to 1}    H_\alpha  \left({\bf p}\right).\] 
Using this fact, we define the generalized  Shannon and R\'enyi entropies as 
\[ S   \left( \rho\right)     : =   H\left( {\bf p}\right)   \qquad S_\alpha  \left( \rho\right)   :  =   H_\alpha  \left( {\bf p}\right)  . \]  
Note that one has  the obvious bounds  
\[
0  \le    S_\alpha  \left(\rho\right)  \le    \log d,     \qquad \forall \rho  \in \St_1 \left(\rA\right)  ,  \forall \alpha \geq0 .     
\]
where $d$ is the dimension of  the system.

It is worth noting that the marginals of a pure bipartite state have the same entropy, and, more generally, the same value of the monotone $M_f$, for every possible  Schur-concave function  $f$:
\begin{cor}
Let $\Psi$ be a pure state of $\rA\otimes\rB$ and let $\rho_\rA$ and $\rho_\rB$ be its marginals on systems $\rA$ and $\rB$, respectively.  Then, one has  
\[
   M_f  \left(\rho_\rA\right) =   M_f\left(\rho_\rB\right) ,
\]
   for every  Schur-concave function $f$.
\end{cor}

The proof is immediate from the Schmidt decomposition (theorem \ref{theo:schmidt}), which ensures that the marginals of a pure bipartite state have the same spectrum. 
   
\subsection{Preparation and measurement monotones} 
In every sharp theory with purification, the mixedness monotones defined in the previous paragraph have a nice characterization in terms of optimal measurements, or, dually, in terms of optimal ensemble decompositions.  

Let us start from two definitions:   given a Schur-concave function, the \emph{measurement monotone}  $M_f^{\rm meas}$ is defined as 
\[
M_f^{\rm meas}\left(\rho\right)  :  =  \inf_{  {\bf a} \in  \Pur  \Obs  \left(\rA\right)  } f    \left(  {\bf q} \right), 
\]
where $ q_i : =   \left(a_i |\rho\right)  $, and the infimum is over all pure observation-tests $ \mathbf{a}\equiv\left\lbrace a_i \right\rbrace  $ of system $\rA$.  

Similarly, the \emph{preparation monotone}  $M_f^{\rm prep}$ is defined as 
\[
M_f^{\rm prep}\left(\rho\right)  :  =  \inf_{\begin{array}{c}
	{\scriptstyle \left\{ \varphi_{i}\right\} \subset\mathsf{PurSt}_{1}\left(\mathrm{A}\right)}\\
	{\scriptstyle \pi_{i}\geq0,\sum_{i}\pi_{i}\varphi_{i}=\rho}
	\end{array}}    f    \left(   {\bs \pi}  \right)  ,
\]
where the infimum is over all pure-state decompositions of the state $\rho$. 
In words, the measurement monotone $M_f^{\rm meas}$ is the smallest amount of mixedness (as measured by the function $f$) present in the probability distributions generated by pure measurements on $\rho$.  Dually,  the preparation monotone $M_f^{\rm  prep}$ is the smallest amount of mixedness  present in the prior probabilities of the pure state ensembles for $\rho $.    Examples of preparation and measurement monotones are the preparation and measurement entropy defined in   Refs.~\cite{Entropy-Barnum,Entropy-Short,Entropy-Kimura}.

In every sharp theory with purification,  preparation and measurement monotones for a particular class of Schur-concave functions coincide.

A vector $ \widetilde{\bf x}\in\mathbb{R}^n $ is called \emph{reducible} if it has some vanishing entries. In this case we can ``extract'' a sub-vector  $ \bf x\in\mathbb{R}^d $, with $ d< n $ made only of the non-vanishing components. We will call this operation ``reduction''.
\begin{defi}
	A Schur-concave function $ f $ is called \emph{reducible} if for every reducible vector $ \widetilde{\bf x}\in\mathbb{R}^n $ one has
	\[
	f\left(\widetilde{\bf x} \right)= f\left(\bf x \right),
	\]
	where $ \bf x $ is the reduced vector extracted from $ \widetilde{\bf x} $.
\end{defi}
In words, a reducible Schur-concave function is a Schur-concave function for which the vanishing entries of a vector do not matter. Examples of reducible Schur-concave functions are Rényi entropies and Shannon-von Neumann entropy. Not all Schur-concave functions are reducible: given a vector of probabilities $ \mathbf{p} $ of dimension $ d $, consider the function $ V\left(\mathbf{p}\right) =\frac{1}{d} \left( 1-\sum_{i=1}^{d}p_i^2\right)  $. $ V $ is Schur-concave, but it is \emph{not} reducible. Indeed, consider the vectors $  \mathbf{p}  =\left(\begin{array}{cc}
\frac{1}{2} & \frac{1}{2}\end{array}\right) $, and $ \widetilde{\mathbf{p}}=\left(\begin{array}{ccc}
\frac{1}{2} & \frac{1}{2} & 0\end{array}\right) $; we have $ V\left(\mathbf{p}\right)=\frac{1}{4}  $, whereas $ V\left(\widetilde{\mathbf{p}}\right)=\frac{1}{6}  $, whence $ V $ is not reducible.

Now we can state the following 
\begin{theo} \label{thm:measurement=preparation}
In every sharp theory with purification one has  
\[   M_{f}^{\rm meas}  \left(\rho\right)   =   M_f^{\rm prep}   \left(\rho\right) =  M_f   \left(\rho\right)  ,\]
for every reducible Schur-concave function $f$ and for every state $\rho$.   
\end{theo}
The proof is in appendix~\ref{app:measurement=preparation}.

\section{The Shannon-von Neumann entropy}\label{sec:vN}

\subsection{Basic properties} 

As  seen in subsection~\ref{sub:functional calculus},  the Shannon-Von Neumann entropy can be expressed as 
\[  S\left( \rho\right)    =   \left( -  \log  \rho^\dag   |  \rho   \right) , \]
meaning that  $S\left( \rho\right) $ is the expectation value of the \emph{surprisal observable}  $-  \log   \rho^\dag $.  
This alternative formulation is useful because it suggests a  generalization of  the Kullback-Leibler divergence in GPTs  satisfying our axioms:  

\begin{defi}
	Let $\rho$ and $\sigma$ be two normalized states. The \emph{Kullback-Leibler divergence} of $\rho$ to $\sigma$ is
	\[
	S \left(\rho \parallel \sigma\right):=\left(\log \rho^\dag-\log \sigma^\dag|\rho\right).
	\]
\end{defi}
The key property of the Kullback-Leibler divergence is the Klein inequality, which can be easily extended to the general case: 
\begin{lemma}[Klein's inequality]\label{lem:Klein}
	Let $\rho$ and $\sigma$ be two normalized states. One has $S\left(\rho\parallel\sigma\right)\geq0$
	and $S\left(\rho\parallel\sigma\right)=0$ if and only if $\rho=\sigma$.\end{lemma}
The proof follows the same steps as the one in  quantum theory (see e.g.~\cite{Nielsen-Chuang}), and is reported in appendix~\ref{app:Klein}.

Like in quantum theory,  the GPT version of Klein's inequality  allows one to prove a number of important properties. The easiest application  is   the subadditivity of Shannon-Von Neumann entropy, expressed  by the following
\begin{prop}[Subadditivity]
	\label{prop:subadditivity}Let $\rho_{\mathrm{AB}}$ be a bipartite
	state of system $\mathrm{A}\otimes\mathrm{B}$, and let $\rho_{\mathrm{A}}$
	and $\rho_{\mathrm{B}}$ be its marginals on system $\mathrm{A}$
	and $\mathrm{B}$ respectively. Shannon-Von Neumann is \emph{subadditive},
	namely
	\[
	S\left(\rho_{\mathrm{AB}}\right)\leq S\left(\rho_{\mathrm{A}}\right)+S\left(\rho_{\mathrm{B}}\right) .
	\]
	The equality holds if and only if $\rho_{\mathrm{AB}}$ is a product
	state.\end{prop}
The proof follows form the application of Klein's inequality to the states $\rho  : =\rho_\rA\otimes \rho_\rB$ and $\sigma : =\rho_{\rA\rB}$.    
The subadditivity of the entropy guarantees that the \emph{mutual information}, defined as 
\[   I \left(  \rA  :  \rB  \right)_{\rho_{\rA\rB}}  :  =   S\left( \rho_\rA\right)    +  S\left( \rho_\rB\right)   -  S\left( \rho_{\rA\rB}\right)    \]
is a positive quantity and vanishes if and only if $\rho_{\rA\rB}$ is a product state.  

Another consequence of Klein's inequality 
is the
\emph{triangle inequality}: 
\begin{prop}[Triangle inequality]
	For every bipartite state $\rho_{\rA\rB}$ one has 
	\[
	S\left(\rho_{\mathrm{AB}}\right)\geq\left|S\left(\rho_{\mathrm{A}}\right)-S\left(\rho_{\mathrm{B}}\right)\right|,
	\]
	where $\rho_\rA$ and $\rho_\rB$ are the marginals of $\rho_{\rA\rB}$.  \end{prop}
	 The proof is the same as in the quantum case (see e.g.~\cite{Nielsen-Chuang}). Combining subadditivity and the triangle inequality, one obtains the bound 
\[
\left|S\left(\rho_{\mathrm{A}}\right)-S\left(\rho_{\mathrm{B}}\right)\right|\leq S\left(\rho_{\mathrm{AB}}\right)\leq S\left(\rho_{\mathrm{A}}\right)+S\left(\rho_{\mathrm{B}}\right),
\]
valid in all sharp theories with purification.

\section{Operational reconstruction of Landauer's principle}\label{sec:landauer}

\subsection{The second law lemma}  

Consider the evolution of a system in interaction with the surrounding environment, under the assumption that the system and the environment are uncorrelated at the initial time.    Consistently with the Purification Principle, here we assume that, by suitably enlarging the environment, the interaction can be modelled by a reversible channel $\map U$.    We denote   the initial states of the system and the environment by $\rho_\rS$ and $\rho_\rE$  respectively, so that the initial state of the composite system is  $\rho_{\rS\rE}  =  \rho_\rS  \otimes \rho_\rE$.  Primed states will denote the states after the interaction.  

The result of the interaction is typically to create correlations between the system and the environment, thus increasing the mutual information from the initial zero value to a  final non-zero value.  The creation of correlations can be equivalently phrased as an increase of the   sum of the system and environment entropies.   Indeed, the positivity of the mutual information gives the bound 
\begin{align*}
0   &  \le   I\left(   A:E \right) _{  \rho_{\rA\rE}'}  \\
   &   =    S  \left( \rho_{\rS}'\right)   +  S  \left( \rho_\rE'\right)   -   S\left(   \rho_{\rA\rE}' \right) \\
    &   =    S  \left( \rho_{\rS}'\right)   +  S  \left( \rho_\rE'\right)   -   S\left(   \rho_{\rA\rE} \right) \\
&  =  S  \left( \rho_{\rS}'\right)   +  S  \left( \rho_\rE'\right)   -   S  \left( \rho_{\rS}\right)   - S  \left( \rho_\rE\right)  ,
\end{align*}
the third line coming from the fact that reversible transformations do not change the entropy.  The resulting bound 
\begin{align}\label{almost2law} S  \left( \rho_{\rS}'\right)   +  S \left( \rho_\rE'\right)   \ge    S  \left( \rho_{\rS}\right)   +S  \left( \rho_\rE\right)    \end{align} 
 is sometimes  regarded as an elementary instance of the second law of thermodynamics \cite{Preskill}.   It is important, however, not to confuse the sum of the entropies $S  \left( \rho_\rS'\right)  +  S \left( \rho_\rE'\right) $ with the total entropy $S\left( \rho_{\rS\rE}'\right) $, which remains unchanged due to the reversibility of the global evolution.    The best reading of Eq.~\eqref{almost2law} is probably that a decrease in the entropy of the system must be accompanied by an increase  of the entropy of the environment.  
   Following Reeb and Wolf \cite{Reeb-Wolf} we will refer to Eq.~\ref{almost2law}) as the  \emph{second law lemma}.  

Operationally, the second law lemma is  the statement that uncorrelated systems can only become more correlated as the result of reversible interactions.   The interesting part of it is that ``correlations'' here are measured in terms of entropies: the existence of an entropic measure of correlations is a non-trivial consequence of the axioms. 
\subsection{Gibbs states}  

We have seen that every element of the effect vector space $\Eff_\R  (\rA)$ can be regarded as an observable.      Now, suppose that the only information we have about the state of system $\rA$ is the expectation value of a certain  observable $H$.    Which state should we assign to the system?   The maximum entropy principle \cite{Jaynes1,Jaynes2} posits that, among the states with the given expectation value,  we should choose the state that maximizes the (Shannon) entropy---in formula: 
\[
\rho_{\max}   =    \arg \max  \left\{   S \left( \rho\right)  ~|~   \left\langle   H \right\rangle_\rho   =    E    \right\}.         
\]
The maximum entropy state can be characterized in every sharp theory with purification, using the entropic techniques developed in the last section.  
As in quantum theory, it turns out that the maximum entropy states are the \emph{Gibbs states}, defined as  
\[
\rho_{\beta}    :=     \frac {  \mathrm{e}^{ -  \beta  H^\dag}}{  \Tr  \left(  \mathrm{e}^{-\beta  H^\dag}\right)}  ,   \qquad \beta \in \left[-\infty,+\infty\right] .
\]  
More explicitly, Gibbs states can be expressed as 
\[  \rho_\beta  =    \sum_{i=1}^d  \frac{ \mathrm{e}^{- \beta  E_i}}Z      \varphi_i  , \qquad Z:  =  \sum_{i=1}^d   \mathrm{e}^{-\beta E_i}   ,   \]
where the $E_i$'s are the eigenvalues of $H$ and  each $\varphi_i$ is a pure state such that $\left( H|\varphi_i\right)   = E_i$, namely the corresponding eigenstate. 
The expectation value of $H$ on the Gibbs state is given by 
\[     E\left( \beta\right)    :  =    \left\langle  H\right\rangle_{\rho_\beta}    =   -  \frac{ \d  }{\d \beta}    \ln   Z ,\]
and can assume all values between $E_{\mathrm{min}}$ and $E_{\mathrm{max}}$ (the minimum and maximum eigenvalue of $H$).   
If $ E_{\mathrm{min}}<E_{\mathrm{max}} $, namely $ H $ is not fully degenerate, the function  $E\left( \beta\right) $ is invertible \cite{Reeb-Wolf}. Denoting the inverse by $\beta  \left( E\right) $, we now show that  the Gibbs state $\rho_{\beta \left( E\right) }$ is the maximum entropy state with expectation value $E$.    The proof is based on an argument by Preskill \cite{Preskill}.  First, note that the entropy of a Gibbs state is  
\[  S\left( \rho_\beta\right)     =   \beta    E  \left( \beta\right)     +  \ln   Z .   \]
 Then use Klein's inequality
 \begin{align*}
 0   &\le  S\left(  \rho\parallel  \rho_\beta \right)   \\
   &  =  \left(  \ln \rho^\dagger    -   \ln \rho^\dagger_{\beta \left( E\right) }|\rho\right) \\
   & =   \left( \ln \rho^\dagger   +  \beta  \left( E\right)    H   + u\ln Z |\rho\right)   \\
   & =   -S\left( \rho\right)     +    \beta  \left( E\right)   E    +\ln Z  \\
   & =   -S  \left( \rho\right)   +  S  \left( \rho_{\beta\left( E\right) }\right),  
   \end{align*}
   which yields the bound 
   \[S\left( \rho\right)  \le S\left( \rho_{\beta  \left( E\right) }\right) , \] 
for every $ \rho $ such that $ \left\langle H\right\rangle_{\rho}=E $.

Motivated by the characterization  of the Gibbs states as maximum entropy states, we regard the Gibbs state $\rho_{\beta \left( E\right) }$ as the equilibrium state of a system with fixed expectation value $E$ of the observable $H$.  
  In the following we will focus on the case where  $H$ is the ``energy of the system''.    In this case, we will write the parameter  $\beta$ as $\beta=  1/k_\rB  T$, where $k_\rB$ is Boltzmann's constant and $T$ is interpreted as the ``temperature''.  Consistently, we will regard $\rho_\beta$ as the ``equilibrium state at temperature $T$''.

\subsection{An operational derivation of Landauer's principle}

The entropic tools constructed from the axioms allow us to prove an operational version of Landauer's principle, based on a recent argument by Reeb and Wolf \cite{Reeb-Wolf}.  The scenario considered here is that of a system $\rS$ that interacts reversibly with an environment, initially in the equilibrium state at temperature $T$.     In this context, Landauer's principle amounts to the statement that a decrease in the entropy of the system must be accompanied by an increase in the expected energy of the environment.  
More formally, we have the following
\begin{prop}[\cite{Reeb-Wolf}]
  Suppose that the system and the environment are initially in the product state $\rho_{\rS\rE}  =   \rho_\rS  \otimes \rho_{\rE,\beta}$ where $\rho_{\rE,\beta}   =   \mathrm{e}^{ - \beta    H_\rE^\dag}/{\Tr \left(  \mathrm{e}^{-\beta   H_\rE^\dag} \right) }$ is the equilibrium state at inverse temperature $\beta$ and $H_\rE$ is the energy of the environment. 
After a reversible interaction  $\map U$, the system and the environment will satisfy the equality  
\begin{align}
\nonumber \left\langle  H_\rE'  \right\rangle   -\left\langle   H_\rE  \right\rangle      &=  k_\rB  T  \left[    S\left( \rho_\rS\right) -   S  \left( \rho_\rS'\right)   \right.    \\
\label{landauer}   &    \quad \left. +   I\left(   \rS:\rE\right) _{\rho_{\rS\rE}'}   +  S \left(     \rho_\rE'  \parallel  \rho_{\rE, \beta}\right)   \right]  ,      
\end{align}  
where $\left\langle H_\rE\right\rangle   =   \left( H_\rE  |\rho_{\rE\beta}\right) $ and $\left\langle H_\rE'\right\rangle    =   \left( H_\rE|  \rho_\rE'\right) $ are the expectation values of the environment energy at the initial and final times, respectively.  
\end{prop} 

The equality follows from the definitions, specialized to the case where one state is the Gibbs state.     The key point is, again, Klein's inequality, which implies that the terms in the second line of Eq.~\eqref{landauer} are always non-negative and therefore one has the bound  
\begin{align}\label{landauerbound}
\left\langle  H_\rE'  \right\rangle   -\left\langle   H_\rE  \right\rangle     \ge  k_\rB  T  \left[    S\left( \rho_\rS\right) -   S  \left( \rho_\rS'\right)  \right]     ,
\end{align}
stating that it is impossible to reduce the entropy of the system without heating up the environment.   
Furthermore, the equality condition  in  Klein's inequality implies that the  lower bound \eqref{landauerbound} is attained if and only if \emph{i)} the system and the environment remain uncorrelated  after the interaction and \emph{ii)}  the environment remains in the equilibrium state.

\section{Conclusions\label{sec:conclusion}}
In this paper we proposed sharp theories with purification as an axiomatic foundation of statistical mechanics. In this class of theories, the Purification axiom guarantees that every mixed state can be regarded as  the marginal state of a composite system,  opening the way to the derivation of the equilibrium states from typicality arguments \cite{Popescu-Short-Winter,Canonical-typicality}  or dynamical symmetries of entanglement \cite{Zurek}.   
  The class of sharp theories with purification is fairly broad: it includes quantum theory with complex and real amplitudes, as well as a number of quantum theories with superselection rules.  Sharp theories also include  an extension of classical probability theory, where classical bits are complemented by coherent bits, a type of systems that can be entangled with classical bits.  The extended theory built on classical bits and cobits offers a new possibility for the foundations of classical statistical mechanics, allowing one to view  classical ensembles  as arising from pure joint  states of classical bits and cobits.      The example of classical theory motivates us to the following 
  \begin{conj}
  Every theory with a  ``well-behaved'' thermodynamics must be a subtheory of a sharp theory with purification. 
  \end{conj}  
As we  currently lack a formal  definition of  ``well-behaved'' thermodynamics, our conjecture is not a mathematical statement for the time being, but rather an open research programme.     The implementation of this programme is likely to proceed in two steps: The first step is to reconstruct the key structures of thermodynamics directly from the axioms of sharp theories with purification. This is the type of work initiated in the present paper with the derivation of the von Neumann entropy, the Gibbs state, and Landauer's principle.     The most urgent problem that remains open is an information-theoretic derivation of the strong subadditivity  \cite{Strong-subadditivity} and of the monotonicity of the quantum Kullback-Leibler divergence \cite{Petz}.  The proof of these results is notoriously difficult even in ordinary quantum theory, but the motivation is extremely strong, for these results are the key to the derivation of the second law of thermodyanamics  \cite{Preskill} and to its quantum generalizations \cite{2ndlaws}.   Another direction is the derivation of quantitative bounds on entanglement typicality, along the lines of \cite{Dahlsten,Muller-blackhole}, and  the derivation of the equilibrium ensembles from dynamical symmetries, thus achieving an axiomatic version of the approach of Daffner and Zurek \cite{Zurek}.     Results in these directions would bring further evidence that sharp theories with purification provide the appropriate ground for the construction of a well-founded statistical mechanics.  The second  and final step to the proof of our conjecture  is to rigorously formulate a set of desiderata about thermodynamics, and to \emph{derive} from there  the requirements that the underlying physical theory has to meet.   An example of desiderata is provided by Lieb-Yngvason axioms \cite{Lieb-Yngvason}, which capture the fundamental structures at the basis of the second law of thermodynamics. Connecting the GPT  with the Lieb-Yngvason is a promising route to approach our conjecture.

\begin{acknowledgments}
This work is supported by   the Foundational Questions Institute through the large grant ``The fundamental principles of information dynamics'' (FQXi-RFP3-1325), by the Canadian Institute for Advanced Research (CIFAR), by the 1000 Youth Fellowship Program of China, and by the HKU Seed Funding for Basic Research.   CMS acknowledges the support by a scholarship from ``Fondazione Ing.\ Aldo Gini'', by the Chinese Government Scholarship, by EPSRC Doctoral Training Grant and by Oxford-Google DeepMind Graduate Scholarship. 

We are indebted to M Krumm, H Barnum, J Barrett, and M Mueller for coordinating with us the arXiv posting  of their work and giving us the opportunity to complete our work.  
   GC thanks the organizers of the conference ``Information-Theoretic Foundations of Physics''  and  the participants  L Hardy, C Brukner, R Spekkens, G
't Hooft, R Oeckl, A Caticha,  H Barnum, M
M\"uller, G De Las Cuevas, R Schack, and C Beny for valuable comments on an early version of these results.    CMS thanks  M Krumm, A Wilce, L Hardy, J Barrett, and M Hoban for valuable discussions. Parts of this research have been done during visits to  the Simons Center for the Theory of Computation and Perimeter Institute. 
Research at Perimeter Institute for Theoretical Physics is supported in part by the Government of Canada through NSERC and by the Province of Ontario through MRI.
\end{acknowledgments}

\bibliographystyle{apsrev4-1}
\bibliography{Majorization}

\appendix

\section{Proof of the results of section~\ref{sec:pre-diagonalization}}

\subsection{Proof of corollary~\ref{cor:pupstar}\label{app:pupstar}}
By proposition 11 of Ref.~\cite{QPL15}, we already know that $ 1\Rightarrow2 $. Let us prove the converse implication $ 2\Rightarrow1 $.
Suppose that $  \left( a|\rho\right)   =  p_*$.  Then, for every purification of $\rho$, say $\Psi\in\mathsf{PurSt}_1 \left( \rA\otimes \rB\right)  $,  one has  
\begin{equation}\label{11}
\begin{aligned}\Qcircuit @C=1em @R=.7em @!R { & \multiprepareC{1}{\Psi}    & \qw \poloFantasmaCn{\rA} &  \measureD{a}   \\  & \pureghost{\Psi}    & \qw \poloFantasmaCn{\rB}  &   \qw }\end{aligned}~=q\!\!\!\!\begin{aligned}\Qcircuit @C=1em @R=.7em @!R { & \prepareC{\beta}    & \qw \poloFantasmaCn{\rB} &  \qw   }\end{aligned} ~, 
\end{equation}  
where $\beta$ is a normalized state, which is pure by Purity Preservation.    Now, 
\[
q=q\!\!\!\!\begin{aligned}\Qcircuit @C=1em @R=.7em @!R { & \prepareC{\beta}    & \qw \poloFantasmaCn{\rB} &  \measureD{u}  }\end{aligned} ~=\!\!\!\!\begin{aligned}\Qcircuit @C=1em @R=.7em @!R { & \multiprepareC{1}{\Psi}    & \qw \poloFantasmaCn{\rA} &  \measureD{a}   \\  & \pureghost{\Psi}    & \qw \poloFantasmaCn{\rB}  &   \measureD{u} }\end{aligned}~=\!\!\!\!\begin{aligned}\Qcircuit @C=1em @R=.7em @!R { & \prepareC{\rho}    & \qw \poloFantasmaCn{\rA} &  \measureD{u}  }\end{aligned}~=p_{*}~.
\]
Hence Eq.~\eqref{11} becomes
\begin{equation}\label{12}
\begin{aligned}\Qcircuit @C=1em @R=.7em @!R { & \multiprepareC{1}{\Psi}    & \qw \poloFantasmaCn{\rA} &  \measureD{a}   \\  & \pureghost{\Psi}    & \qw \poloFantasmaCn{\rB}  &   \qw }\end{aligned}~=p_{*}\!\!\!\!\begin{aligned}\Qcircuit @C=1em @R=.7em @!R { & \prepareC{\beta}    & \qw \poloFantasmaCn{\rB} &  \qw   }\end{aligned} ~. 
\end{equation}
This condition implies that $\beta$ is an eigenvector of the marginal state $\widetilde{\rho}=  \Tr_\rA\Psi$. The implication $ 1\Rightarrow2 $ guarantees that $\left( b|\widetilde{\rho}\right)   = p_*$, for every pure effect $ b $ such that $\left( b|\beta\right) =1$.    
The last condition implies an equation very similar to Eq.~\eqref{12}:
\begin{equation}\label{22}
\begin{aligned}\Qcircuit @C=1em @R=.7em @!R { & \multiprepareC{1}{\Psi}    & \qw \poloFantasmaCn{\rA} &  \qw   \\  & \pureghost{\Psi}    & \qw \poloFantasmaCn{\rB}  &   \measureD{b} }\end{aligned}~=p_{*}\!\!\!\!\begin{aligned}\Qcircuit @C=1em @R=.7em @!R { & \prepareC{\alpha'}    & \qw \poloFantasmaCn{\rA} &  \qw   }\end{aligned} ~.
\end{equation}
for some pure state $\alpha'$.  Hence, $\alpha'$ is an eigenvector of $\rho$ with eigenvalue $p_*$.  To conclude the proof, it is enough to observe that $\alpha'  =  \alpha$.     Indeed, combining Eqs.~\eqref{11} and \eqref{22}  we have
\[    \left(   a|  \alpha'\right)     =  \frac{1}{p_*} \!\!\!\!\begin{aligned}\Qcircuit @C=1em @R=.7em @!R { & \multiprepareC{1}{\Psi}    & \qw \poloFantasmaCn{\rA} &  \measureD{a}   \\  & \pureghost{\Psi}    & \qw \poloFantasmaCn{\rB}  &   \measureD{b} }\end{aligned}~= \left( b|\beta\right)   =  1 ,     \]
which implies $\alpha'  = \alpha$ by proposition~\ref{prop:uniqueness of state}.  
\qed

\subsection{Proof of proposition~\ref{prop:pstar=pstar}\label{app:pstar=pstar}}
It is enough we show that $ p^*  \le p_*  $. Pick a pure effect $a\in\Pur\Eff\left( \rA\right) $ such that $\left( a|\rho\right)  \neq    0$. Such a pure effect exists because any pure effect $ a $ such that $ \left(a|\alpha \right) =1 $, where $ \alpha $ is an eigenstate of $ \rho $ with maximum eigenvalue $ p_* $, has the property $\left(a|\rho \right)=p_* \neq 0$. Now consider a purification $ \Psi \in \mathsf{PurSt}_{1}\left(\mathrm{A}\otimes\mathrm{B} \right) $ of $ \rho $, and define the pure state  $\beta$ as   
\[
p\!\!\!\!\begin{aligned}\Qcircuit @C=1em @R=.7em @!R { & \prepareC{\beta}    & \qw \poloFantasmaCn{\rB} &  \qw   }\end{aligned}~=\!\!\!\!\begin{aligned}\Qcircuit @C=1em @R=.7em @!R { & \multiprepareC{1}{\Psi}    & \qw \poloFantasmaCn{\rA} &  \measureD{a}   \\  & \pureghost{\Psi}    & \qw \poloFantasmaCn{\rB}  &   \qw }\end{aligned}~.
\]
Note that $ p $ is non-vanishing because it is given by $ p=\left( a|\rho\right) $. So $ \beta $ arises in a convex decomposition of the marginal of $ \Psi $ on $ \mathrm{B} $ with probability $ p $. By construction $ p\leq p_* $, namely $ \left( a|\rho\right) \leq p_* $. Taking the supremum over $a$, we finally obtain $p^* \le p_{*}$, thus proving that $ p^*=p_* $.
\qed

\subsection{Proof of proposition~\ref{prop:steerpstar}\label{app:steerpstar}}
Let $\rho_\rB$ be the marginal of $\Phi$ on system $\rB$, written as $\rho_\rB   =  p_* \beta_0   +   \left( 1-p_*\right) \tau$, for some pure state $\beta_0$ and some state $\tau$, and let $b_0$ be a pure effect such that $\left(b_0 |\beta_0\right)=1 $.        Then, corollary~\ref{cor:pupstar} implies  $\left( b_0| \rho_\rB\right) =  p_*$.  
Now, let us apply $b_0$ on the pure state $\Phi$.  By Purity Preservation, we must have 
\[
\begin{aligned}\Qcircuit @C=1em @R=.7em @!R { & \multiprepareC{1}{\Phi}    & \qw \poloFantasmaCn{\rA} &  \qw   \\  & \pureghost{\Phi}    & \qw \poloFantasmaCn{\rB}  &   \measureD{b_0} }\end{aligned}~=p_{*}\!\!\!\!\begin{aligned}\Qcircuit @C=1em @R=.7em @!R { & \prepareC{\alpha_0}    & \qw \poloFantasmaCn{\rA} &  \qw   }\end{aligned} 
\]  
for some pure state $\alpha_0$.   By transitivity, there exists a reversible channel $\mathcal  U$ such that $\mathcal U  \alpha_0  =  \alpha$.    Moreover, since the states $\left( \mathcal U  \otimes \mathcal I_\rB\right)  \Phi$ and $\Phi$ are both purifications of the invariant state $\chi_\rA$, the uniqueness of purification implies that there exists another reversible transformation $\mathcal   V$ such that 
\[
\begin{aligned}\Qcircuit @C=1em @R=.7em @!R { & \multiprepareC{1}{\Phi}    & \qw \poloFantasmaCn{\rA} &  \gate{\mathcal U}  & \qw \poloFantasmaCn{\rA} &  \qw \\  & \pureghost{\Phi}    & \qw \poloFantasmaCn{\rB}  &   \qw&\qw&\qw}\end{aligned} ~= \!\!\!\!  \begin{aligned}\Qcircuit @C=1em @R=.7em @!R { & \multiprepareC{1}{\Phi}    & \qw \poloFantasmaCn{\rA} &  \qw  & \qw  &  \qw \\  & \pureghost{\Phi}    & \qw \poloFantasmaCn{\rB}  &   \gate{\mathcal V}&\qw \poloFantasmaCn{\rB}&\qw}\end{aligned}  \, .
\] 
Defining the pure effect $ b:=   b_0   \mathcal V$ we then obtain the desired equality:  
\[
\begin{aligned}\Qcircuit @C=1em @R=.7em @!R { & \multiprepareC{1}{\Phi}    & \qw \poloFantasmaCn{\rA} &  \qw   \\  & \pureghost{\Phi}    & \qw \poloFantasmaCn{\rB}  &   \measureD{b} }\end{aligned}~ = \!\!\!\! \begin{aligned}\Qcircuit @C=1em @R=.7em @!R { & \multiprepareC{1}{\Phi}    & \qw \poloFantasmaCn{\rA} &  \qw   &\qw &\qw  \\  & \pureghost{\Phi}    & \qw \poloFantasmaCn{\rB}  &\gate{\mathcal V} & \qw \poloFantasmaCn{\rB}   &    \measureD{b_0} }\end{aligned} ~=
\]
\[ 
=\!\!\!\!  \begin{aligned}\Qcircuit @C=1em @R=.7em @!R { & \multiprepareC{1}{\Phi}    & \qw \poloFantasmaCn{\rA} &  \gate{\mathcal U}  & \qw \poloFantasmaCn{\rA} &\qw\\  & \pureghost{\Phi}    & \qw \poloFantasmaCn{\rB}   &\qw &\qw &\measureD{b_0} }  \end{aligned} ~
= p_{*} \!\!\!\! \begin{aligned}\Qcircuit @C=1em @R=.7em @!R { & \prepareC{\alpha_0}    & \qw \poloFantasmaCn{\rA} &   \gate{\mathcal U}  &  \qw \poloFantasmaCn{\rA} & \qw   }\end{aligned}~=
\]
\[   
= p_{*} \!\!\!\! \begin{aligned}\Qcircuit @C=1em @R=.7em @!R { & \prepareC{\alpha}    & \qw \poloFantasmaCn{\rA} & \qw   }
\end{aligned}  ~ .
\]  
This proves Eq.~\eqref{daprovare}. To conclude the proof, since there is a pure state $ \alpha $ associated with every normalized pure effect $ a $ such that $ \left(a|\alpha \right)=1 $, and every pure state is an eigenstate of $ \chi_{\mathrm{A}} $ with maximum eigenvalue, by corollary~\ref{cor:pupstar}, we have $\left( a|\chi_\rA\right)=p_{*} $. \qed

\subsection{Proof of proposition~\ref{prop:atmostoneeffect}\label{app:atmostoneeffect}}
Suppose that $a$ and $a'$ are two pure effects such that  $\left(a| \alpha\right) =\left(a'| \alpha\right)=1 $. Then, let $\Phi\in\mathsf{PurSt }\left( \rA\otimes \rB\right)$ be a purification of the invariant state   $\chi_\rA$.  By proposition~\ref{prop:steerpstar}, there exists a pure effect $b$ such that    
\begin{equation}\label{uffissima}
\begin{aligned}\Qcircuit @C=1em @R=.7em @!R { & \multiprepareC{1}{\Phi}    & \qw \poloFantasmaCn{\rA} &  \qw   \\  & \pureghost{\Phi}    & \qw \poloFantasmaCn{\rB}  &   \measureD{b} }\end{aligned}~=p_{*}\!\!\!\!\begin{aligned}\Qcircuit @C=1em @R=.7em @!R { & \prepareC{\alpha}    & \qw \poloFantasmaCn{\rA} &  \qw   }\end{aligned} ~,
\end{equation}
and the two effects $a$ and $a'$ must satisfy
\begin{equation}\label{achiaprimchi} 
\left( a| \chi_\rA\right)   =  p_*    =   \left( a'|\chi_\rA\right)   .
\end{equation}
Now, let us define the pure states $\beta$ and $\beta'$ via the relations  
\[
\begin{aligned}\Qcircuit @C=1em @R=.7em @!R { & \multiprepareC{1}{\Phi}    & \qw \poloFantasmaCn{\rA} &  \measureD{a}   \\  & \pureghost{\Phi}    & \qw \poloFantasmaCn{\rB}  &   \qw }\end{aligned}~  =:q\!\!\!\!\begin{aligned}\Qcircuit @C=1em @R=.7em @!R { & \prepareC{\beta}    & \qw \poloFantasmaCn{\rB} &  \qw   }\end{aligned} ~ ,
\]
\[
\begin{aligned}\Qcircuit @C=1em @R=.7em @!R { & \multiprepareC{1}{\Phi}    & \qw \poloFantasmaCn{\rA} &  \measureD{a'}   \\  & \pureghost{\Phi}    & \qw \poloFantasmaCn{\rB}  &   \qw }\end{aligned}~  =:q'\!\!\!\!\begin{aligned}\Qcircuit @C=1em @R=.7em @!R { & \prepareC{\beta'}    & \qw \poloFantasmaCn{\rB} &  \qw   }\end{aligned}	~,
\]
where $q$ and $q'$ are suitable probabilities.   Applying the deterministic effect on both sides and using  Eq.~\eqref{achiaprimchi} one obtains the equality $  q  =    p_* =   q' $.    Hence, Eqs.~\eqref{uffissima}, \eqref{achiaprimchi}, and \eqref{definitions}  lead to the equalities
\begin{align*}
\left( b  |  \beta\right)     &=     \frac {1} {p_*}\!\!\!\! \begin{aligned}\Qcircuit @C=1em @R=.7em @!R { & \multiprepareC{1}{\Phi}    & \qw \poloFantasmaCn{\rA} &  \measureD{a}   \\  & \pureghost{\Phi}    & \qw \poloFantasmaCn{\rB}  &   \measureD{b} }\end{aligned} ~=  \left( a|\alpha\right)     =  1 \\
\left( b  |  \beta'\right)      &=     \frac {1} {p_*} \!\!\!\! \begin{aligned}\Qcircuit @C=1em @R=.7em @!R { & \multiprepareC{1}{\Phi}    & \qw \poloFantasmaCn{\rA} &  \measureD{a'}   \\  & \pureghost{\Phi}    & \qw \poloFantasmaCn{\rB}  &   \measureD{b} }\end{aligned}         =     \left( a'|\alpha\right)     =  1 .
\end{align*}
By proposition~\ref{prop:uniqueness of state} we conclude that $\beta$ and $\beta'$ must be equal.   Recalling the definitions of $\beta$ and $\beta'$, we then obtain the relation  
\begin{equation}\label{definitions}
\begin{aligned}\Qcircuit @C=1em @R=.7em @!R { & \multiprepareC{1}{\Phi}    & \qw \poloFantasmaCn{\rA} &  \measureD{a}   \\  & \pureghost{\Phi}    & \qw \poloFantasmaCn{\rB}  &   \qw }\end{aligned}~ = \!\!\!\!
\begin{aligned}\Qcircuit @C=1em @R=.7em @!R { & \multiprepareC{1}{\Phi}    & \qw \poloFantasmaCn{\rA} &  \measureD{a'}   \\  & \pureghost{\Phi}    & \qw \poloFantasmaCn{\rB}  &   \qw }\end{aligned}~,
\end{equation}
which implies $a=a'$ because the state $\Phi$ is dynamically faithful on system $\rA$  (proposition~\ref{prop:faithful-effects}).  
\qed  

\section{From minimally disturbing transformations to the distinguishability of states}
\subsection{Proof of proposition~\ref{prop:non-disturbing measurement}\label{app:non-disturbing measurement}}
The starting point of the proof is a result of Ref.~\cite{Chiribella-informational}, which guarantees that every effect $a  \in\mathsf{Eff}\left( \rA\right) $ can be written as 
\begin{equation}\label{effdecomp} 
a = u_\rB  \mathcal A ,
\end{equation}
where $\mathcal A$ is a pure transformation from $\rA$ to $\rB$ and $\rB$ is a suitable system.  

Now, let $\Psi\in\mathsf{PurSt}_{1}\left(\mathrm{A}\otimes\mathrm{A}'\right)$
be a purification of $\rho$.  By Eq.~\eqref{effdecomp}, we have
\begin{equation}\label{uno}
\begin{aligned} \Qcircuit @C=1em @R=.7em @!R { & \multiprepareC{1}{\Psi} & \qw \poloFantasmaCn{\rA} & \gate{\mathcal A} & \qw \poloFantasmaCn{\rB} &\measureD{u} \\ & \pureghost{\Psi} & \qw \poloFantasmaCn{\rA'} & \qw &\qw &\qw }\end{aligned} ~= \!\!\!\! \begin{aligned}\Qcircuit @C=1em @R=.7em @!R { & \multiprepareC{1}{\Psi} & \qw \poloFantasmaCn{\rA} & \measureD{a} \\ & \pureghost{\Psi} & \qw \poloFantasmaCn{\rA'} & \qw }\end{aligned}
\end{equation}
Now,   since $(a|\rho)=1$, we have $a =_\rho  u_\rA$.   Hence, proposition \ref{prop:purifications -> input} implies  \begin{align}\label{due}
\begin{aligned}\Qcircuit @C=1em @R=.7em @!R { & \multiprepareC{1}{\Psi} & \qw \poloFantasmaCn{\rA} & \measureD{a} \\ & \pureghost{\Psi} & \qw \poloFantasmaCn{\rA'} & \qw }\end{aligned}~= \!\!\!\! \begin{aligned}\Qcircuit @C=1em @R=.7em @!R { & \multiprepareC{1}{\Psi} & \qw \poloFantasmaCn{\rA} & \measureD{u} \\ & \pureghost{\Psi} & \qw \poloFantasmaCn{\rA'} & \qw }\end{aligned}~.
\end{align}
Combining Eqs.~\eqref{uno}  and \eqref{due},  we obtain \[
\begin{aligned} \Qcircuit @C=1em @R=.7em @!R { & \multiprepareC{1}{\Psi} & \qw \poloFantasmaCn{\rA} & \gate{\mathcal A} & \qw \poloFantasmaCn{\rB} &\measureD{u} \\ & \pureghost{\Psi} & \qw \poloFantasmaCn{\rA'} & \qw &\qw &\qw }\end{aligned} ~=\!\!\!\! \begin{aligned}\Qcircuit @C=1em @R=.7em @!R { & \multiprepareC{1}{\Psi} & \qw \poloFantasmaCn{\rA} & \measureD{u} \\ & \pureghost{\Psi} & \qw \poloFantasmaCn{\rA'} & \qw }\end{aligned}~,
\] meaning that the two pure states $\left(\mathcal{A}\otimes\mathcal{I}_{\mathrm{A}'}\right)\Psi$
and $\Psi$ have the same marginal on system $\mathrm{A}'$. By the
uniqueness of purification, for fixed pure states $\alpha_{0}\in\mathsf{PurSt}_{1}\left(\mathrm{A}\right)$
and $\beta_{0}\in\mathsf{PurSt}_{1}\left(\mathrm{B}\right)$, there
must exist a reversible transformation $\mathcal{U}$ on $\mathrm{A}\otimes\mathrm{B}$,
such that
\[
\begin{aligned} \Qcircuit @C=1em @R=.7em @!R { & \prepareC{\alpha_0} & \qw \poloFantasmaCn{\rA} & \qw & \qw & \multigate{1}{\mathcal U} & \qw \poloFantasmaCn{\rB} &\qw \\ & \multiprepareC{1}{\Psi} & \qw \poloFantasmaCn{\rA} & \gate{\mathcal A} & \qw \poloFantasmaCn{\rB} &\ghost{\mathcal U} & \qw \poloFantasmaCn{\rA} &\qw \\ & \pureghost{\Psi} & \qw \poloFantasmaCn{\rA'} & \qw &\qw &\qw &\qw &\qw} \end{aligned}~=\!\!\!\! \begin{aligned}\Qcircuit @C=1em @R=.7em @!R { & \prepareC{\beta_0} & \qw \poloFantasmaCn{\rB} & \qw \\ & \multiprepareC{1}{\Psi} & \qw \poloFantasmaCn{\rA} & \qw \\ & \pureghost{\Psi} & \qw \poloFantasmaCn{\rA'} & \qw }\end{aligned}~.
\]
Applying $ \beta_{0}^\dagger $ to both sides, we obtain
\[
\begin{aligned} \Qcircuit @C=1em @R=.7em @!R { & \multiprepareC{1}{\Psi} & \qw \poloFantasmaCn{\rA} & \gate{\mathcal A} & \qw \poloFantasmaCn{\rB} & \gate{\mathcal P} & \qw \poloFantasmaCn{\rA} &\qw \\ & \pureghost{\Psi} & \qw \poloFantasmaCn{\rA'} & \qw &\qw &\qw &\qw &\qw}\end{aligned} ~= \!\!\!\! \begin{aligned}\Qcircuit @C=1em @R=.7em @!R { & \multiprepareC{1}{\Psi} & \qw \poloFantasmaCn{\rA} & \qw \\ & \pureghost{\Psi} & \qw \poloFantasmaCn{\rA'} & \qw }\end{aligned}~,
\]where $\mathcal{P}$ is the pure transformation defined as\[
\begin{aligned} \Qcircuit @C=1em @R=.7em @!R { & \qw \poloFantasmaCn{\rB} & \gate{\mathcal P} & \qw \poloFantasmaCn{\rA} &\qw }\end{aligned} ~:= \!\!\! \begin{aligned} \Qcircuit @C=1em @R=.7em @!R { & \prepareC{\alpha_0} & \qw \poloFantasmaCn{\rA} & \multigate{1}{\mathcal U} & \qw \poloFantasmaCn{\rB} &\measureD{\beta_0^{\dagger}} \\ && \qw \poloFantasmaCn{\rB} & \ghost{\mathcal U} & \qw \poloFantasmaCn{\rA} &\qw } \end{aligned}~.
\]
Let us define  the transformation $\mathcal{T}:=\mathcal{P}\mathcal{A}$, which is pure by Purity Preservation. With this choice,  we  have
\[
\begin{aligned} \Qcircuit @C=1em @R=.7em @!R { & \multiprepareC{1}{\Psi} & \qw \poloFantasmaCn{\rA} & \gate{\mathcal T} & \qw \poloFantasmaCn{\rA} &\qw \\ & \pureghost{\Psi} & \qw \poloFantasmaCn{\rA'} & \qw &\qw &\qw }\end{aligned} ~= \!\!\!\! \begin{aligned}\Qcircuit @C=1em @R=.7em @!R { & \multiprepareC{1}{\Psi} & \qw \poloFantasmaCn{\rA} & \qw \\ & \pureghost{\Psi} & \qw \poloFantasmaCn{\rA'} & \qw } \end{aligned} \, , 
\]
which implies $\mathcal{T}=_{\rho}\mathcal{I}$ by proposition~\ref{prop:purifications -> input}.
Finally, for all states $ \sigma \in \mathsf{St} \left( \rA\right) $ we have  the inequality
\[
\left(  u_\rA  \left|    \mathcal{T}   \right|\sigma\right)=\left (  u _\rA  \right|   \mathcal{P}\mathcal{A}  \left| \sigma\right)\le \left(  u_\rB  \left|   \mathcal{A} \right|\sigma\right)=\left(a|\sigma\right).
\]
Here, the inequality follows from proposition~\ref{prop:norm-transformations} applied to the norm of $ \mathcal{A}\sigma $ under the action of the transformation $ \mathcal{P} $, while   the last equality follows from Eq.~\eqref{effdecomp}.  
\qed

\subsection{Distinguishability of pure states\label{app:distinguishability of states}}

Using the existence of non-disturbing transformations we can now give a sufficient condition for the perfect distinguishability of pure states. 

\begin{lemma}
	\label{lem:distinguishable}
	Let  $\left\{ \rho_{i}\right\} _{i=1}^{n}$ be a set of normalized states. If there exists a set of effects (not necessarily an observation-test or a subset of an observation-test) $\left\{a_i\right\}_{i=1}^n$  such that  
	\[  \left(a_i |\rho_{j}\right)=0   \qquad \forall i  \in  \left\{1,\ldots, n\right\} ,  \forall j>i \]
	then the states  $\left\{\rho_{i}\right\} _{i=1}^{n}$  are perfectly distinguishable.
\end{lemma}
\Proof 
By hypothesis, the binary observation-test $\left\{ a_i,u-a_i\right\} $
distinguishes perfectly between $\rho_{i}$ and all the other states $\rho_{j}$ with $j>i$. Equivalently, this observation-test distinguishes
perfectly between $\rho_{i}$ and the state $\widetilde{\rho}_{i}:=\frac{1}{n-i} \sum_{j>i}\rho_{j}$.
Specifically, $\left(u-a_i|\widetilde{\rho}_{i}\right)=1$.
Applying proposition~\ref{prop:non-disturbing measurement}, we can
construct a pure transformation $\mathcal{A}_{i}^{\perp}$ such that
$\mathcal{A}_{i}^{\perp}=_{\widetilde{\rho}_{i}}\mathcal{I}$, and, specifically,
\begin{equation}\label{eq:orthogonal non disturbing}
\mathcal{A}_{i}^{\perp}\rho_{j}=\rho_{j}\qquad\forall j>i.
\end{equation}
Moreover, proposition~\ref{prop:non-disturbing measurement}  implies 
\[
\left(u  \right|  \mathcal{A}_{i}^{\perp}  \left|\rho_{i}\right)\le  \left(u-a_{i}|\rho_{i}\right)=0,
\]
meaning that the transformation $\mathcal{A}_{i}^{\perp}$ never occurs
on the state $\rho_{i}$. Let us define the effect $  a_{i,0}  :  =   u-  a_i  -  u  \mathcal A_i^\perp$. Note that this effect is well-defined, because $ \left( a_{i,0}|\sigma\right) \geq0 $, for all $ \sigma\in\mathsf{St}_{1}\left( \mathrm{A}\right)  $. Indeed, by proposition~\ref{prop:non-disturbing measurement}, we have $ \left(u  \right|  \mathcal{A}_{i}^{\perp}  \left|\sigma\right)\le  \left(u-a_{i}|\sigma\right) $, for all $ \sigma \in\mathsf{St}_{1}\left( \mathrm{A}\right)$, whence $ \left(u-  a_i  -  u  \mathcal A_i^\perp |\sigma \right)\geq 0 $. Note that $ a_{i,0} $ never occurs on the states $\rho_k$ with $k\ge i$.    

Now, define the transformations    $\mathcal A_i    =      \rho_i a_i$  and  $\mathcal A_{i,0}   =   \rho_0 a_{i,0} $, where $\rho_0$ is a fixed normalized state.    
By proposition~\ref{prop:sufficientfortest},  the transformations $\left\{\mathcal A_i, \mathcal A_i^\perp, \mathcal A_{i,0}\right\}$ form a test. Summarizing the above observations,  the test satisfies the properties 
\begin{align}\label{eq:propertiesA}
\nonumber\mathcal A_i   \rho_i   & =   \rho_i \\
\nonumber\mathcal A_i   \rho_j   & =   0 \qquad \forall j>i  \\
\nonumber\mathcal A_i^\perp   \rho_i&  =  0 \\
\nonumber\mathcal A_i^\perp   \rho_j&  = \rho_j  \qquad \forall j>i  \\
\mathcal A_{i,0}  \rho_k   &   =  0   \qquad \forall k  \ge i. 
\end{align}  
By construction, the test distinguishes without error between the
state $\rho_{i}$ and all the states $\rho_{j}$ with $j>i$,
in such a way that the latter are not disturbed. Indeed, by construction $ \mathcal{A}_{i} $ can only occur if the state is $ \rho_{i} $, instead $ \mathcal{A}_{i}^{\perp} $ never occurs on $ \rho_{i} $, but it occurs with probability 1 if the state is any of the $ \rho_{j} $'s, with $ j>i $, and it leaves them unchanged. Finally, $ \mathcal A_{i,0} $ never occurs on the states $ \rho_k $'s with $ k  \ge i $, so it does not play a role in the discrimination process. Essentially, $ \mathcal A_{i,0} $ only plays the role of making  $\left\{\mathcal A_i, \mathcal A_i^\perp, \mathcal A_{i,0}\right\}$ a test.

Using the tests $\left\{ \mathcal{A}_{i},\mathcal{A}_{i}^{\perp} , \mathcal A_{i,0}\right\} $
it is easy to construct a protocol that distinguishes perfectly between
the states $\left\{ \rho_{i}\right\} _{i=1}^{n}$. The protocol
works as follows: starting from $i=1$ perform the
test $\left\{ \mathcal{A}_{i},\mathcal{A}_{i}^{\perp},\mathcal A_{i,0}\right\} $.
If the transformation $\mathcal{A}_{i}$ takes place, then the state
is $\rho_{i}$. If the transformation $\mathcal{A}_{i}^{\perp}$
takes place, then perform the test $\left\{ \mathcal{A}_{i+1},\mathcal{A}_{i+1}^{\perp} ,  \mathcal A_{i+1,0}\right\} $ (this can be done because $\mathcal{A}_{i}^{\perp}$ is non-disturbing). 
Using this protocol, every state in the set $\left\{\rho_i\right\}_{i=1}^n$ will be identified without error in at most $n$ steps.    Overall, the protocol is described by a test with $2n+1$ outcomes, corresponding to the transformations  
\begin{align}\label{eq:distinguishing test}
\nonumber\mathcal T_1   &  =   \mathcal A_1  \\
\nonumber\mathcal T_2  &  =   \mathcal A_2  \mathcal A_1^\perp  \\
\nonumber&~~\vdots  \\
\nonumber\mathcal T_n  &  =  \mathcal A_n   \mathcal A_{n-1}^\perp  \ldots  \mathcal A_1^\perp  \\
\nonumber\mathcal T_{n+1}  &  =   \mathcal A_{1,0}  \\
\nonumber\mathcal T_{n+2}  &  =   \mathcal A_{2,0}\mathcal A_{1}^{\perp}  \\
\nonumber&~~\vdots  \\
\nonumber\mathcal T_{2n}  &  =   \mathcal A_{n,0}   \mathcal A_{n-1}^\perp  \ldots \mathcal A_1^\perp \\
\mathcal T_{2n+1}  &  =   \mathcal A_{n}^\perp  \ldots \mathcal A_1^\perp
\end{align}
To show that these transformations form a test, we use proposition~\ref{prop:sufficientfortest}:  $ \left\lbrace \mathcal{T}_i\right\rbrace_{i=1}^{2n+1} $ is a test if and only if $ \sum_{i=1}^{2n+1}u\mathcal{T}_{i}=u $. An easy check shows that this is the case.

To complete the proof, we need to construct a perfectly distinguishing test $ \left\lbrace e_{i} \right\rbrace_{i=1}^{n} $ for the states $ \left\lbrace \rho_{i} \right\rbrace_{i=1}^{n} $. Discarding the output of the transformations of Eq.~\eqref{eq:distinguishing test}, we get an observation-test $\left\lbrace t_{i} \right\rbrace_{i=1}^{2n+1}$ with $ 2n+1 $ outcomes and effects $t_i   : =  u   \mathcal T_i$. We claim that the observation-test \begin{equation}\label{eq:observation-test}
\left\lbrace e_{i} \right\rbrace_{i=1}^{n}=\left\lbrace t_{1},\ldots, t_{n-1},u-t_{1}-\ldots-t_{n-1} \right\rbrace 
\end{equation}
is perfectly distinguishing for the states $ \left\lbrace \rho_{i} \right\rbrace_{i=1}^{n} $. First of all, since $ t_{1},\ldots,t_{n-1} $ coexist in a ($ 2n+1 $)-outcome test, the effect $ u-t_{1}-\ldots-t_{n-1} $ is well-defined. Now let us prove that the observation-test~\eqref{eq:observation-test} perfectly distinguishes between the states $ \rho_{i} $'s. We start from $ t_1=u\mathcal{A}_1 $; from Eq.~\eqref{eq:propertiesA} we get
\[
\left(t_1|\rho_j \right)=\left( u\right|\mathcal{A}_1\left|\rho_j \right)=\delta_{1j}\left( u|\rho_j\right) =   \delta_{1j}.
\]
If now $ i>1 $,
\[   t_i=u \mathcal{A}_{i}\mathcal{A}_{i-1}^{\perp}\ldots\mathcal{A}_{1}^{\perp}.
\] If we wish to calculate $ \left( t_i | \rho_{j}\right)  $, by Eq.~\eqref{eq:propertiesA}, $ \rho_{j} $ is left invariant by all the $ \mathcal{A}_{k}^\perp $ with $ k<j $. If $ i\neq j $, then
\[
\left( t_i | \rho_{j}\right)  =\left( u\right|  \mathcal{A}_{i}\mathcal{A}_{i-1}^{\perp}\ldots\mathcal{A}_{j}^{\perp}\left|\rho_{j} \right)=0, 
\]
again by Eq.~\eqref{eq:propertiesA}. If, instead $ j=i $,
\[
\left( t_i | \rho_{i}\right)=\left( u\right|\mathcal{A}_i\left|\rho_i \right)=\left( u|\rho_i\right) =1.
\]
As a consequence of these result
\[
\left( u-t_{1}-\ldots-t_{n-1}|\rho_j\right) =\delta_{nj}.
\]
We conclude that $ \left\lbrace e_{i} \right\rbrace_{i=1}^{n} $ is really a perfectly distinguishing test, because we have $ \left( e_i | \rho_{j}\right)=\delta_{ij} $.	  	
\qed

In the case when the states are pure and the effects in the statement of lemma~\ref{lem:distinguishable} are the daggers of those pure states, we can prove something more.
\begin{cor}
	\label{cor:dagger distinguishable}
	Let  $\left\{ \alpha_{i}\right\} _{i=1}^{n}$ be a set of normalized \emph{pure} states such that
	\[  \left(\alpha_i ^{\dagger}| \alpha_{j}\right)=0   \qquad \forall i  \in  \left\{1,\ldots, n\right\} ,  \forall j>i \]
	then the states  $\left\{ \alpha_{i}\right\} _{i=1}^{n}$  are perfectly distinguishable, and the pure effects $\left\{\alpha_i^{\dagger} \right\}_{i=1}^n$ coexist in an observation-test, which distinguishes perfectly between the states $\left\{ \alpha_{i}\right\} _{i=1}^{n}$. As a consequence, $ \left( \alpha_i ^{\dagger}| \alpha_{j}\right)=\delta_{ij} $.
\end{cor}	      
\Proof If we take $ a_i:=\alpha_i^\dagger $, by lemma~\ref{lem:distinguishable}, we know that the states $\left\{ \alpha_{i}\right\} _{i=1}^{n}$ are perfectly distinguishable. Referring to the proof of lemma~\ref{lem:distinguishable}, note that, since $\mathcal A_i^\perp$ is pure, we have  
\begin{equation}\label{usadopo}  \alpha_j^\dag \mathcal A_i^\perp   =   \alpha_j^\dagger \qquad \forall j>i.\end{equation}
Indeed, the effect $      \alpha_j^\dag \mathcal A_i^\perp$ is pure by Purity Preservation, and satisfies \[ \left( \alpha_j^\dag \right| \mathcal A_i^\perp \left|\alpha_j\right)   =  \left( \alpha_j^\dag|\alpha_j\right) =1,\]
where we have used Eq.~\eqref{eq:orthogonal non disturbing}.
This means that it must be equal to $\alpha_j^\dag$  (proposition~\ref{prop:atmostoneeffect}). Let us construct the perfectly distinguishing observation-test like in the proof of lemma~\ref{lem:distinguishable}, by considering the effects $ t_i=u\mathcal{A}_i $. One has, recalling that $ \mathcal{A}_i = \alpha_{i}\alpha_{i}^{\dagger}$,
\begin{align}\label{eq:t_i}
\nonumber t_1  &  =   u  \mathcal  A_1   =  \alpha_1^\dag  \\
\nonumber t_2  &  =       u \mathcal A_2 \mathcal  A^\perp_1   =  \alpha_2^\dag A^\perp_1=\alpha_2^\dag\\
\nonumber& ~~\vdots\\
t_n  &  =   u  \mathcal A_n  \mathcal A_{n-1}^\perp  \ldots \mathcal A_1^\perp  = \alpha_n^\dag ,
\end{align}
having used Eq.~\eqref{usadopo}.  This proves that the effects $\left\{ \alpha^\dag_{i}\right\} _{i=1}^{n}$ coexist in a ($ 2n+1 $)-outcome observation-test. As a consequence, as shown in the proof of lemma~\ref{lem:distinguishable}, we have that
\[
\left\lbrace \alpha_{1}^{\dagger},\ldots,\alpha_{n-1}^{\dagger},u-\alpha_{1}^{\dagger}-\ldots-\alpha_{n-1}^{\dagger}\right\rbrace
\]
is perfectly distinguishing, and specifically $ \left( \alpha_i ^{\dagger}| \alpha_{j}\right)=\delta_{ij} $.
\qed

\section{Proof of the results of section~\ref{sec:Diagonalization-of-states}}
\subsection{Proof of theorem~\ref{thm:diago}\label{app:diago}}
The proof consists of a constructive procedure for diagonalizing
arbitrary states. In order to diagonalize the state $\rho$, it is
enough to proceed along the following steps:
\begin{enumerate}
	\item Set $\rho_{1}=\rho$ and $p_{*,0}=0$
	\item For $i$ starting from $i=1$, decompose $\rho_{i}$ as $\rho_{i}=p_{*,i}\alpha_{i}+\left(1-p_{*,i}\right)\sigma_{i}$,  where $p_{*,i}$ is the maximum eigenvalue of $\rho_i$.  
	Set $\rho_{i+1}=\sigma_{i}$,
	$p_{i}=p_{*,i}\prod_{j=0}^{i-1}\left(1-p_{*,j}\right)$. If $p_{*,i}=1$,
	then terminate, otherwise continue to the step $i+1$.
\end{enumerate}
Recall that theorem~\ref{thm:probability balance}
guarantees the condition $\left(\alpha_{i}^{\dagger}|\sigma_{i}\right)=0$ at every step of the procedure.
Since by construction every state $\alpha_{j}$ with $j>i$ is contained
in $\sigma_{i}$, we also have $\left(\alpha_{i}^{\dagger}|\alpha_{j}\right)=0$
for every $j>i$. Hence, corollary~\ref{cor:dagger distinguishable}
implies that the states $\left\{ \alpha_{k}\right\} _{k=1}^{i}$,
generated by the first $i$ iterations of the protocol, are perfectly
distinguishable, for any $i$. For a finite-dimensional system, this
means that the procedure must terminate in a finite number of iterations.  
Once the procedure has been completed, the state $\rho$ is decomposed 
as $\rho=\sum_{i=1}^r  p_{i} \alpha_{i}$ where $r$ is some finite integer and  $\left\{\alpha_i\right\}_{i=1}^r$  are perfectly distinguishable pure states.
\qed

\subsection{Proof of theorem~\ref{theo:schmidt}\label{app:schmidt}}
Let $\rho_\rA$ be the marginal of $\Psi$ on system $\rA$ and let $  \rho_\rA  =  \sum_{i  =1}^r  p_i   \alpha_i  $  be a diagonalization of $\rho_\rA$, with the probabilities $\left\{p_i\right\}$ arranged in non-increasing order and with $p_r>0$.    By Pure Steering, there exists an observation-test on $\rB$, call it $\left\{\widetilde {b}_i\right\}_{i=1}^r$, such that 
\[
\begin{aligned}\Qcircuit @C=1em @R=.7em @!R { & \multiprepareC{1}{\Psi}    & \qw \poloFantasmaCn{\rA} &  \qw   \\  & \pureghost{\Psi}    & \qw \poloFantasmaCn{\rB}  &   \measureD{\widetilde {b}_i} }\end{aligned}~=p_{i}\!\!\!\!\begin{aligned}\Qcircuit @C=1em @R=.7em @!R { & \prepareC{\alpha_i}    & \qw \poloFantasmaCn{\rA} &  \qw   }\end{aligned} \qquad \forall i\in \left\{ 1,\ldots, r\right\}  .
\]
On the other hand, by corollary~\ref{cor:computeegv}, the pure observation-test $ \left\{\alpha_i^\dag\right\}_{i=1}^{n_\rA} $,  induces pure states on system $\rB$,  as follows  
\begin{equation}\label{sterb}
\begin{aligned}\Qcircuit @C=1em @R=.7em @!R { & \multiprepareC{1}{\Psi}    & \qw \poloFantasmaCn{\rA} &  \measureD{\alpha_i^\dagger}   \\  & \pureghost{\Psi}    & \qw \poloFantasmaCn{\rB}  &   \qw}\end{aligned}~=p_{i}\!\!\!\!\begin{aligned}\Qcircuit @C=1em @R=.7em @!R { & \prepareC{\beta_i}    & \qw \poloFantasmaCn{\rB} &  \qw   }\end{aligned}~,
\end{equation}
where each state $\beta_i$ is pure and normalized. Combining the two equations above, we obtain
\[
\left( \widetilde{b}_j  |   \beta_i\right)=\frac 1 {p_i}\!\!\!\!\begin{aligned}\Qcircuit @C=1em @R=.7em @!R { & \multiprepareC{1}{\Psi}    & \qw \poloFantasmaCn{\rA} &  \measureD{\alpha_i^\dagger}   \\  & \pureghost{\Psi}    & \qw \poloFantasmaCn{\rB}  &  \measureD{\widetilde{b}_j}}\end{aligned}=\frac {p_j} {p_i} \left(\alpha_i^\dag| \alpha_j\right)=\delta_{ij},
\]
for all $ i,j  \in \left\{1,\ldots, r\right\}  $.
Hence, the states $\left\{ \beta_i \right\}_{i=1}^r$ are perfectly distinguishable.   
If $\rho_\rB$ is the marginal of $ \Psi $ on system $ \rB $, we have seen that the pure observation-test $ \left\{\alpha_i^\dag\right\}_{i=1}^{n_\rA} $ induces a diagonalisation of $\rho_\rB$ in terms of the states  $\left\{ \beta_i \right\}_{i=1}^r$. By corollary~\ref{cor:dagger distinguishable}, we know that the effects  $\left\{ \beta_i^\dagger \right\}_{i=1}^r$ are such that
\[
\delta_{ij}=\left(  \beta_j^\dag|  \beta_i\right)=\frac 1 {p_i}\!\!\!\!\begin{aligned}\Qcircuit @C=1em @R=.7em @!R { & \multiprepareC{1}{\Psi}    & \qw \poloFantasmaCn{\rA} &  \measureD{\alpha_i^\dagger}   \\  & \pureghost{\Psi}    & \qw \poloFantasmaCn{\rB}  &  \measureD{\beta_j^\dagger}}\end{aligned}~.
\]

Choosing $\left\{a_i\right\}$ and $\left\{b_j\right\}$  to be  pure measurements with 
\[  a_i   := \alpha_i^\dag  \qquad \textrm {and}  \qquad  b_i  :  = \beta_i^\dag   ,\]
respectively, one obtains Eq.~\eqref{eq:schmidt}.
Recall  that $\rho_A  =   \sum_i p_i \alpha_i$ is a diagonalization of $\rho_\rA$.  Moreover,  Eq.~\eqref{sterb} implies the equality  $\rho_\rB  =  \sum_i p_i  \beta_i$. Since the states $\left\{\beta_i\right\}$ are pure and perfectly distinguishable, this is a diagonalization  of $\rho_\rB$. \qed   

\subsection{Proof of proposition~\ref{prop:maximal}\label{app:maximal}}
The proof is by contradiction. Suppose that  $\left\{\alpha_i\right\}_{i=1}^{d+1}$ is a set of perfectly distinguishable pure states and let $\left\{a_i\right\}_{i=1}^{d+1}$ be the observation-test that distinguishes between them.  Then, one must have $\left( a_{d+1} |\chi\right)     =    \frac 1 d \sum_{i=1}^{d}\left( a_{d+1}|  \alpha_i\right)    =  0$.     But $\left( a_{d+1} |\chi\right) =0$ implies $\left( a_{d+1}|\rho\right) =0$ for every $\rho$, since every state is contained in the invariant state, which is complete.  This is in contradiction with the hypothesis $\left( a_{d+1}|\alpha_{d+1}\right) =1$.   Hence, we have proved  that the set     $\left\{\alpha_i\right\}_{i=1}^{d}$ is maximal.

To prove  that     $\left\{\alpha_i^\dag\right\}_{i=1}^d$   is maximal, it is enough to prove that $\left\{\alpha_i^\dag\right\}_{i=1}^d$ is an observation-test, namely $ \sum_{i=1}^{d} \alpha_i^\dag=u $. By proposition~\ref{prop:diagonalization chi d-level} and corollary~\ref{cor:computeegv}, we have
\[
\sum_{i=1}^{d} \left( \alpha_i^\dagger|\chi\right) = 	\sum_{i=1}^{d} \frac{1}{d}=1.
\]
Since $ \chi $ is complete, this means that $ \sum_{i=1}^{d} \left( \alpha_i^\dagger|\rho\right)=1 $ for every normalized state $ \rho\in \mathsf{St}_1 \left(\mathrm{A} \right)  $, whence 	$ \sum_{i=1}^{d} \alpha_i^\dag $ is the deterministic effect $ u $. \qed

\subsection{Proof of proposition~\ref{prop:diagonalization chi d-level 2}\label{app:diagonalization chi d-level 2}}
We know that every pure state is an eigenstate of $\chi$ with maximum eigenvalue. Specifically, we must have  
\begin{equation}\label{sigma1decomp}
\chi  =  \frac 1 d  \alpha_1    +     \frac{d-1}d  \sigma_1
\end{equation} 
for a state $\sigma_1$ that is perfectly distinguishable from $\alpha_1$ (cf.\ theorem~\ref{thm:probability balance}).   The proof proceeds by induction: for $n<  r$, we assume that the invariant state can be decomposed as 
\begin{equation}\label{inductionbasis}  \chi  =  \frac 1 d  \left(     \sum_{i=1}^n      \alpha_i\right)    +     \frac{ d-n}d  \sigma_n, 
\end{equation}
where the states $\left\{ \alpha_i\right\}_{i=1}^n  \cup\left\{\sigma_n\right\}$ are perfectly distinguishable,  and we prove that a decomposition of the same form can be found  for $n+1$.  

To this purpose, we use the relations
\begin{align}\label{same}  \left(\alpha_i^\dag|\chi\right)  =  \frac 1d  \qquad \forall i \in  \left\{1,\dots, r\right\} ,\end{align}    
following from proposition~\ref{prop:steerpstar} and valid for all normalized pure effects,  and 
\begin{align}\label{alfalfa}  \left(\alpha_i^\dag|\alpha_j\right)  =  \delta_{ij}  \qquad \forall i,j  \in  \left\{1,\dots, r\right\} \end{align}
following from the assumption that the states $ \left\lbrace \alpha_1,\ldots,\alpha_r\right\rbrace  $ are perfectly distinguishable  (cf.\ corollary~\ref{cor:dagger distinguishable}). 
Eqs.~\eqref{inductionbasis},  \eqref{same}, and \eqref{alfalfa} yield the relation
\[ \frac 1d   =  \left(\alpha^\dag_{n+1} |  \chi  \right)    =      \frac{d-n}d    \left(\alpha_{n+1}^\dag | \sigma_n\right) ,
\]
or equivalently,  
\begin{align}\label{sigman}     \left(\alpha_{n+1}^\dag | \sigma_n\right)        =  \frac 1{d-n} .  
\end{align}
Hence, by proposition~\ref{prop:pstar=pstar}, the maximum eigenvalue of $\sigma_n$ is greater than or equal to  $\frac 1{d-n}$.  In fact, it must be equal to  $\frac 1{d-n} $, because otherwise the corresponding eigenstate $\alpha$ would lead to the contradiction, recalling Eq.~\eqref{inductionbasis}   
\[     \frac 1d   =   \left( \alpha^\dag   | \chi\right)     \ge   \frac{d- n}d  \left(\alpha^\dag | \sigma_n\right)    >  \frac 1d .  \]
Hence, Eq.~\eqref{sigman} and corollary~\ref{cor:pupstar} imply that $\alpha_{n+1}$ is an eigenstate of $\sigma_n$ with maximum eigenvalue. Therefore, $\sigma_n$ can be decomposed as $\sigma_n  =  \frac1{d-n}  \alpha_{n+1}  +    \frac{ d-n-1}{d-n}   \sigma_{n+1} $, where the states $\alpha_{n+1}$ and $\sigma_{n+1}$ are perfectly distinguishable.   
Inserting this relation into Eq.~\eqref{inductionbasis} we obtain  
\[\chi    =  \frac 1 d    \left(   \sum_{i=1}^{n+1} \alpha_i  \right)    +    \frac{ d-n-1}{d}     \sigma_{n+1}   .   \]
Now, since the states $\left\{\alpha_i\right\}_{i=1}^n  \cup\left\{\sigma_{n}\right\}$ are perfectly distinguishable, so are the states $\left\{\alpha_i\right\}_{i=1}^{n+1}  \cup\left\{\sigma_{n+1}\right\}$.   This proves the validity of Eq.~\eqref{inductionbasis} for every $n   < r$.      To conclude the proof, consider Eq.~\eqref{inductionbasis} for $n=  r$.    The condition that set $\left\{\alpha_i\right\}_{i=1}^r$ is maximal implies that the state $\sigma_{r}$ should not arise in the decomposition---which is  possible only if the corresponding probability is zero---that is, if  one has $  r  =d$.      Let us now prove the converse:   suppose that the invariant state can be decomposed as $\chi=\frac{1}{d}\sum_{i=1}^{d}\alpha_{i}$.   Then, one has $ \left(\alpha_j^\dag|\chi\right)   = \frac{1}{d}$ and   consequently  $ \left(\alpha_j^\dag|\alpha_i\right)   = \delta_{ij}$. By corollary~\ref{cor:dagger distinguishable}, the states $\left\{\alpha_i\right\}_{i=1}^d$ are perfectly distinguishable.  Since they are $d$ states, they  form a maximal set, because we have just proved that all maximal sets of perfectly distinguishable pure states have the same cardinality $ d $.  
\qed

\subsection{Proof of proposition~\ref{prop:naimark}\label{app:naimark}}

Let $\rB$ be a system of dimension  $d_\rB =   \left| \set X\right| $  (this condition can always be met by fictitiously adding dummy outcomes to the original test $\left\{a_i\right\}_{i\in\set X}$).      Consider the channel  $\map C  \in  \Transf  \left( \rA,   \rB\right) $ defined by  
\[  \map C      :  =      \sum_{i\in\set X}      \beta_i      a_i  ,   \] 
where  $\left\{\beta_i\right\}_{i\in\set X}$ are perfectly distinguishable pure states of $\rB$.  Now, Purification implies that the channel $\map C$ can be realized as 

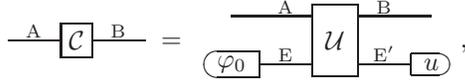
\[
\begin{aligned}\Qcircuit @C=1em @R=.7em @!R { & & \qw \poloFantasmaCn{\rA} & \gate{\map C}& \qw \poloFantasmaCn{\rB}& \qw}\end{aligned} ~= \!\!\!\! \begin{aligned}\Qcircuit @C=1em @R=.7em @!R { & & \qw \poloFantasmaCn{\rA} & \multigate{1}{\map U} & \qw \poloFantasmaCn{\rB} &\qw \\ & \prepareC{\varphi_0} & \qw \poloFantasmaCn{\rE} &\ghost{\map U} & \qw \poloFantasmaCn{\rE'} & \measureD{u} }\end{aligned}~,
\]
where $\varphi_0$ is a pure state of a suitable system $\rE$ and $\map U $ is a reversible channel from $\rA\otimes \rE $ to $\rB\otimes\rE'$, $\rE'$ being a suitable system.  Now, define  the transformations

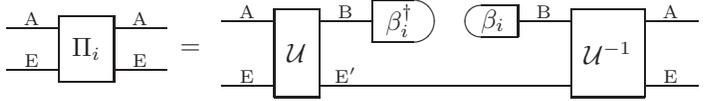
\[
\begin{aligned}\Qcircuit @C=1em @R=.7em @!R { & & \qw \poloFantasmaCn{\rA} & \multigate{1}{\Pi_i}& \qw \poloFantasmaCn{\rA}& \qw\\& & \qw \poloFantasmaCn{\rE} & \ghost{\Pi_i}& \qw \poloFantasmaCn{\rE}& \qw}\end{aligned} ~= \!\!\!\! \begin{aligned}\Qcircuit @C=1em @R=.7em @!R { && \qw \poloFantasmaCn{\rA} & \multigate{1}{\map U} & \qw \poloFantasmaCn{\rB} &\measureD{\beta_i^\dagger} &\prepareC{\beta_i} & \qw \poloFantasmaCn{\rB}  & \multigate{1}{\map U^{-1}} & \qw \poloFantasmaCn{\rA} &\qw\\ &  & \qw \poloFantasmaCn{\rE} &\ghost{\map U} & \qw \poloFantasmaCn{\rE'} &\qw &\qw & \qw  &\ghost{\map U^{-1}} & \qw \poloFantasmaCn{\rE} & \qw}\end{aligned}~,
\] 
Thanks to Purity Preservation, each $\Pi_i$ is a pure transformation.   Moreover, one has 
\[
\Pi_i\Pi_j=\delta_{ij}\begin{aligned}\Qcircuit @C=1em @R=.7em @!R { && \qw \poloFantasmaCn{\rA} & \multigate{1}{\map U} & \qw \poloFantasmaCn{\rB} &\measureD{\beta_j^\dagger} &\prepareC{\beta_i} & \qw \poloFantasmaCn{\rB}  & \multigate{1}{\map U^{-1}} & \qw \poloFantasmaCn{\rA} &\qw\\ &  & \qw \poloFantasmaCn{\rE} &\ghost{\map U} & \qw \poloFantasmaCn{\rE'} &\qw &\qw & \qw  &\ghost{\map U^{-1}} & \qw \poloFantasmaCn{\rE} & \qw}\end{aligned}~=\delta_{ij}\Pi_i,
\]
and
\[
\begin{aligned}\Qcircuit @C=1em @R=.7em @!R { & & \qw \poloFantasmaCn{\rA} & \multigate{1}{\Pi_i} & \qw \poloFantasmaCn{\rA} &\measureD{u} \\ & \prepareC{\varphi_0} & \qw \poloFantasmaCn{\rE} &\ghost{\Pi_i} & \qw \poloFantasmaCn{\rE} & \measureD{u} }\end{aligned}~=
\] 
\[
=\!\!\!\!\begin{aligned}\Qcircuit @C=1em @R=.7em @!R { && \qw \poloFantasmaCn{\rA} & \multigate{1}{\map U} & \qw \poloFantasmaCn{\rB} &\measureD{\beta_i^\dagger} &\prepareC{\beta_i} & \qw \poloFantasmaCn{\rB}  & \multigate{1}{\map U^{-1}} & \qw \poloFantasmaCn{\rA} &\measureD{u}\\ &\prepareC{\varphi_0}  & \qw \poloFantasmaCn{\rE} &\ghost{\map U} & \qw \poloFantasmaCn{\rE'} &\qw &\qw & \qw  &\ghost{\map U^{-1}} & \qw \poloFantasmaCn{\rE} & \measureD{u}}\end{aligned}~=
\]
\[
=\!\!\!\!\begin{aligned}\Qcircuit @C=1em @R=.7em @!R { & & \qw \poloFantasmaCn{\rA} & \multigate{1}{\map U} & \qw \poloFantasmaCn{\rB} &\measureD{\beta_i^\dagger} \\ & \prepareC{\varphi_0} & \qw \poloFantasmaCn{\rE} &\ghost{\map U} & \qw \poloFantasmaCn{\rE'} & \measureD{u} }\end{aligned}~=\beta_i^{\dagger}\mathcal{C}=a_i
\]
\qed 
   
\subsection{Proof of proposition~\ref{prop:completely-mixed eigenvalues}\label{app:completely-mixed eigenvalues}}
Consider a complete state $\omega$ and one of its diagonalizations
$\omega=\sum_{i=1}^{r}p_{i}\alpha_{i}$, where $ r\leq d $, and the $ p_i $'s are non-vanishing, for all $ i\in\left\lbrace 1,\ldots,r\right\rbrace  $. Suppose by contradiction that $ r<d $; this means that the states $ \left\lbrace\alpha_{i} \right\rbrace_{i=1}^{r} $ do not form a maximal set, and therefore we can complete it by adding $ d-r $ states $ \left\lbrace \alpha_{i}\right\rbrace_{i=r+1}^d$. In this way we can rewrite the diagonalization of $ \omega $ as  $\omega=\sum_{i=1}^{d}p_{i}\alpha_{i}$, where $ p_i=0 $ for $ i\in\left\lbrace r+1,\ldots,d\right\rbrace  $, and the states $\left\lbrace \alpha_i\right\rbrace_{i=1}^{d}$ are a maximal set of perfectly distinguishable pure states. Take any $ \alpha_i $ with $ i\in\left\lbrace r+1,\ldots,d\right\rbrace  $; we have
\begin{equation}\label{eq:omega}
0=p_{i}=\left( \alpha_i^\dagger|\omega\right) .
\end{equation}
On the other hand, $ \omega $ is complete, therefore there exists a non-vanishing $ \lambda_i\in\left( 0,1\right]  $ such that
\begin{equation}\label{eq:omega_lambda}
\omega= \lambda_i\alpha_i+\left( 1-\lambda_i\right) \rho_i,
\end{equation}
where $ \rho_i $ is a suitable normalized state. Hence, applying $ \alpha_i^\dagger $ to Eq.~\eqref{eq:omega_lambda}, the LHS vanishes by Eq.~\eqref{eq:omega}, so we have the contradiction
\[
0=\left(\alpha_{i}^\dagger|\omega\right)=\lambda_i+\left(1-\lambda_i\right)\left(\alpha_{i}^\dagger|\rho_i\right),
\]
where the RHS is strictly positive. Hence we conclude that $ r=d $, thus the pure states arising in any diagonalization of $ \omega $ form a maximal set.
\qed

\subsection{Proof of proposition~\ref{prop:completely mixed converse}\label{app:completely mixed converse}}
Consider a convex combination of the $ \alpha_i $'s with non-vanishing coefficients $ p_i $: $ \omega=\sum_{i=1}^{d}p_i\alpha_i $. Let $p_{\mathrm{min}} = \min \left\{p_i: i = 1,\ldots,d\right\}$, and note that $ p_{\mathrm{min}}\neq1 $. Define
\[
\sigma:=\frac{1}{1-p_{\mathrm{min}}}\sum_{i=1}^{d}\left( p_i-\frac{p_{\mathrm{min}}}{d}\right) \alpha_i
\]
It is easy to check that $ \omega=p_{\mathrm{min}}\chi+ \left( 1-p_{\mathrm{min}}\right) \sigma$, where $ \chi $ is the invariant state. Since $ \omega $ contains $ \chi $, which is complete, in its convex decomposition, we conclude that $ \omega $ is complete too.
\qed

\subsection{Proof of proposition~\ref{prop:pure test chi}\label{app:pure test chi}}
Let  $\left\{ \alpha_{i}\right\} _{i=1}^{d}$ be a pure maximal set.  
By proposition~\ref{prop:diagonalization chi d-level 2}, we know
that $\frac{1}{d}\sum_{i=1}^{d}\alpha_{i}$ is a diagonalization of the invariant state $ \chi $.   Then, proposition~\ref{prop:maximal} implies that $\left\{ \alpha_{i}^\dag\right\} _{i=1}^{d}$ is a maximal set of coexisting effects, in which every $\alpha_i^\dag$ is pure and normalized.

Conversely, suppose that $\left\{a_i\right\}_{i=1}^n$ is a pure sharp measurement.    Since each $a_i$ is pure there is a unique pure state  $\alpha_i$ associated with it---in other words, $ a_i  =  \alpha_i^\dag  $.
Clearly, the  measurement  $\left\{a_i\right\}_{i=1}^n$  distinguishes perfectly between the states $\left\{\alpha_i\right\}_{i=1}^n$.  
Moreover, the states $\{\alpha_i\}_{i=1}^n$  must form a pure maximal set. This can be proved by contradiction:  suppose   the set $\left\{\alpha_i\right\}_{i=1}^n$  is not maximal, and 
extend it to a maximal set  $\left\{\alpha_i\right\}_{i=1}^{d}$.    Then,  by the first part of this proof we have that $\left\{\alpha_i^{\dag}\right\}_{i=1}^d$ is an observation-test.   By Causality, we then obtain  
\[
\sum_{i=1}^d  \alpha_i^\dag     =  u   =      \sum_{i=1}^n   a_i      =    \sum_{i=1}^n    \alpha_i^\dag    , 
\]  
having used the equality $a_i  = \alpha_i^\dag$  following from the condition $\left(a_i|\alpha_i\right)=1$.    In conclusion, we obtained the relation $\sum_{i=n+1}^d  \alpha_i^\dag  =  0$, which can be satisfied  only if $n=d$.  Hence, the states  $\left\{\alpha_i\right\}_{i=1}^d$ form a pure maximal set. 
\qed

\subsection{Proof of theorem~\ref{thm:uniqueness diago}\label{app:uniqueness diago}}
Let the two diagonalizations be $ \rho=\sum_{i}p_i\alpha_{i} $ and $ \rho=\sum_{j}q_j\alpha'_j $. By definition, we have $\lambda_1   =  \lambda_1'  =  p_*$, the maximum eigenvalue of $\rho$.   Let us define the degeneracies  
\[  d_1  =    \left|  \left\{  i  :     p_i   =  \lambda_1\right\} \right| , \qquad d'_1  =    \left|  \left\{  j  : q_j   =  \lambda_1\right\} \right|,\]
and assume $d_1  \ge d_1'$ without loss of generality.  
By definition, we have for $ i \in\left\{1,\ldots , d_1\right\} $
\[
p_*  =   \left(\alpha^\dag_i|  \rho\right)   = \sum_{j}    q_j   \left(\alpha_i^\dag|\alpha_j' \right)   =   \sum_j  T_{ij}   q_j     \le p_* ,
\]
having used the fact that the transition matrix $T_{ij}=\left(\alpha_i^\dag|\alpha_j' \right)$ is  doubly stochastic. The above relation implies the equality
\[  \sum_{j=1}^{d_1'}   \left(\alpha_i^\dag  |  \alpha_j'\right)  =1   \qquad \forall i\in\left\{1,\dots ,  d_1\right\}  , \] 
or, equivalently,
\[       \left(\alpha_i^\dag  |  \chi_1'\right)   =  \frac 1{d_1'}      \qquad \forall i\in\left\{1,\dots ,  d_1\right\} , \] 
where $\chi_1'  :  =  \frac{1}{d_1'} \Pi'_1 $.  Note that $\frac{1}{d_1'} $ is the maximum eigenvalue of $\chi_1'$ because the states $ \left\lbrace\alpha'_j \right\rbrace_{j=1}^{d'_1}  $ are perfectly distinguishable and, therefore corollary~\ref{cor:pupstar} implies that $\alpha_i$ is an eigenstate with maximum eigenvalue.      In particular, choosing $i=1$  we obtain  the decomposition  
\[  \chi_1'   =    \frac 1 {d_1'}   \alpha_1     +  \frac {d_1'-1} {d_1'}  \sigma_1  ,\]
where $\sigma_{1}$ is a suitable state, perfectly distinguishable from $\alpha_1$.       We are now in the position  to  repeat the argument in the proof of proposition~\ref{prop:diagonalization chi d-level 2} for the states $ \left\lbrace \alpha_i\right\rbrace_{i=1}^{d_1}  $, to find that $ d_1= d_1'$ and  
\[  \chi_1'   =  \frac1 {d_1}  \sum_{i=1}^{d_1}   \alpha_1    \equiv     \frac1{d_1}\Pi_1.  \]  
Hence, we proved the equality $\Pi_1'  =  \Pi_1$.     We can now define the state 
\begin{align*}\rho_2  &:=      \frac1{1-\lambda_1}   \left(  \rho   -   \lambda_1  \Pi_1\right)  =   \frac1{1-\lambda_1}  \left( \sum_{k=2}^s   \lambda_k   \Pi_k \right)   =\\& =   \frac1{1-\lambda_1}     \left( \sum_{l=2}^{s'}   \lambda'_l   \Pi_l'   \right).    	\end{align*}
Repeating the above argument, we can prove the equalities $ \lambda_2  =  \lambda_2' $ and $\Pi_2  =  \Pi_2'$.  Once all distinct eigenvalues have been scanned, the normalization of the probability distribution implies the condition $s=s'$.     \qed

\subsection{Proof of proposition~\ref{prop:well-defined}\label{app:well-defined}}
Let us extend   $\left\{\alpha_i\right\}_{i=1}^r$ and $\left\{\alpha_j'\right\}_{j=1}^r$  to two maximal   sets  $\left\{\alpha_i\right\}_{i=1}^d$ and $\left\{\alpha_j'\right\}_{j=1}^d$.  Then, the invariant state has the two diagonalizations $ \chi  =  \frac  1d  \sum_{i=1}^d \alpha_i  $ and $ \chi=    \frac 1d  \sum_{j=1}^d \alpha'_j $
(proposition~\ref{prop:diagonalization chi d-level 2}).   Using this fact and the condition  $ \sum_{i=1}^r \alpha_i  =  \sum_{j=1}^r \alpha_j' $, we obtain  $  \sum_{i=r+1}^d \alpha_i  =  \sum_{j=r+1}^d \alpha_j' $.  
Hence, the invariant state can be decomposed as \[\chi   =  \frac 1d     \left(  \sum_{j=1}^r \alpha'_j      +   \sum_{i=r+1}^d    \alpha_i \right) .  \] 
By proposition \ref{prop:diagonalization chi d-level 2}, this implies that the states $\left\{\alpha'_j\right\}_{j=1}^r  \cup \left\{\alpha_i\right\}_{i=r+1}^d$ form a maximal set of perfectly distinguishable pure states.  Now, the correspondence between maximal sets of pure states and pure sharp measurements implies that the effects    $\left\{\alpha^{\prime \dag}_j\right\}_{j=1}^r  \cup \left\{\alpha^\dag_i\right\}_{i=r+1}^d$ form a measurement.   Causality implies 
\[    \sum_{j=1}^r \alpha^{\prime \dag}_j      +   \sum_{i=r+1}^d   \alpha^\dag_i     =        u  . \]
On the other hand,  the normalization of the measurement $\left\{  \alpha_i^\dag\right\}_{i=1}^d$ reads  
\[  \sum_{i=1}^d  \alpha_i^\dag    =  u  .\]
Comparing the two equalities we obtain  the desired relation  
$   \sum_{i=1}^r \alpha^\dag_i  =  \sum_{j=1}^r \alpha^{\prime \dag}_j $.  
\qed

\section{Operational characterization of the eigenvalues\label{app:operational eigenvalues}}
First we need a lemma on the structure of pure observation-tests.
\begin{lemma}
	\label{lem:characterization of pure tests}Let $\left\{ a_{i}\right\} _{i=1}^{n}$
	be a pure observation-test. Then, for every $ i\in\left\lbrace 1,\ldots,n\right\rbrace  $, $a_{i}=\lambda_{i}\alpha_i^\dagger$, for some normalized pure state $\alpha_i$, and  $\lambda_{i}\in\left(0,1\right]$. 
	Moreover $\sum_{i=1}^{n}\lambda_{i}=d$, and $n\geq d$. One has
	and $n=d$ if and only if $\left\{ a_{i}\right\} _{i=1}^{n}$ is a perfectly
	distinguishing test.\end{lemma}
\Proof 	We can always write a pure effect $ a_i $ as $ a_i=\left\| a_i\right\| a'_i $, where $ a'_i $ is a \emph{normalized} pure effect Hence we can find a pure state $ \alpha_i $ such that $ a'_i=\alpha_i^\dagger $. Recall that for all physical effects $ \left\| a_i\right\|\in\left( 0,1\right]  $, so we can take $ \lambda_i:= \left\| a_i\right\|$. Now let us prove that $\sum_{i=1}^{n}\lambda_{i}=d$.
Indeed, by Causality, $\sum_{i=1}^{n}\lambda_{i}\alpha_{i}^{\dagger}=u$. Now consider
\[
1=\mathrm{Tr}\:\chi=\sum_{i=1}^{n}\lambda_{i}\left(\alpha_{i}^{\dagger}|\chi\right)=\sum_{i=1}^{n}\lambda_{i}\cdot\frac{1}{d},
\]
whence $\sum_{i=1}^{n}\lambda_{i}=d$. Since $\lambda_{i}\leq1$,
we have 
\[
d=\sum_{i=1}^{n}\lambda_{i}\leq\sum_{i=1}^{n}1=n,
\]
so $ n\geq d $.

Now let us prove that $ n=d $ if and only if $\left\{ a_{i}\right\} _{i=1}^{n}$ is a perfectly distinguishing test.
Suppose $\left\{ a_{i}\right\} _{i=1}^{n}$ is a pure perfectly distinguishing
test, then $n=d$, otherwise by proposition~\ref{prop:pure test chi} there would be $n>d$ perfectly
distinguishable pure states. Conversely, suppose we know that $n=d$.
Then in this case, the only possibility of having $\sum_{i=1}^{n}\lambda_{i}=d$
is when $\lambda_{i}=1$ for every $i$. Therefore all the effects
are normalized and the observation-test can be rewritten as $ \left\{ \alpha_{i}^\dagger\right\} _{i=1}^{d} $ for some pure states $\left\{ \alpha_{i}\right\} _{i=1}^{d}$, which are perfectly distinguished by the test considered.
\qed

\subsection{Proof of proposition~\ref{prop:majorization measurement}}
By lemma~\ref{lem:characterization of pure tests}, for each $ a_i\in \left\lbrace a_i\right\rbrace_{i=1}^{n} $, we have $ a_i=\lambda_i\alpha_i^\dagger $, for some $ 0<\lambda_i \leq 1$, and for some pure state $ \alpha_i $. Consider a diagonalization of $\rho=\sum_{j=1}^{d}p_{j}\alpha'_{j}$.
We have 
\[
q_{a,i}:=\left(a_{i}|\rho\right)=\sum_{j=1}^{d}p_{j}\left(a_{i}|\alpha'_{j}\right)=\sum_{j=1}^{d}\lambda_{i}p_{j}\left(\alpha_{i}^\dagger|\alpha'_{j}\right)
\]
Now, $M_{ij}:=\lambda_{i}\left(\alpha_{i}^\dagger|\alpha'_{j}\right)$ is a
$n\times d$ matrix such that $ q_{a,i}=\sum_{j=1}^{d}M_{ij}p_j $. Clearly $M_{ij}\geq0$ for all $i$, $j$. Calculating $\sum_{i=1}^{n}M_{ij}$, we have
\begin{equation}\label{eq:sum column M}
\sum_{i=1}^{n}M_{ij}=\sum_{i}^{d}\left(\lambda_{i}\alpha_{i}^\dagger|\alpha'_{j}\right)=\mathrm{Tr}\:\alpha'_{j}=1,
\end{equation}
whence the column of the matrix $M$ sum to 1. Now let us move to
$\sum_{j=1}^{d}M_{ij}$.
\begin{equation}\label{eq:sum row M}
\sum_{j=1}^{d}M_{ij}=\lambda_{i}\sum_{j=1}^{d}\left(\alpha_{i}^\dagger|\alpha'_{j}\right)=\lambda_{i}d\left(\alpha_{i}^\dagger|\chi\right)=\lambda_{i}d\cdot\frac{1}{d}=\lambda_{i}\leq1
\end{equation}
We wish to construct an $n\times n$ doubly stochastic matrix $D$
from $M$, such that we can write $q_{a,i}=\sum_{j=1}^{n}D_{ij}\widetilde{p}_{j}$,
where $\widetilde{p}$ is the vector of probabilities defined as
\[
\widetilde{p}_{j}:=\begin{cases}
p_{j} & 1\leq j\leq d\\
0 & d+1\leq j\leq n
\end{cases}.
\]
Let us define $D$ as
\[
D:=\left(\begin{array}{c|c}
M & \frac{1-\lambda_{i}}{n-d}\end{array}\right).
\]
Now, $D$ is doubly stochastic. Indeed each entry is non-negative, because $ \lambda_i\leq1 $ for all $ i=1,\ldots,n $ and $ n\geq d $. Furthermore,
\[
\sum_{i=1}^{n}D_{ij}=\begin{cases}
\sum_{i=1}^{n}M_{ij} & 1\leq j\leq d\\
\frac{n-\sum_{i=1}^{n}\lambda_{i}}{n-d} & d+1\leq j\leq n
\end{cases}=1
\]
by Eq.~\eqref{eq:sum column M}, and because $\sum_{i=1}^{n}\lambda_{i}=d$ (by lemma~\ref{lem:characterization of pure tests}). 
Finally
\[
\sum_{j=1}^{n}D_{ij}=\sum_{j=1}^{d}M_{ij}+\sum_{j=d+1}^{n}\frac{1-\lambda_{i}}{n-d}=1,
\]
having used Eq.~\eqref{eq:sum row M}.
Clearly now we have $q_{a,i}=\sum_{j=1}^{n}D_{ij}\widetilde{p}_{j}$,
because, by construction of $\widetilde{\mathbf{p}}$ and $D$, 
\[
\mathbf{q_a}=\left(\begin{array}{c|c}
M & \frac{1-\lambda_{i}}{n-d}\end{array}\right)\left(\begin{array}{c}
\mathbf{p}\\
\hline \mathbf{0}
\end{array}\right).
\]
Therefore $\mathbf{q_a}\preceq\widetilde{\mathbf{p}}$. \qed

\section{Proof of theorem~\ref{thm:measurement=preparation}\label{app:measurement=preparation}}
Let us prove that $M_{f}^{\rm meas}  $ coincides with $M_f$.     Let $\rho=\sum_{i=1}^{d}p_{i}\alpha_{i}$ be a diagonalization  of $ \rho $.   If we take the pure sharp measurement $ \left\lbrace \alpha_{i}^\dagger\right\rbrace_{i=1}^{d}  $, we have $ \left( \alpha_i^{\dagger}|\rho\right) =p_i $.  Hence,  
\[
M_{f}^{\rm meas}  \left( \rho\right)   \le   f\left(   {\bf p}\right)   =   M_f \left( \rho\right) .
\] 
To prove the converse, recall proposition~\ref{prop:majorization measurement}:  for every pure observation-test $ \left\lbrace a_i\right\rbrace  $, one has $ \mathbf{q}\preceq  \widetilde{ \mathbf{p}} $, where $ \mathbf{q} $ is the vector of probabilities $q_i  = \left( a_i|  \rho\right)     $ and $ \widetilde{ \mathbf{p}} $ is the vector of the eigenvalues of $ \rho $  (with additional zeros appended, if needed). Since $ f $ is Schur-concave, we  have   $f\left( \bf q\right)   \ge f\left( \widetilde{ \mathbf{p}} \right) $ and, taking the infimum over all pure measurements
\[  M_{f}^{\rm meas}  \left( \rho\right)      \ge   f\left( \widetilde{ \mathbf{p}}\right) =f\left(  \mathbf{p}\right)  =  M_f \left( \rho\right),\]
where we have used the fact that $ f $ is reducible.
Summarizing,   we obtained the equality $M_{f}^{\rm meas}    = M_f$.  

We now prove  the equality $M_{f}^{\rm prep}    = M_f$.  By definition,  we have 
\[   M_f^{\rm prep}  \left( \rho\right)   \le  f\left( \st p\right)    = M_f \left( \rho\right) ,\]
because the diagonalization is a special case of pure state decomposition.  The converse inequality follows from Pure Steering.      Consider a purification of $\rho\in \mathsf{St}_1\left( \mathrm{A}\right)$, say $\Psi\in \mathsf{PurSt}_1\left( \mathrm{A}\otimes\mathrm{B}\right)$.   Consider a \emph{pure} observation-test  $\left\{  b_i\right\}$ on system $ \mathrm{B} $, it will induce a decomposition of $ \rho_{\mathrm{A}} $ into pure states $\rho_{\mathrm{A}}  =  \sum_i  \pi_i \alpha_i$.
\[
\begin{aligned}\Qcircuit @C=1em @R=.7em @!R { & \multiprepareC{1}{\Psi}    & \qw \poloFantasmaCn{\rA} &  \qw   \\  & \pureghost{\Psi}    & \qw \poloFantasmaCn{\rB}  &   \measureD{b_i}}\end{aligned}~=\pi_{i}\!\!\!\!\begin{aligned}\Qcircuit @C=1em @R=.7em @!R { & \prepareC{\alpha_i}    & \qw \poloFantasmaCn{\rA} &  \qw   }\end{aligned}~,
\] 
Discarding system $\rA$  on both sides we obtain 
\[       \left(   b_i|  \rho_\rB\right)        =\pi_i ,  \]
where $\rho_\rB$  is the marginal state on system  $\rB$.    In other words,  $\bs  \pi$ is the vector of the outcome probabilities for the pure measurement   $\{b_i\}$.    By definition of measurement monotone, we must have 
\[   f\left( {\bs \pi}\right)    \ge  M_f^{\rm meas}  \left( \rho_\rB\right)        \]
and, taking the infimum over all pure state decompositions  
\[  M_f^{\rm prep}   \left( \rho_\rA\right)   \geq  M_f^{\rm meas}  \left( \rho_\rB\right) . \]  
To conclude, it is enough to recall the equalities   $M_f^{\rm meas}  \left( \rho_\rB\right)      =  M_f \left( \rho_\rB\right) $ and $M_f  \left( \rho_\rB\right)   =  M_f\left( \rho_\rA\right)   \equiv M_f \left( \rho\right) $.   \qed

\section{Proof of lemma~\ref{lem:Klein}\label{app:Klein}}
Let $\rho=\sum_{i=1}^{d}p_{i}\alpha_{i}$
and $\sigma=\sum_{i=1}^{d}q_{i}\alpha'_{i}$ be   diagonalizations of $\rho$ and $\sigma$ respectively. Now, let
us compute $S\left(\rho\parallel\sigma\right)$ explicitly.   
Assume that  all the eigenvalues of $\rho$ and $\sigma$ are non-zero, as the result in the general case can be obtained by using the continuity of the logarithm function.    
Hence, 
\[
\left(\log \rho^\dagger|\rho\right)=\sum_{i=1}^{d}p_{i}\log p_{i},
\]
and
\begin{align*}
\left(  \log \sigma^\dagger|\rho\right)  &=\sum_{i,j=1}^{d}\left(\alpha_{j}'^{\dagger}|\alpha_{i}\right)p_{i}\log q_{j} \\
&  =  \sum_{i,j=1}^d   T_{ij}  \log   q_j  ,
\end{align*}
where $T_{ij}:=\left(\alpha_{j}'^{\dagger}|\alpha_{i}\right ) $ are the entries of a doubly stochastic matrix (lemma~\ref{lem:doubly stochastic}). 
Then
\begin{align}
\nonumber 	S\left(\rho\parallel\sigma\right)  &=\sum_{i=1}^{d}p_{i}\left(\log p_{i}-\sum_{j=1}^{d}T_{ij}\log q_{j}\right) \\
\label{KLbound}	&\le \sum_{i=1}^{d}p_{i}\left(\log p_{i}-\log r_{i}\right) , \qquad \st r  :  =   T  \st q    ,
\end{align}
having used the concavity of the logarithm.    The RHS of the last equality is the classical Kullback-Leibler divergence $D\left( \st p  \parallel \st r\right) $.  Since $D\left( \st p  \parallel \st r\right) $ is always non-negative, we obtained the bound    
\begin{align}
S \left( \rho \parallel \sigma\right)     \ge   D\left( \st p\parallel \st r\right)   \ge 0 .
\end{align}

Moreover, since the classical Kullback-Leibler divergence vanishes if and only if $\st p =  \st r$,  the condition $S \left( \rho \parallel \sigma\right) =0$ implies
\[   p_i  =  \sum_j  T_{ij} q_j ,\] 	for all $ i \in  \left\{1,\ldots, d\right\}$. 
Inserting this equality into Eq.~\eqref{KLbound} we obtain the relation  
\[  0   =  \sum_i  p_i  \left[    \log  \left(  \sum_j  T_{ij}  q_j \right)    -   \sum_j   T_{ij} \log q_j  \right].\] 
Since the logarithm is a strictly concave function, the equality implies that   the entries of $\st p$ are a permutation of the entries of $\st q$, namely 
\[p_i  =  q_{\pi \left( i\right) } \qquad \forall i\in  \left\{1,\ldots, d\right\},\]
where $\pi$ is a suitable permutation.    Since the entries of $\st p$ and $\st q$ are all distinct, the above condition implies that $T$ is a permutation matrix (recall that doubly stochastic matrices are mixtures of permutation matrices).  Hence, we have   
\[   T_{ij}  =  \delta_{j,  \pi\left( i\right) } , \qquad \forall i,j\in\left\{1,\ldots,  d\right\} . \]
Recalling the definition of $T$, we obtain  
\[
T_{ij}  =  \left(\alpha_{j}'^{\dagger}|\alpha_{i}\right)  =\delta_{j,\pi\left(i\right)},
\]
which in turn implies    
\[\alpha_{i}=\alpha'_{\pi\left( i\right) }  ,\] 
for all $ i\in  \left\{1,\ldots, d\right\} $ due to the ``pure state certification'' result (proposition~\ref{prop:uniqueness of state}). 
In conclusion, we have obtained 
\[
\rho  =  \sum_{i}   p_i \alpha_i    =\sum_{i} q_{\pi \left( i\right) } \alpha'_{\pi\left( i\right) }   =   \sigma  .	\]
\qed

\end{document}